\documentclass[preprint,12pt,eqsecnum,nofootinbib,amsmath,amssymb]{revtex4}

\newcommand{\datechange}{7/20/2020}

\newcommand{\prepnumber}{}

\newcommand{\mytitle}{BPS Explained III: Dimensional Leveraging\\
or\\
The Leading Order Behavior of the BBGKY Hierarchy\\
in a Plasma}
\usepackage{graphicx}  % Include figure files
\usepackage{dcolumn}  % Align table columns on decimal point
\usepackage{bm}             % Bold math: $\bm{\alpha}$
\usepackage{latexsym}  % Several additional symbols
\usepackage{fancyhdr}  % Fancy header package
\usepackage{wrapfig}
\usepackage{comment}
\usepackage{dsfont}
\usepackage{mathtools}
\usepackage{datetime}
\newcommand{\smA}{{\scriptscriptstyle \rm A}}
\newcommand{\smB}{{\rm\scriptscriptstyle B}}
\newcommand{\smN}{{\rm\scriptscriptstyle N}}
\newcommand{\smX}{{\rm\scriptscriptstyle X}}

\newcommand{\smL}{{\rm\scriptscriptstyle L}}

\newcommand{\smC}{{\rm\scriptscriptstyle C}}

\newcommand{\smR}{{\rm\scriptscriptstyle R}}

\newcommand{\smD}{{\rm\scriptscriptstyle D}}

\newcommand{\smLT}{{\rm\scriptscriptstyle <}}
\newcommand{\smGT}{{\rm\scriptscriptstyle >}}

\newcommand{\bodyskip}{\baselineskip 18pt plus 1pt minus 1pt}
\newcommand{\bibskip}{\baselineskip16pt plus 1pt minus 1pt}
\newcommand{\tableofcontentsskip}{\baselineskip 14pt plus 1pt minus 1pt}
\newcommand{\footnoteskip}{\baselineskip 12pt plus 1pt minus 1pt}
\newcommand{\abstractskip}{\baselineskip 13pt plus 1pt minus 1pt}
\newcommand{\titleskip}{\baselineskip 18pt plus 1pt minus 1pt}
\newcommand{\affiliationskip}{\baselineskip 15pt plus 1pt minus 1pt}
\newcommand{\captionskip}{\footnotesize \baselineskip 12pt plus 1pt minus 1pt}

\begin{document}

%%% notes info page
%%\hfill{\prepnumber}
%%\vskip0.3cm
%\centerline{{ \Large\bf \projname: \fname}}
%\vskip0.25cm 
%\centerline{\bf \mytitle}
%\vskip0.25cm
%\centerline{\myauthors}
%\vskip0.75cm 
%\baselineskip 14pt plus 1pt minus 1pt
%\begin{flushright}
%Research Notes   \\[3pt]
%{\it Project}:          \\
%\projname                      \\
%  {\it Path of TeX Source}:          \\
%\dirname/\fname                      \\[3pt]
%{\it Last Modified By}:            \\
%\whochange                         \\
%\datechange                        \\[3pt]
%{\it Date Started:}                \\
%\datestart                         \\[3pt]
%{\it Date:}                \\
%\draftverson~ \today ~\currenttime \\
%\end{flushright}

\baselineskip 20pt plus 1pt minus 1pt

%% mini abstract
%\abstractskip
%\noindent
%These are notes on Logic from Ref.~\cite{ref_chang}.  
%\bodyskip

%% title page
\pagebreak
\preprint{\prepnumber}

% publication title page
\title{\titleskip
  \mytitle
}

\author{Robert L Singleton Jr}

\affiliation{\affiliationskip
University of Leeds\\
School of Mathematics\\
UK, LS2 9JT
}

\date{\datechange}

\begin{abstract}
\abstractskip
\vskip0.3cm 
\noindent
This is the third in a series of lectures on the technique of dimensional 
continuation, employed by Brown, Preston and Singleton (BPS), for 
calculating Coulomb energy exchange rates in a plasma. Two important
examples of such processes are the charged particle stopping power 
and the temperature equilibration rate between different plasma 
species. The first lecture was devoted to understanding the machinery 
of dimensional continuation, and the second concentrated on calculating 
the electron-ion temperature equilibration rate in the extreme quantum 
limit where the Born approximation is fully justified.
In this lecture, I will 
examine one of the main theoretical underpinnings of the BPS theory, 
namely, the {\em dimensional reduction} of the BBGKY hierarchy.  There 
are two broad classes of kinetic equations, applicable in complementary 
regimes, represented by the Boltzmann equation (BE) and the 
Lenard-Balescu equation (LBE). The BE describes the short-distance
effects of 2-body scattering, while the LBE models 2-point long-distance 
correlations.  It is well known that the BE suffers a long-distance 
logarithmic divergence (in three spatial dimensions), confirming that 
it is indeed missing long-distance physics (correlations are being 
ignored). Conversely, the LBE suffers from a short-distance 
logarithmic divergence (in three dimensions), another indication that 
relevant physics is being overlooked (the scattering physics). There 
are multiple industries in plasma physics devoted to regulating these 
infinities, thereby giving mathematical and physical meaning to the 
various calculations. To my knowledge, BPS is the only formalism 
that applies a regularization scheme systematically in a perturbative 
expansion of the dimensionless plasma coupling parameter $g$, 
while simultaneously treating short- and long-distance scales 
consistently and in the same manner. A novel aspect of the BPS 
formalism is that it employs dimensional continuation to regulate 
the divergent integrals in the kinetic equations, a procedure first used 
in quantum field theory to regulate divergent integrals during the 
renormalization program. The idea of dimensional continuation is that 
one should perform the integrals in an arbitrary number of spatial
dimensions $\nu$, where, remarkably, the integrals become finite 
(except for $\nu=3$, where we happen to live). The only remembrance 
of the three dimensional divergences are simple poles of the form 
$1/(\nu-3)$. The BPS formalism hinges on the leading order in $g$ 
behavior of the BBGKY hierarchy as a function of the spatial dimension 
$\nu$, both above and below the critical dimension $\nu=3$. In these 
notes, I will prove that to {\em leading} order in~$g$, the BBGKY 
hierarchy reduces to the BE for $\nu > 3$ and to the LBE for $\nu<3$. We 
must eventually return to three dimensions, and the BPS formalism shows 
that the simple poles associated with the BE and the LBE exactly cancel, 
rendering the $\nu \to 3$ limit finite.  Furthermore, the leading order
behavior of the LBE becomes next-to-leading order when $\nu$ is analytically
continued from $\nu < 3$ to $\nu > 3$. This provides the leading and 
next-to-leading order terms in $g$ exactly, which is equivalent to an 
exact calculation of the so-called {\em Coulomb logarithm} with no
use of an integral cut-off. Therefore, in this way, BPS takes all Coulomb 
interactions into account to leading and next-to-leading order in $g$.

\end{abstract}

\maketitle
%%

% to change page settings
%\thispagestyle{empty}
%\pagestyle{empty}
%\setcounter{page}{0}

\pagebreak
\tableofcontentsskip
\tableofcontents
%\thispagestyle{empty}

%\pagebreak
\newpage
\bodyskip

\pagebreak
\clearpage

\section{Introduction}

\begin{figure}[t!]
\begin{picture}(120,120)(-75,0)
\put(-20,80){\oval(100,50)}
\put(-40,90){\text{BBGKY}}
\put(-55,70){(\text{arbitrary } $\nu$) }

\put(-50,55){\vector(-1,-1){30}}
\put(14,55){\vector(1,-1){30}}
\put(-100,40){$\nu\!<\!3$}\
\put(38,40){$\nu\!>\!3$}

\put(50,5){\text{Boltzmann}}
\put(43,-10){\text{Equation (BE)}}
\put(78,0){\oval(100,50)}

\put(-150,7){\text{Lenard-Balescu}}
\put(-150,-8){\text{Equation (LBE)}}
\put(-113,0){\oval(100,50)}

\put(-20,55){\vector(0,-1){80}}
\put(-32,-35){$\nu\!=\!3$}
\put(-45,-60){\text{Break Down}}
\put(-17,-60){\oval(80,35)}

\end{picture}
\vskip3cm 
\caption{\baselineskip12pt plus 1pt minus 1pt For $\nu\!>\!3$ the
``textbook derivation'' of the Boltzmann equation for a Coulomb
potential is rigorous; furthermore, the BBGKY hierarchy reduces to the
Boltzmann equation to leading order in $g$. A similar reduction from
the BBGKY hierarchy holds for the Lenard-Balescu equation in $\nu
\!<\! 3$, and the ``textbook derivation'' is also rigorous in these
dimensions. In $\nu=3$, the derivations of the Boltzmann and
Lenard-Balescu equations break down for the Coulomb potential.  }
\label{fig_bbgky_breakdown}
\end{figure}
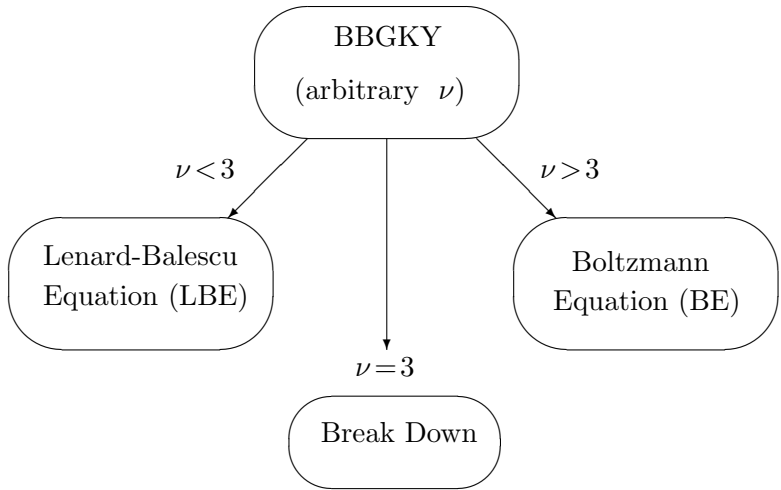

This is the third lecture on a novel technique for calculating the charged 
particle stopping power and the temperature equilibration rate in a weakly 
coupled fully-ionized plasma\,\cite{lfirst, bps}.  The method is exact to
leading and next-to-leading order in the plasma coupling $g$, and therefore 
calculates the Coulomb logarithm exactly. In Lecture~I~\cite{bps1} of 
this series, I discussed the basic theoretical machinery of dimensional 
continuation, and in Lecture~II~\cite{bps2}, as an example of the method, 
I calculated the energy exchange rate between electrons and ions in a 
hot plasma in using the BPS formalism. This formalism can be viewed
in the light of convergent kinetic equations, and to my knowledge, it is 
the only formalism in the literature that uses a systematic expansion in 
powers of $g$. It is quite gratifying, therefore, that the BPS stopping 
power has recently been verified experimentally\,\cite{bps_confirmed}, 
and this has provided impetus for another lecture.  The purpose of these 
notes is to prove one of the primary claims upon which BPS  is based, 
namely,  that to leading order in the plasma coupling $g$, the 
BBGKY hierarchy\,\cite{bbgky_ref} reduces to (i) the Boltzmann equation 
in dimensions $\nu>3$, and to (ii) the Lenard-Balescu equation\,\cite{len,bal}  
in dimensions $\nu<3$. However, for $\nu=3$ (the dimension of interest), 
things are not so clean: the Boltzmann equation (BE) suffers a long-distance 
divergence, and the Lenard-Balescu equation (LBE) contains a short-distance 
divergence. In both cases, the divergences are logarithmic, and this is a
crucial observation in regularizing them. Denoting the $\nu$-dimensional 
Coulomb potential by $\phi_\nu(r)$, we see that the divergences in $\nu=3$ 
arise because the potential  $\phi_3(r) \sim 1/r$  is the only potential 
$\phi_\nu(r)$ whose integral contains  both a short- and a long-distance 
divergence. The dimensional reduction of BBGKY is illustrated schematically 
in Fig.~\ref{fig_bbgky_breakdown}. 

The kinetic equations for systems interacting via the Coulomb force 
diverge in three spatial dimensions, and there have been many attempts 
to rectify this problem. In these notes, I will concentrate 
on the method of Brown, Preston, and Singleton (BPS) of Ref.~\cite{bps}.  
The method relies on dimensional continuation, which is a regularization 
technique adopted from quantum field theory calculations in arbitrary 
spatial dimensions $\nu$. I will prove rigorously that the BBGKY hierarchy 
collapses to the Lenard-Balescu equation for $\nu<3$ to leading order 
in the plasma coupling~$g$. This is quite an involved calculation, and 
Clemmow and Dougherty\,\cite{cd} is my primary reference.  Their 
calculation breaks down in $\nu=3$ dimensions, but goes through 
unscathed in dimensions less than three. For completeness, I will also 
prove that to leading order in $g$, the BBGKY hierarchy reduces to the 
Boltzmann equation for $\nu > 3$. I will base this calculation on that of 
Huang in Ref.~\cite{huang}, which breaks down in three dimensions, but 
becomes rigorous in dimensions greater than three. 

As we are concerned with short- and long-distance divergences, we must
be clear in our nomenclature. In keeping with the standard usage of quantum 
mechanics, I will call a short-distance divergence an {\em ultra-violet} (UV) 
divergence, and a long-distance divergence an {\em infra-red} (IR) divergence. 
This nomenclature arises from the well known episode in the history of physics 
in which classical physics spectacularly failed to calculate the observed 
black-body spectrum. The classical calculation captured the long-distance 
infra-red part of the spectrum correctly, but it predicted that the short-distance 
ultra-violet part of the spectrum would diverge, which is absurd (and contrary 
to observation).  In text books this episode is now called the {\em  ultraviolet
catastrophe}\,\cite{quantum1}, although more precisely it might be called 
the Rayleigh-Jeans catastrophe. As we will see, the Lenard-Balescu equation 
in three dimensions suffers its own UV catastrophe, and for similar reasons. 
Conversely, it turns out that the Boltzmann equation in three dimensions 
suffers from an IR divergence. Both are related to the $1/r$ behavior of
the Coulomb potential. 

It might be of interest to review the history of the UV catastrophe in
more details, and  to bring out its role in the development of quantum 
mechanics. The UV catastrophe was indeed a catastrophe for classical 
physics, and in retrospect can be marked as the birth of quantum mechanics, 
although in a round about fashion\,\cite{quantum1}. The classical calculation 
of the spectral output of a black body is quite simple,  involving a single 
integral over all black-body frequencies~$\omega$. It was supposed to 
be a  triumph of classical physics, but embarrassingly, the spectral integral 
turned out to diverge at small wavelengths or high frequencies. In other 
words, the classical integral possessed a UV divergence. This was completely 
unexpected, and is, as we know, cured by the discrete nature of quantum 
particles of light. Max Planck was examining the divergent classical integral 
in 1900, and noticed that it became finite if the integral were replaced by 
a sum over discrete energy states $E_n = n \hbar\omega$, where the angular 
frequency $\omega$ is that of the light or the
electromagnetic radiation emitted from the black body. One of the most 
radical things in Planck's scheme is that it required a new physical constant 
$h$, sometime written as $\hbar = h/2\pi$, with units of action (energy times 
time, or equivalently momentum times distance). The constant
$h$ is now called Planck's constant, but at the time, as far as I know, Planck 
attached no fundamental significance to it. We can, in a certain sense. think 
of Planck's method as  just another attempt at regulating a divergence, in
this case a UV divergence. Then, in 1905, in his work on the photo-electric
effect,  Einstein proposed that Planck's energy quanta be taken literally. 
Einstein reasoned that light with frequency~$\nu$ is composed of discrete
particles, now called {\em photons}, with energy $E=h \nu = \hbar\omega$. 
The history of physics is rich in attempts to regulate infinite integrals, and 
the unexpected consequences from doing so. Dimensional continuation
is just one of many regularization schemes, and is interesting that 
quantum mechanics has its roots in one such regularization attempt. 

These notes are organized as follows. In 
Section~\ref{sec_Coulomb_Plasma_Arbitrary} we discuss the
Coulomb plasma in arbitrary dimensions, showing that the Coulomb
force is short-range in dimensions $\nu>3$ and long-range in $\nu<3$.
The dimension $\nu=3$ is the critical dimension in which long- and
short-range contributions are comparable. This section also discusses 
the distribution function, and as a warm-up exercise we derive the 
standard result for the dielectric function in a multi-component plasma. 
In Section~\ref{sec_coulomb_energy_xfer}  we discuss how to find 
energy transfer rates using dimensional continuation, and in 
Section~\ref{sec_bbgky_arbitrary} we derive the BBGKY hierarchy 
in an arbitrary number of dimensions. We show how to define
perturbation theory in the plasma coupling constant $g$, and we
calculate the BBGKY hierarchy accurate to order $g^2$. We show 
that a complementary collection of 2-point correlations are 
dominant in $\nu >3$ compared to $\nu <3$, and this leads to 
the qualitative differences between the Boltzmann equation and 
the Lenard-Balescu equation.
In Section~\ref{sec_be_fr_bbgky} we derive the Boltzmann equation 
from BBGKY in $\nu > 3$, and in Section~\ref{sec_LBE} we derive 
the Lenard-Balescu equation in $\nu<3$. We conclude with 
Section~\ref{sec_conclusions}, and cover some supplementary 
material in the appendices.

\pagebreak
\section{The Coulomb Plasma in Arbitrary Dimensions}
\label{sec_Coulomb_Plasma_Arbitrary}

We start with a  plasma composed of multiple species labeled by an 
index $a$, the various species being delineated by of a common electric 
charge $e_a$ and a common mass $m_a$. Each species is assumed 
to be in thermal equilibrium with itself at temperature $T_a$, with 
a spatially uniform number density $n_a$.   Since we are working in 
$\nu$ spatial dimensions, the engineering units of the number density 
are $L^{-\nu}$, and the units of charge $e_a$ are energy times $L^\nu$. 
We will measure temperature in units of energy, setting Boltzmann's 
constant to unity  $k_\smB=1$, and we will employ the notation 
$\beta_a = 1/T_a$ for the inverse temperature.

\subsection{The Coulomb Potential}
\label{sec_Coulomb_Potential}

Before considering a plasma in a general number of dimensions, 
it is instructive to look at the Coulomb field of a single point charge
$e$ in $\nu$ dimensions. Since Gauss's law holds in an arbitrary 
number of dimensions, then for a point particle at the origin with 
charge $e$ we have (in cgs rationalized units)
\begin{eqnarray}
  {\bm\nabla}\!\cdot\!{\bf E} = e \, \delta^\nu({\bf x}) \ ,
\label{Gaussdiff}
\end{eqnarray}
where ${\bf E}=(E_1, \cdots, E_\nu)$ is the electric field vector, with
${\bm\nabla}=(\partial/\partial x_1, \cdots , \partial/\partial
x_\nu)$ being the \hbox{$\nu$-dimensional} spatial gradient, and
$\delta^\nu({\bf x})$  being the $\nu$-dimensional Dirac $\delta$-function
centered at the origin. This can be expressed in an integral fashion
by integrating any spatial region $\Sigma$ containing the charge, 
\begin{eqnarray}
  \int_\Sigma d^\nu x \, {\bm\nabla}\!\cdot\!{\bf E} 
  = 
  e 
  \ .
\label{Gaussint}
\end{eqnarray}
To find the electric field we will use Gauss's theorem, 
\begin{eqnarray}
  \int_\Sigma d^\nu x \,
  {\bm\nabla}\!\cdot\!{\bf E}
  =
  \int_{\partial\Sigma} d {\bf A} \cdot {\bf E}
  \ , 
\end{eqnarray}
and exploit the usual symmetry arguments. Let $\Sigma=B_r$ be the 
$\nu$-dimensional ball of radius $r$ centered on the point charge $e$, 
and therefore the $(\nu\!-\!1)$-dimensional hyperspherical boundary 
is $\partial \Sigma = S_r$. By symmetry, the field ${\bf E}({\bf x})$ points 
radially outward with a magnitude $E(r)$, along the direction $\hat{\bf x}$ 
normal to $S_r$. The length $E(r)$ depends only upon the radial distance 
$r = \vert {\bf x} \vert$ and not upon its angular location along $S_r$, and 
therefore (\ref{Gaussint}) gives
\begin{eqnarray}
  e
  =
    \int_{B_r} d^\nu x \, {\bm\nabla}\!\cdot\!{\bf E}
  = 
  \oint_{S_r}\! d{\bf A}\!\cdot\! {\bf E}
  =
  \Omega_{\nu-1}\,r^{\nu-1} \cdot  E(r) 
  ~~~~~ \text{with}~~~\Omega_{\nu-1} = 
  \frac{2\pi^{\nu/2}}{\Gamma(\nu/2)}
   \ .
\label{eintGamma}
\end{eqnarray}
The relation for the solid angle comes from (\ref{OmegaMinusOne}),
and the electric field of a point particle at the origin becomes
\begin{eqnarray}
  {\bf E}({\bf x}) 
  = 
  e\,\frac{\Gamma(\nu/2)}{2 \pi^{\nu/2}}\, \frac{\hat{\bf x} }{r^{\nu-1}}
  \ ,
\label{Enu}
\end{eqnarray}
where $\hat{\bf x}$ is a unit vector pointing in the direction 
of ${\bf x}$, and the radial variable is $r = \vert {\bf x} \vert$.
Note that  ${\bf x}= r\, \hat{\bf x}$,  which allows us to write 
(\ref{Enu}) in a frequently used alternative form,
\begin{eqnarray}
  {\bf E}({\bf x}) 
  = 
  e \,\frac{\Gamma(\nu/2)}{2 \pi^{\nu/2}}\, \frac{{\bf x}}{r^{\nu}}\, 
  \ .
\label{EnuAlt}
\end{eqnarray}

\vskip0.2cm
\noindent
It will often be more convenient to work with the electric potential
$\phi$ defined by
\begin{eqnarray}
  {\bf E} = -{\bm \nabla}\phi  
  \ ,
\label{eq_phi_def}
\end{eqnarray}
and upon integrating (\ref{Enu}), we find
\begin{eqnarray}
  \phi({\bf x})
  = 
  \frac{\Gamma(\nu/2-1)}{4 \pi^{\nu/2}} \,
  \frac{e}{r^{\nu-2}} 
  \ ,
\label{Vnu}
\end{eqnarray}
where we have chosen the constant of integration so that the potential 
vanishes at radial infinity (for $\nu > 2$). Generalizing to a 
multi-component plasma, the electric field of a particle of type $b$ at 
the origin is
\begin{eqnarray}
  {\bf E}_b({\bf x}) 
  = 
  e_b \,\frac{\Gamma(\nu/2)}{2 \pi^{\nu/2}}\, \frac{\hat{\bf x}}{r^{\nu-1}}\, 
  \ ,
\label{EnuAlta}
\end{eqnarray}
and the corresponding potential is
\begin{eqnarray}
  \phi_b({\bf x})
  = 
  e_b \,\frac{\Gamma(\nu/2-1)}{4 \pi^{\nu/2}}\, \frac{1}{r^{\nu-2}}\, 
  \ .
\label{EnuAltaPot}
\end{eqnarray}
Note that ${\bf E}_b({\bf x}_a - {\bf x}_b)$ is the electric field at 
${\bf x}_a$ produced by a point charge $e_b$ at ${\bf x}_b$, and 
consequently, the force acting on charge $a$ from charge $b$ is
\begin{eqnarray}
  {\bf F}_a^{(b)}
  =
  e_a {\bf E}_b({\bf x}_a - {\bf x}_b) 
  = 
  e_a e_b \,\frac{\Gamma(\nu/2)}{2 \pi^{\nu/2}}\, \frac{{\bf x}_a - {\bf x}_b}
  {\vert {\bf x}_a - {\bf x}_b \vert^{\nu}}
  \ .
\label{EnuAltaF}
\end{eqnarray}
This form will appear in the BBGKY kinetic equations for many-body
Coulomb systems. 

The prefactor of the electric field in these units, known as cgs rationalized 
units, depends upon the spatial dimension $\nu$. In three dimensions we 
find the usual factor of $4\pi$, 
\begin{eqnarray}
  {\bf E}_3({\bf x})
  &=&
  \frac{e}{4\pi}\, \frac{\hat{\bf x} }{r^2}
\\[8pt]
  \phi_3({\bf x})
  &=& 
  \frac{e}{4\pi}\, \frac{1}{r}
  \ ,
\label{VnuThree}
\end{eqnarray}
where the numerical subscript denotes $\nu=3$. This potential will turn out 
to be special, in that its integral diverges logarithmically at both small and  short 
distances. It is the only potential whose integral diverges in the IR and the UV. 
It will be useful for our intuition to look at the electric field and its 
potential for dimensions on either side of three. For example, in $\nu=4$, 
we have
\begin{eqnarray}
  {\bf E}_4({\bf x})
  &=&
  \frac{e}{2\pi^2}\, \frac{\hat{\bf x} }{r^3}
\\[8pt]
  \phi_4({\bf x})
  &=& 
  \frac{e}{4\pi^2}\, \frac{1}{r^2}
  \ ,
\label{VnuFour}
\end{eqnarray}
and we see that the potential converges more quickly than $\phi_3$ 
for large values of $r$. The case $\nu=2$ must be handled with a 
little care, as we cannot simply substitute $\nu=2$ into (\ref{Vnu}). 
The electric field in $\nu=2$ dimensions is proportional to $1/r$, 
which integrates to a logarithm for the potential, so that
\begin{eqnarray}
{\bf E}_2({\bf x})
&=&
 % \phantom{-}
  \frac{e}{2 \pi }\,\frac{\hat{\bf x} }{r}
\\[8pt]
  \phi_2({\bf x})
  &=& 
 - \frac{e}{2 \pi} \, \ln(r/r_0) 
  \ \ ,
\label{VnuTwo}
\end{eqnarray}
where $r_0$ is an arbitrary integration constant at which the potential 
is chosen to vanish. In \hbox{$\nu=2$} something different has happened. 
The potential no longer asymptotes to a constant value at large $r$, but 
diverges logarithmically. We must therefore choose a finite but arbitrary 
radius $r_0$ along which the potential vanishes. Note that the 2-dimensional 
potential also diverges logarithmically at small $r$. Since a logarithmic 
divergence is an integrable divergence, it is not as severe as the $1/r$ 
divergence in $\nu=3$, and this is why the Lenard-Balescu equation in 
$\nu<3$ does not contain a short distance divergence, as it does for
$\nu \ge 3$. We can also arrive at (\ref{VnuTwo}) by performing an analytic 
continuation in $\nu$ near the region $\nu=2$.  In other words, define 
the small (continuous) parameter $\epsilon=\nu-2$, and note that
 (\ref{Vnu}) takes the form 
\begin{eqnarray}
  \phi({\bf x})
  &=& 
  \frac{e}{4 \pi} \,
  \frac{\Gamma(\epsilon/2)}{(\sqrt{\pi}\, r)^\epsilon} 
  \ .
\end{eqnarray}
Using the expansions
\begin{eqnarray}
  \Gamma(\epsilon/2) 
  &=&
  \frac{2}{\epsilon}- \gamma + {\cal O}(\epsilon)
  \\[5pt]
  a^{-\epsilon} 
  &=& 
  e^{-\epsilon\ln a} = 1 - \epsilon \,\ln a + {\cal O}(\epsilon)
  \ \ ,
\end{eqnarray}
and dropping linear and higher order terms in $\epsilon$, we find
\begin{eqnarray}
  \phi_2({\bf x})
  &=& 
  \frac{e}{4\pi}
   \Bigg(\frac{2}{\epsilon}- \gamma\Bigg)\Bigg(1 - \epsilon \,\ln \pi^{1/2} r\Bigg)
  =
    \frac{e}{4\pi}\left[\, 
    \frac{2}{\epsilon} - \gamma - \ln \pi   - 2 \ln r 
    \, \right]
  \ \ .
\end{eqnarray}
In the limit $\nu \to 2$, this potential contains the infinite constant 
$2/\epsilon -\gamma - \ln \pi$. The physical reason for this (harmless) 
divergence is that the zero of potential energy in (\ref{Vnu}) vanishes 
in the asymptotic limit $r \to \infty$, while the logarithmic potential 
for $\nu=2$ does not vanish at large~$r$, but instead diverges. This 
is not a problem, as we are free to subtract a constant (even an infinite 
constant) from any potential. Indeed, as we move from $\nu=3$ to 
$\nu=2$, there is no discontinuity in the $r$-behavior of the electric 
field.  Let us therefore define a shifted potential $\bar\phi_2({\bf x}) 
= \phi_2({\bf x}) - \phi_2({\bf x}_0)$, that is to say, we subtract the 
constant value $\phi_2({\bf x}_0)$ at an arbitrary ${\bf x}_0$, and 
we find $\bar\phi_2({\bf x}) = -(e/2\pi) \ln(r/r_0)$, which is just 
(\ref{VnuTwo}). Thus, there are no physically measurable discontinuities 
as we dimensionally continue from $\nu=3$ to $\nu=2$. Since we 
are eventually interested in analytically continuing $\nu$ to complex 
values in a small neighborhood around $\nu=3$, and then taking the 
limit $\nu \to 3$, we may continue to formally use (\ref{Vnu}).  It is, 
however, important that the electric field and its potential are well 
defined for all positive integer values of $\nu$. Finally, let us examine 
$\nu=1$, for which we obtain
\begin{eqnarray}
  E_1(x)
&=&
  (e/2)\, {\rm sign}(x)
\\[5pt]
  \phi_1(x)
  &=& 
  -(e/2) \, r 
  \ ,
\end{eqnarray}
where $r = \vert x \vert$.  
The unit vector pointing away from the origin, $\hat x$, can point only left
or right, and can therefore be thought of as the sign function ${\rm sign}(x)$:
plus  one for positive $x$ and minus one for negative $x$, located at $x=0$. 
Since $x = \vert x \vert \,{\rm sign}(x)$, we can still express the spatial point
$x$ in the form $x = r \hat x$.

\subsection{The Coulomb Potential and Dimensional Regularization}

Let us  explore  in more detail how dimensional continuation acts as a
{\em regulator} for divergent integrals, rendering them finite and therefore 
algebraically amenable. We shall start by considering the Boltzmann equation 
for a plasma in three dimensions. We do not require the actual equation 
at the moment, but we only need to recall that it suffers a long-distance 
IR divergence.\footnote{
\footnoteskip
As stated in the introduction, we are using the quantum mechanical 
nomenclature in which {\em IR}  stands for {\em infra-red} long-distance 
physics, {\em UV} stand for short-distance {\em ultra-violet} physics. 
} % footnote
In contrast, it is interesting to note that the far more idealized 
{\em hard-sphere} 
scattering model is finite. This hard-sphere model is based on the 
idea that particles in a dilute gas behave like billiard balls. This is of course
incorrect, or at least a highly idealized picture, but the model still provides 
useful insight. The reason that hard-sphere scattering is finite is that  it is  
{\em short-range}: the force acts only during the collision, after which the 
particles move freely in a constant potential (until
the next collision).  There are no long-distance effects in this model. A 
gas of neutral particles acts somewhat like  billiard balls, so we expect 
the Boltzmann equation to be finite for a gas. And indeed it is. 
This is because the force between neutral particles  is short-range, and 
they do not see one another at large distances.  In fact, the Boltzmann
kernel is finite for any short-range force in three dimensions. The irony, 
however, is that we are not interested in short-range forces. Instead, it 
is the Coulomb force that is of relevance to plasma physics, and in three 
dimensions this is a long-range force. Consequently, we find a long-distance
 logarithmic divergence in the Boltzmann scattering kernel. 
This IR divergence essentially arises from the integration of the potential 
$\phi_3(r) \sim 1/r$ at large-$r$. Since the divergence is {\em only} logarithmic,
any potential that falls off faster than $1/r$ at large-$r$ will {\em not} 
produce a divergence in the Boltzmann scattering kernel. We will return 
to this point in the next paragraph. In summary, even though the Boltzmann 
equation gets the short-distance physics right, it gets the long-distance 
physics wrong, and we pay the price through a logarithmic IR divergence. 
Conversely, when we capture long-distance collective effects in a 
plasma using the Lenard-Balescu equation, we find a short-distance UV 
divergence (in three dimensions). 
The Lenard-Balescu equation captures the long-distance physics correctly, 
but models the short-distance physics incorrectly, and once again we 
pay a price, this time introducing a logarithmic UV divergence. In this
case, any potential that diverges less severely than $\phi_3(r) \sim 1/r$
at small-$r$ will not suffer a divergence. But as before, we are not
interested in such forces.  This reasoning, however, will be essential
to understanding why the Coulomb potential in $\nu$ dimensions 
regulates the various kinetic equations. 

\begin{figure}[t!]
\begin{minipage}[c]{0.45\linewidth}
\includegraphics[scale=0.55]{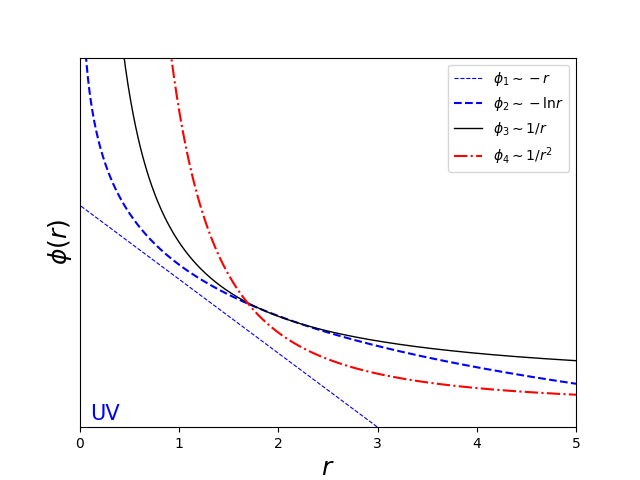} 
%\caption{Image A}
\end{minipage}
\hfill
\begin{minipage}[c]{0.5\linewidth}
\includegraphics[scale=0.55]{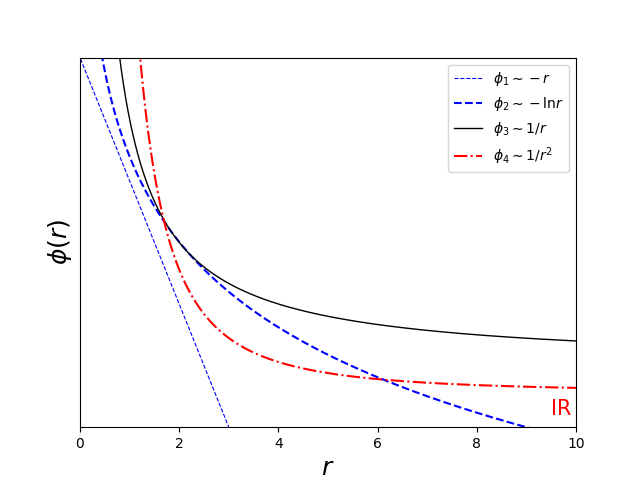} 
%\caption{Image B}
\end{minipage}
%\vskip-1.0cm
\caption{\footnoteskip
  The Coulomb potential $\phi$ for a positive charge at the origin
  as a function of radius $r$, for dimensions \hbox{$\nu=1,2,3,4$}.  
  The zeroes of potential energy have been adjusted for visual clarity. 
  Short-distance ultraviolet (UV) physics is emphasized in dimensions 
  $\nu>3$ (left panel), and long-distance or infrared (IR) physics 
  dominates when $\nu<3$ (right panel). For $\nu=3$,
  the UV and IR physics are equally important, and the energy rate 
  diverges logarithmically at both large and small distances in three
  dimensions. The left panel  illustrates that for $\nu<3$, the Coulomb 
  potential diverges less severely that $1/r$ as $r$ gets small, and 
  consequently the LBE does not suffer a short-distance divergence
  in $\nu<3$. Similarly, the right panel illustrates that the Coulomb 
  force converges to zero more rapidly than $1/r$ as $r$ gets large, 
  and this renders the Boltzmann kernel finite at large distances when 
  $\nu>3$. 
}
\label{fig_coulomb}
\end{figure}

We can now show why working in an arbitrary number of dimensions
$\nu$ acts as a regulator. 
It turns out that the Boltzmann equation (BE)  becomes IR finite for $\nu>3$, 
and that the Lenard-Balescu equation (LBE) becomes UV finite for $\nu<3$, 
with the only reminder of past divergences being simple poles of the form 
$1/(\nu-3)$. Now that we know how the Coulomb force works as a function 
of dimension, we can understand this behavior.
 Figure~\ref{fig_coulomb} shows the  Coulomb potential $\phi_\nu(r) 
\sim 1/r^{\nu-2}$ for a positive charge at the origin for $\nu=1,2,3,4$ 
(remembering that $\nu=2$ is actually logarithmic). For aesthetic 
reasons, the arbitrary integration constants of the potentials have been 
adjusted so  that the graphs for $\nu=2,3,4$ intersect at a common 
point. The Figure shows that by simply dialing the dimension $\nu$, a 
potential $\phi_\nu(r)$ can be selected with the appropriate short- and
long-distance behavior. The potential $\phi_3(r) \sim 1/r$ is special,  in 
that it produces logarithmic divergences in the UV and IR, indicating that 
short- and long-distance physics are equally important in three dimensions. 
Thus, $\phi_3(r)$ is a borderline case, and this is the reason that
the BE and the LBE suffer IR and UV divergences, respectively. 
However,  and this is the key point, for $\nu<3$, the left panel shows 
that  the potential diverges less slowly than $\phi_3(r)$ 
for small-$r$, and this is what renders the LBE finite in the UV.
Conversely, for $\nu>3$ the right panel of the Fig.~\ref{fig_coulomb} 
shows that the potential converges more rapidly than $\phi_3(r)$ 
for large $r$, and this renders the BE finite in the IR. This is the
 reason dimensional continuation works as a regulator.

As we have emphasized,  the divergences in question are {\em only}
logarithmic (rather than linear or higher order), and can therefore be 
rendered finite by {\em slightly}  adjusting the rate of convergence of 
the potential $\phi_\nu(r)$ in the offending region of $r$ (either at large 
or small values of $r$). This takes us into the domain of convergent kinetic 
equations. The integral of any potential that diverges less slowly that $1/r$ 
as $r\to \infty$, even by an infinitesimal  amount, will in fact converge at 
large-$r$. For example, the potential $\phi(r) \sim 1/r^{1 + \delta_1}$ with 
$\delta_1 > 0$, but otherwise $\delta_1$ can be as close to zero as we 
wish, gives a convergent integral in the IR, 
\begin{eqnarray}
 \int dr\,\frac{1}{r^{1 + \delta_1}} \sim r^{-\delta_1} \to 0  
 ~~~\text{as}~ r \to \infty
\ ,
\label{eq_larger}
\end{eqnarray}
and the BE does not possess an IR divergence for such a potential. 
Conversely, the integral of any potential that diverges less slowly that 
$1/r$ as $r\to 0$ will in fact converge at small-$r$. For example, 
consider a potential of the form $\phi(r) \sim 1/r ^{1 -\delta_2}$, 
where $\delta_2 > 0$ (with $\delta_2$ as close to zero as we wish). 
Then the integral of the potential is finite in the UV, 
\begin{eqnarray}
 \int dr\,\frac{1}{r^{1 - \delta_2}} \sim r^{\delta_2} 
\to 0  ~\text{as}~ r \to 0
\ ,
\label{eq_smallr}
\end{eqnarray}
and the LBE does not suffer a UV divergence for such a potential. This is the 
essence of the techniques of convergent kinetic theory, of which the BPS 
formalism is an  example. One must be exceedingly careful, however, as the 
same physical regularization scheme must be used at short- and long-distances. 
The quantities $\delta_1$ of (\ref{eq_larger}) and $\delta_2$ of  (\ref{eq_smallr}) 
are not independent! If they are treated independently, then one can produce 
spurious unphysical constants in the Coulomb logarithm. Another benefit 
of the  BPS formalism is that it treats long and short distances in the same 
manner. See Lecture I regarding the Lamb Shift, in which the significance 
of using the same regularization was first realized.  To my knowledge, BPS 
is the only convergent kinetic scheme that treats the long- and short-distance 
divergences in the same way. 

\pagebreak
\subsection{The Fourier Transform of the Coulomb Potential}

Unlike the spatial representation of the potential, we will show that the
Fourier representation  takes the same form in any dimension. This is quite 
useful for calculations. There a number of conventions for the spatial Fourier 
transform,  and I employ
\begin{eqnarray}
  \phi({\bf x}) 
  &=& 
  \int \!\frac{d^\nu k}{(2\pi)^\nu} \,
  e^{i{\bf x}\cdot{\bf k}}\, \tilde \phi({\bf k})
\label{VxVtilde}
\\[5pt]
  \tilde \phi({\bf k}) 
  &=& 
  \int \! d^\nu x \,e^{-i{\bf x}\cdot{\bf k}}\, 
  \phi({\bf x})
   \ .
\label{VtildeVx} 
\end{eqnarray}
As a general rule, the factors of $2\pi$ will always be placed with the 
${\bf k}$-integral, as this is analogous to placing factors of $2\pi\hbar$ 
with the ${\bf p}$-integral, a convention based in quantum mechanics 
that we shall also follow. As we now show, the $\nu$-dimensional Coulomb 
potential (\ref{Vnu}), which is repeated here for convenience, 
\begin{eqnarray}
  \phi({\bf x})
  = 
  \frac{\Gamma(\nu/2-1)}{4 \pi^{\nu/2}} \,
  \frac{e}{r^{\nu-2}} 
  \ ,
  \label{eq_Vnu_again}
\end{eqnarray}
has the Fourier transform 
\begin{eqnarray}
  \tilde \phi({\bf k})
  =
  \frac{e}{k^2} 
  ~~~~~\text{where}~~~
  k^2
  \equiv
  {\bf k} \cdot {\bf k}
  =
  \sum_{\ell=1}^\nu k_\ell^2
  \ .
\label{CoulombXform}
\end{eqnarray}
As emphasized above, the form of $\tilde \phi({\bf k})$ does not depend 
upon the dimension of space, except in a trivial way though the length 
of $k^2$. 

Expression (\ref{CoulombXform}) for the Fourier transform of the potential
(\ref{eq_Vnu_again}) can be established in a number of ways. Perhaps the easiest 
is just  to use Laplace's equation,
\begin{eqnarray}
  \nabla^2 \phi({\bf x}) 
  = - e \, \delta^{(\nu)}({\bf x}) \ ,
\label{laplaceV}
\end{eqnarray}
which is obtained by substituting (\ref{eq_phi_def}) into (\ref{Gaussdiff}) .
Upon inserting (\ref{VxVtilde}) for $\phi({\bf x})$ into (\ref{laplaceV}), and 
using the integral representation of the $\delta$-function, we can write 
Laplace's equation in the form
\begin{eqnarray}
 - \int \! \frac{d^\nu k}{(2\pi)^\nu} \,
  e^{i{\bf x}\cdot{\bf k}}\, k^2\, \tilde \phi({\bf k})
  =
  - e \, \delta^\nu({\bf x})
  =
  - e \int \!
  \frac{d^\nu k}{(2\pi)^\nu} \, e^{i{\bf x}\cdot{\bf k}} \ ,
\end{eqnarray}
or
\begin{eqnarray}
  \int \! \frac{d^\nu k}{(2\pi)^\nu} \,
  e^{i{\bf x}\cdot{\bf k}}\, \Big[
  k^2\, \tilde \phi({\bf k}) - e
  \Big]
  = 0
  \ .
\end{eqnarray}
The quantity in square brackets must vanish, and solving for 
$\tilde \phi({\bf k})$ indeed gives (\ref{CoulombXform}). 

It is also informative to prove this result by taking the Fourier 
transform directly. The formula to remember is
\begin{eqnarray}
  \frac{1}{a} = \int_0^\infty ds \, e^{-a s}
  \ ,
  \label{eq_magic_one}
\end{eqnarray}
where ${\rm Re}\,a> 0$, from which we can take $p$ derivatives 
to obtain another useful expression, 
\begin{eqnarray}
  \frac{1}{a^p} = \frac{1}{\Gamma(p)}\int_0^\infty ds \, s^{p-1} \, e^{-a s}
  \ .
  \label{eq_magic_two}
\end{eqnarray}
We can analytically continue to complex values of $p$, in particular to
$p=(\nu-2)/2$.  We can also prove (\ref{eq_magic_two}) by changing 
variables in the $s$-integral to $u=a s$, which produces the correct scaling 
$1/a^p$, while the remaining $u$-integral exactly cancels the Gamma-function 
$\Gamma(p)$. We  now use relation (\ref{eq_magic_two}) to rewrite the 
term $1/r^{\nu-2}$ in the potential. First note that 
$r = ({\bf x} \cdot {\bf x})^{1/2}$, which we write as $r= (x^2)^{1/2}$, 
and this allows us to express
\begin{eqnarray}
  \frac{1}{r^{\nu-2}} = \frac{1}{(x^2)^{(\nu-2)/2}}
  =
  \frac{1}{\Gamma(\nu/2-1)}\int_0^\infty ds \, s^{-(\nu-4)/2} \, e^{-s\, x^2}
  \ .
\end{eqnarray}
The Fourier transform of the Coulomb potential is therefore
\begin{eqnarray}
  \tilde \phi({\bf k}) 
  &=& 
  e\,\frac{\Gamma(\nu/2-1)}{4 \pi^{\nu/2}} 
   \int \! d^\nu x \,e^{-i{\bf x}\cdot{\bf k}}\, \frac{1}{r^{\nu-2}}  
  \\[5pt]
  &=&
   \frac{e}{4 \pi^{\nu/2}} \,
    \int \! d^\nu x \, \int_0^\infty ds\,
    s^{(\nu-4)/2} \,e^{-s x^2 -i{\bf x}\cdot{\bf k}}
    \ .
  \end{eqnarray}
We can interchange the $s$ and $x$ integrals because the integrand is
uniformly convergent. We then perform the $x$-integrals by completing 
the square, and we find
\begin{eqnarray}
  \tilde \phi({\bf k}) 
  &=& 
   \frac{e}{4 \pi^{\nu/2}} \
   \int_0^\infty ds\,   s^{(\nu-4)/2} \int \! d^\nu x \, 
   \,e^{-s ({\bf x} + i {\bf k}/2s )^2} \, e^{-k^2/4s}
    \\[5pt]
     &=& 
   \frac{e}{4 \pi^{\nu/2}} \,
   \int_0^\infty ds\,   s^{(\nu-4)/2} \,\left(\frac{\pi}{s}\right)^{\nu/2}
    e^{-k^2/4s}
       \\[5pt]
     &=& 
   \frac{e}{4} \,
   \int_0^\infty ds\,   s^{-2} \, e^{-k^2/4s} 
   =
    \frac{e}{4} \,
   \int_0^\infty dt \, e^{-t k^2/4} 
   =
   \frac{e}{k^2}
   \ .
  \end{eqnarray}
 \subsection{The Distribution Function}

For each component $a$ of the plasma, there is a distribution 
function $f_a$ define by 
\begin{eqnarray}
\nonumber 
  f_a({\bf x},{\bf p},t)\,\frac{d^\nu x\,d^\nu p}{(2\pi\hbar)^\nu}\, 
  &\equiv& 
  \text{number of particles of type $a$ in a hypervolume}
\\[-10pt] && 
  \text{$d^\nu x$ about ${\bf x}$ and $d^\nu p$ about ${\bf p}$ 
  at time $t$} 
  \ ,
\label{fnu}
\end{eqnarray}
where where $h$ is Planck's constant, and $\hbar = h/2\pi$. The 
phase-space factor $(2\pi\hbar)^\nu = h^\nu$ ensures that $f_a$ 
counts  the number of semi-classical quantum states in a phase-space 
volume $d^\nu x \,d^\nu p$. The factor $\hbar$ makes the volume 
element in (\ref{fnu}) dimensionless, thereby rendering $f_a$ 
dimensionless. Using $\hbar$ in place of $h$ is merely a convention. 
More critically, the factor of $\hbar$ makes the classical to quantum
transition more transparent, and the normalization (\ref{fnu}) implies
\begin{eqnarray}
  \int\! \frac{d^\nu p}{(2\pi\hbar)^\nu}\,
  f_a({\bf x}, {\bf p},t) &=& n_a({\bf x},t) \ ,
\label{fbnorm}
\end{eqnarray}
where $n_a({\bf x}, t)$ is the number density of $a$-type particles at 
position ${\bf x}$ and time $t$. That is to say, $n_a\,d^\nu x$ is the 
number of particles of species $a$ in a hypervolume $d^\nu x$ about
position ${\bf x}$ at time $t$. When performing long calculations, it 
is often convenient to  combine the space and momentum variables 
into a single phase-space variable  $X=({\bf x}, {\bf p})$. Distribution 
functions are then written $f=f(X,t)$, and the corresponding integration 
measure becomes
\begin{eqnarray}
  dX = \frac{d^\nu x \, d^\nu p}{(2\pi\hbar)^\nu}
  \ .
  \label{eq_dX_def_X}
\end{eqnarray}
Definition (\ref{fnu}) now means that $f_a(X,t)\,dX$ is the number of 
particles of type $a$ in a volume element $dX$ about phase-space 
location $X$ at time $t$. This notation will be particularly useful when 
considering multi-particle distributions. 

Throughout these notes, we primarily consider plasmas in {\em local} 
thermodynamic equilibrium, so that $n_a$ is a slowly varying function
of {\em space} and {\em time}, with engineering units of $L^{-\nu}$. 
We will often consider special cases in which the plasma is completely
uniform, and the number density is constant in space. As a general
rule, we will leave time dependence implicit. 
From (\ref{fbnorm}),  we see that a normalized Maxwell-Boltzmann 
distribution at temperature $T_a $ and number density of $n_a$
takes the form
\begin{eqnarray}
%  f_a^{\rm\scriptscriptstyle MB}({\bf p}) 
  f_a({\bf p}) 
  =
  n_a
  \left(\frac{2\pi\hbar^2 \beta_a}{m_a}\right)^{\nu/2}
  \exp\left\{-\beta_a\, \frac{p^2}{2m_a}\right\}  
  =
  n_a\, \lambda_a^\nu\, e^{-\beta_a E_a} \ ,
\label{defFiA}
\end{eqnarray}
where  $\beta_a=1/T_a$ is the inverse temperature in energy units, and 
$E_a = p_a^2/2 m_a$ is the kinetic energy of an individual particle of 
species~$a$. The {\em thermal de Broglie wavelength} for particle~$a$
is defined to be
\begin{eqnarray}
  \lambda_a
  &=& 
  \hbar \left( \frac{2\pi \beta_a}{m_a}\right)^{1/2}  
  =
  \frac{h}{\sqrt{2\pi\, m_a T_a}}
  \ .
\label{deflambd}
\end{eqnarray}
Expression 
(\ref{defFiA}) shows explicitly that $f_a$ is dimensionless in accordance 
with our normalization, as $n_a \lambda_a^\nu$ is dimensionless. 
However, one might rightly ask why would a quantum parameter, 
$\hbar$,  appear in a classical distribution? In fact, physical averages 
do not depend on $\hbar$, as the factors of $\hbar$ in the integration 
measure (\ref{eq_dX_def_X})  are exactly canceled by those in the  
normalization term $\lambda_a^\nu$. However, this normalization 
is more than a mere convention, for $n_a \lambda_a^\nu$ counts 
the number of quantum states available to system $a$. Therefore, 
this normalization is the only one that gets questions about entropy 
correct. In fact, upon writing the distribution function in terms of
the chemical potential $\mu_a$, 
\begin{eqnarray}
  f_a = e^{-\beta_a (E_a -\mu_a )}
  \ ,
\end{eqnarray}
we see that 
\begin{eqnarray}
  e^{\beta_a \mu_a} = n_a \lambda_a^\nu
  \ .
\end{eqnarray}
This gives the correct chemical potential of a free gas, 
\begin{eqnarray}
  \mu_a
  =
  T_a \ln n_a \lambda^\nu 
  =
  T_a \ln \left\{ 
  n_a \left[ \hbar \left( \frac{2\pi }{m_a T_a} \right)^{1/2}\right]^\nu 
  \right\}
  \ ,
\end{eqnarray}
a standard result from quantum statistical mechanics, with a trivial
generalization to multiple dimensions. 

Two-particle correlations are described by a two-component
correlation function,  defined by
\begin{eqnarray}
\nonumber 
  f_{ab}({\bf x}_a,{\bf p}_a,{\bf x}_b,{\bf p}_b,t)\,
 \frac{d^\nu x_a\,d^\nu p_a}{(2\pi\hbar)^\nu}\, 
 \frac{d^\nu x_b\,d^\nu p_b}{(2\pi\hbar)^\nu}\, 
  &\equiv& 
  \text{number of particles of type $a$ in a hypervolume}
\nonumber\\[-10pt] && 
  \text{$d^\nu x_a$ about ${\bf x}_a$ and $d^\nu p_a$ about ${\bf p}_a$},
  ~\text{and}~ \text{$d^\nu x_b$}
  \nonumber\\[-10pt] &&
  \text{about ${\bf x}_b$ and $d^\nu p_b$ about ${\bf p}_b,$}~
  \text{at time $t$} \ .
  \nonumber\\
\label{fnuTwoCorr}
\end{eqnarray}
We often write this as $f_2({\bf x}_1, {\bf p}_1, {\bf x}_2, {\bf p}_2,t)$ 
in a single-species plasma. 
We can go on to define higher order distribution functions, and in fact
show that they satisfy a system of connected equations known as
the BBGKY hierarchy. We will discuss this in much more detail in
future sections. For now, let us stick to the basic properties of a
multi-component plasma, and let us see what we can do with just 
the single-particle distribution. 

%%%%%
%\pagebreak
\subsection{The Dielectric Function of a Plasma}

As these notes are also a tutorial, it is useful to take a detour 
and to describe how kinetic theory allows us to calculate the induced
charge density $\rho_{\rm ind}$ of a plasma when a small external 
electric  field ${\bf D}$ is applied. This in turn allows us to calculate
the dielectric function of the plasma. Consider a multi-species 
plasma in which species $b$ has charge $e_b$ and mass $m_b$. 
Furthermore, suppose that each species $b$ is in thermal equilibrium 
with itself at temperature $T_b$ and Maxwell-Boltzmann distribution 
$f_b$. We now use basic kinetic theory to prove that the dielectric
 function in a general number of dimensions takes the form
\begin{equation}
  \epsilon({\bf k},\omega) 
  = 
  1 + {\sum}_b\, \frac{e_b^2}{k^2} 
  \int \! \frac{d^\nu p_b}{(2\pi \hbar)^\nu}  \,
  \frac{1}{\omega - {\bf k}\!\cdot\!{\bf v}_b + i \eta}\, 
  {\bf k} \cdot \frac{\partial f_b({\bf p}_b) }{\partial {\bf p}_b } \,,
\label{epsilonmulti}
\end{equation}
where the limit $\eta \to 0^+$ is understood,  and  ${\bf p}_b = 
m_b {\bf v}_b$. Reference~\cite{lifs}  derives this well known 
result in three spatial dimensions, and this section is a simple 
extension to a general number of spatial dimensions $\nu$. 
It serves mainly as a refresher to the reader who is not an 
expert, and it is very beautiful physics. It is quite astounding 
that one can get so much from so little.

%%
%\pagebreak
\subsubsection{The Induced Charge Density and the Dielectric Function}

Let us start with a neutral equilibrium plasma, and apply a small external 
electric field ${\bf D}$. Since the charges in the plasma are free to move, 
even the smallest field creates an induced charge density $\rho_{\rm ind}$,
as the ions and electrons are separated by the field. The charge separation
creates an induced field ${\bf P}$ that weakens the applied field. The 
observed field ${\bf E}$ is a combination of both of these\,\cite{lifs},
\begin{eqnarray}
\label{dpinduceda}
  {\bf E} &=& {\bf D} - {\bf P} ~~~\text{where}
  \label{eq_E_defhere}
\\[5pt]
\label{Dext}
  {\bm\nabla}\!\cdot\!{\bf D} &=& \rho_{\rm ext}  
\\[5pt]
  {\bm\nabla}\!\cdot\!{\bf P} &=& -\rho_{\rm ind}
  \label{dpinducedb}
  \ .
\end{eqnarray}
We are free to set the density $\rho_{\rm ext}$ to anything that creates the 
desired applied field ${\bf D}$. The observed field ${\bf E}$ satisfies Gauss's 
law, 
\begin{eqnarray}
\label{maxeqexta}
  &&
  {\bm\nabla}\!\cdot\!{\bf E} = \rho
 \ ,
\end{eqnarray}
where $\rho$ is the {\em total} charge density of the medium, which
consists of the external charge density and the induced charge density
\begin{eqnarray}
\label{rhoJa}
  &&
  \rho = \rho_{\rm ext} + \rho_{\rm ind}
   \ .
\end{eqnarray}
The field ${\bf D}({\bf x}, t)={\bf E}({\bf x}, t) + {\bf P}({\bf x}, t)$ depends 
not just on the  electric field ${\bf E}({\bf x},t)$ at time $t$,  but also on the 
electric fields at earlier times as well, through the polarization 
\begin{eqnarray}
  {\bf P}({\bf x},t) 
  =
  \int_{-\infty}^t \!dt^\prime \int \!d^\nu x^\prime \, 
  \chi({\bf x} - {\bf x}^\prime, t - t^\prime) \, {\bf E}({\bf x}^\prime,t^\prime) 
  \ .
%  \label{eq_K_def}
\end{eqnarray}
It is understood that the kernel satisfies causality,  $\chi({\bf x}, t)=0$
when $t < 0$, and we can therefore extend the $t^\prime$ integral 
from $t$ to $\infty$, giving
\begin{eqnarray}
  {\bf D}({\bf x},t) 
  = 
  {\bf E}({\bf x},t) 
  + 
  \int_{-\infty}^\infty \!dt^\prime \int \!d^\nu x^\prime \, 
  \chi({\bf x} - {\bf x}^\prime, t - t^\prime) {\bf E}({\bf x}^\prime,t^\prime) 
  \ .
  \label{eq_K_def}
\end{eqnarray}
 Using the convolution theorem, the Fourier transform 
takes a particularly simple form,
\begin{eqnarray}
\label{edefa}
  \tilde {\bf D}({\bf k},\omega) 
  &=& 
  \epsilon({\bf k},\omega)
  \tilde {\bf E}({\bf k},\omega)
\\[8pt]
\label{edefb}
  \epsilon({\bf k},\omega) 
  &=& 
  1 + \chi({\bf k}, \omega)
  \ ,
\end{eqnarray}
where the spatial and temporal Fourier transform of 
the susceptibility is
\begin{eqnarray}
  \tilde\chi({\bf k},\omega) 
  &=& 
  \int_{-\infty}^\infty  dt \! \int \!d^\nu x\, 
  \chi({\bf x},t) \,
   e^{ - i {\bf k}\cdot {\bf x} + i \omega t} 
   \label{f1ft_convention}
  \ ,
\end{eqnarray}
and the inverse transform is
\begin{eqnarray}
  \chi({\bf x},t) 
  &=& 
  \int_{-\infty}^\infty \frac{d\omega}{2\pi} \! 
  \int \! \frac{d^\nu k}{(2\pi)^\nu}\, 
  \tilde \chi({\bf k},\omega) \,  
  e^{i {\bf k}\cdot {\bf x} - i \omega t} 
  \ .
  \label{f1ft_inv_convention}
\end{eqnarray}
%%%
This sign convention  is consistent with the sign conventions of quantum 
mechanics: $-i\omega t$ and $i{\bf k}\cdot{\bf x}$ in
 (\ref{f1ft_inv_convention}).\,\footnote{
\footnoteskip
The quantum energy and momentum operators are $\hat E = i \hbar \,
\partial/\partial t$ and $\hat {\bf p} = - i \hbar \,\partial/\partial {\bf x}$, 
so that $\hat E\, e^{-i\omega t} = \hbar \omega \, e^{-i\omega t}$ and
$\hat {\bf p}\, e^{i {\bf k} \cdot {\bf x}} = \hbar {\bf k} \,  e^{i {\bf k} 
\cdot {\bf x}}$; therefore,  the energy and momentum Eigenvalues 
are $E=\hbar \omega$ and ${\bf p} = \hbar {\bf k}$. 
} % footnote 
Returning to  (\ref{eq_E_defhere}), we can write the Fourier transform
of the polarization vector as
\begin{eqnarray}
  \tilde{\bf P}({\bf k},\omega) 
  =
  \tilde{\bf D}({\bf k},\omega) - \tilde{\bf E}({\bf k}, \omega)
    =
  \chi({\bf k},\omega) \tilde {\bf E}({\bf k},\omega)
  \ .
\end{eqnarray}
The spatial form of Gauss's law ${\bm\nabla} \cdot {\bf P} =-\rho_{\rm ind}$ 
translates into $i {\bf k}\cdot \tilde{\bf P} = - \tilde\rho_{\rm ind}$ in Fourier 
space, and this allows us to write
\begin{eqnarray}
  \tilde\rho_{\rm ind}({\bf k},\omega) 
  = 
  - i \chi({\bf k},\omega) \,  {\bf k} \cdot  \tilde {\bf E}({\bf k},\omega) 
  =
  - \chi({\bf k},\omega)\,  k^2 \,\tilde\phi({\bf k}, \omega)
  =
  -\chi({\bf k},\omega) \, \tilde\rho({\bf k}, \omega)
  \ .
\end{eqnarray}
The last form has been expressed in terms of the potential $\phi$,
defined by ${\bf E} = - {\bm\nabla}\phi$, which in Fourier space becomes
$\tilde {\bf E}= - i {\bf k} \,\tilde\phi({\bf k}, \omega)$.
The Fourier transform of the dielectric is therefore
\begin{eqnarray}
  \chi({\bf k},\omega)
  &=&
  - \frac{\tilde\rho_{\rm ind}({\bf k},\omega)}{k^2  \tilde\phi({\bf k},\omega)}
  \ ,
  \label{eq_epsilon_calc}
\end{eqnarray}
and the problem reduces to calculating $\tilde\rho_{\rm ind}({\bf k},\omega)$.
The tool  for performing such calculations is kinetic theory. 

%%
%\pagebreak
\subsubsection{Calculation of the Dielectric Function of a Plasma}

Let us concentrate on an individual plasma component~$a$. In the absence 
of an applied field, we assume that species $a$ is in thermal equilibrium 
with itself, specified by a Maxwell-Boltzmann distribution $\bar f_a({\bf p})$ 
with inverse temperature $\beta_a = 1/T_a$ and charge $e_a$. When 
an external electric field is applied, this induces a charge 
density $\rho_{\rm ind}$.  The distribution 
function consequently departs from equilibrium, and the system is then 
specified by a new distribution $f_a({\bf x}, {\bf p}, t)$. Because collisions
are unimportant to the induced charges,  the distribution  $f_a$ satisfies 
the collisionless  Maxwell-Boltzmann equation
\begin{eqnarray}
  \frac{\partial f_a }{\partial t}~
  +
  {\bf v}\,\!\cdot\!
  \frac{\partial f_a}{\partial {\bf x}}
  +
  e_a{\bf E}  \cdot 
  \frac{\partial f_a}{\partial {\bf p}}
  = 0 
  \ ,
  \label{beqnocol}
\end{eqnarray}
where ${\bf v}={\bf p}/m_a$, and ${\bf E} = {\bf D} - {\bf P}$ is 
the total electric field seen by $a$. The electric field ${\bf E}$ 
is the sum of the applied field ${\bf D}$ and an induced  
contribution $-{\bf P}$. 
Note that the kinetic equation (\ref{beqnocol}) implies charge 
conservation. That is to say, if we define the charge density and 
electric current by
\begin{eqnarray}
  \rho_a({\bf x},t) 
  &=& \int  \!\frac{d^\nu p}{(2\pi \hbar)^\nu}\, e_a\, f_a({\bf x},{\bf p},t)
\\[5pt]
  {\bf J}_a({\bf x},t) 
  &=& \int  \!\frac{d^\nu p}{(2\pi \hbar)^\nu}\, e_a {\bf v}_a \, f_a({\bf x},
  {\bf p},t) 
  \ ,
\end{eqnarray}
then these quantities satisfy the continuity equation
\begin{eqnarray}
  \frac{\partial \rho_a }{\partial t}~
  +
  {\bm \nabla}\!\cdot\! {\bf J}_a   = 0 \ .
\end{eqnarray}
This is because the last term in the kinetic equation (\ref{beqnocol}) 
is a total divergence in momentum space, and therefore integrates 
to zero by the divergence theorem. It is reassuring to see  charge 
conservation arising directly from the kinetic equation. Indeed, all 
of hydrodynamics can be recovered from kinetic theory, although 
this would take us well beyond the scope of these notes. 
For this calculation, we start by expressing $f_a$ in terms of a 
{\em small} perturbation~$h_a$, 
\begin{eqnarray}
  f_a({\bf x},{\bf p},t) 
  &=&
   \bar f_a({\bf p}) 
  +
  h_{a}({\bf x},{\bf p},t) 
  \ . 
  \label{eq_f_plus_h}
\end{eqnarray}
Upon substituting (\ref{eq_f_plus_h}) back into (\ref{beqnocol}) 
and work to first order. Since the induced electric field ${\bf E}$ 
is first order, we can neglect the small second-order term 
$e {\bf E} \cdot \partial h/\partial {\bf p}$, and write the kinetic 
equation as
\begin{eqnarray}
  \frac{\partial h_a }{\partial t}~
 +
  {\bf v}\,\!\cdot\!
   \frac{\partial h_a}{\partial {\bf x}}
+
  e_a \, {\bf E}  \cdot
   \frac{\partial \bar f}{\partial {\bf p}}
  = 0 
  \ .
\label{formone}
\end{eqnarray}
It is often convenient to express the electric field in terms of
a potential, 
\begin{eqnarray}
  {\bf E} 
  &=&
  - {\bm \nabla} \phi 
  \equiv
  -\frac{\partial \phi}{\partial {\bf x }}
  \ ,
\end{eqnarray}
in which case the transport equation (\ref{formone}) takes the form
\begin{eqnarray}
  \frac{\partial h_a}{\partial t}~
  +
  {\bf v}\!\cdot\!
  \frac{\partial h_a}{\partial {\bf x }}
  =
  e_a \, \frac{\partial \phi}{\partial {\bf x }} \cdot 
  \frac{\partial \bar f}{\partial {\bf p }} 
   \ .
\label{eq_first_order_h}
\end{eqnarray}
To solve this equation we take the space and time Fourier, thereby giving
\begin{eqnarray}
\label{fonesolft}
  -i \omega \tilde h_a
  +
  i {\bf k}\!\cdot\!{\bf v} \tilde h_a
  =
  e_a \,\tilde\phi({\bf k},\omega)\, i {\bf k} \cdot 
  \frac{\partial \bar f }{\partial {\bf p}} 
  \ ,
\end{eqnarray}
where $\tilde h_a$  is a function of ${\bf k}$, {\bf p}, and $p$. We can 
now solve (\ref{fonesolft}) for the perturbation, 
\begin{eqnarray}
 \tilde h_a({\bf k}, {\bf p}, \omega)
  &=&
  - \frac{e_a \,\tilde\phi({\bf k},\omega) }{\omega - {\bf k} \cdot {\bf v}} 
  \, {\bf k} \!\cdot\! 
  \frac{\partial \bar f({\bf p})}{\partial {\bf p}} 
  \ .
\label{fonesolft_sol}
\end{eqnarray}
Note that $\tilde\phi({\bf k})$ is the Fourier transform of 
the {\em applied} potential, and it not given by (\ref{CoulombXform}). 
In performing the inverse Fourier transform to recover the
correlation function $h({\bf x}, {\bf p}, t)$ in space and time, 
we must integrate over ${\bf k}$ and $\omega$. By convention,
we hold ${\bf k}$ fixed and integrate over the variable $\omega$
first. The integration contour for $\omega$ lies in the complex 
$\omega$-plane slightly above  the real axis. This avoids the 
pole at $\omega = {\bf k} \cdot {\bf v}$ when integrating over 
$\omega$, and establishes the proper causality for $h({\bf x}, 
{\bf p}, t)$. This choice of contour is equivalent to integrating 
over real values of $\omega$, but adding a small complex term 
$i\eta$ to the numerator in (\ref{fonesolft_sol}).  We can therefore 
write the Fourier transform of the correlation function as 
\begin{eqnarray}
 \tilde h_a({\bf k}, {\bf p}, \omega)
  &=&
  - \frac{e_a \,\tilde\phi({\bf k}, \omega) }
  {\omega - {\bf k} \cdot {\bf v} + i \eta} 
  \, {\bf k} \cdot \frac{\partial \bar f({\bf p})}{\partial {\bf p}} 
  \ ,
\label{fonesolft_sol_eta}
\end{eqnarray}
where limit $\eta \to 0^+$ is understood. We can always restore
the correlation function to space and time variables by performing 
the inverse Fourier transform, 
\begin{eqnarray}
  h_{a}({\bf x},{\bf p},t)
  &=& 
  -\int \frac{d^\nu k}{(2\pi)^\nu} \,
  \frac{d\omega}{2\pi} \,
  e^{ i{\bf k}\cdot {\bf x} -i \omega t} ~
  \frac{e_a}{k^2}\,
  \frac{1}{\omega - {\bf k}\!\cdot\!{\bf v}_a + i\eta} ~
  {\bf k}\!\cdot\! \frac{\partial \bar f_a({\bf p})}{\partial {\bf p}}
  {\scriptscriptstyle \times}\, k^2 \tilde\phi({\bf k},\omega)
  \ ,
  \label{eq_f1aDist}
\end{eqnarray}
where we have factored out the term $k^2 \tilde \phi({\bf k})$ 
for convenience. Note that the form of $\tilde \phi({\bf k})$ is
unknown, but it will cancel from the dielectric function. The 
induced charge density therefore becomes
\begin{eqnarray}
  \rho_{\rm ind}({\bf x},t) 
  &=& 
  {\sum}_b \int  \!\frac{d^\nu p_b}{(2\pi \hbar)^\nu}\, e_b \, 
  h_b({\bf x},{\bf p}_b,t)
   \ .
\end{eqnarray}
It is actually more convenient to continue working in Fourier space, 
and using (\ref{fonesolft_sol_eta}) allows us to express the induced 
charge density as
\begin{eqnarray}
  \tilde\rho_{\rm ind}({\bf k},\omega) 
  &=& 
  {\sum}_b\int  \!\frac{d^\nu p_b}{(2\pi \hbar)^\nu}\, e_b \,
  \tilde h_b ({\bf k},{\bf p}_b, \omega) 
  \\[5pt]
  &=&
  -{\sum}_b\int  \!\frac{d^\nu p_b}{(2\pi \hbar)^\nu}\, 
  \frac{e_b^2}{k^2}\,
  \frac{1}{\omega - {\bf k}\!\cdot\!{\bf v}_b + i\eta} \,  {\bf k}\!\cdot\! 
  \frac{\partial \bar f_b({\bf p}_b)}{\partial {\bf p}_b} 
  {\scriptscriptstyle \times}\,  k^2 \tilde\phi({\bf k}, \omega)
   \ .
\end{eqnarray}
The susceptibility is therefore
\begin{eqnarray}
  \tilde\chi({\bf k}, \omega)
  =
  -
  \frac{\tilde\rho_{\rm ind}({\bf k},\omega)}{k^2 \tilde\phi({\bf k},\omega)}
   =
    {\sum}_b\, \frac{e_b^2}{k^2} 
  \int \! \frac{d^\nu p_b}{(2\pi \hbar)^\nu}  \,
  \frac{1}{\omega - {\bf k}\!\cdot\!{\bf v}_b + i \eta}\, {\bf k} \cdot 
  \frac{\partial \bar f_b({\bf p}_b) }{\partial {\bf p}_b } 
  \ ,
\end{eqnarray}
which gives (\ref{epsilonmulti}) for the dielectric function
$\epsilon({\bf k}, \omega) = 1 + \chi({\bf k}, \omega)$. 

\pagebreak
\section{Coulomb Energy Transfer Rates in Arbitrary Dimensions}
\label{sec_coulomb_energy_xfer}

This section is a review of the basic BPS formalism, presented here for 
completeness, with an emphasis on the role of analytic continuation
of the spatial dimension $\nu$. We turn now to calculating Coulomb 
energy exchange rates in the multi-component plasma described in the 
previous section. The charged particle stopping power and the temperature
equilibration rate between plasma species of different temperatures are 
the two canonical examples I have in mind.

\subsection{Coulomb Energy Exchange}

The single-particle distribution function for plasma species $a$ 
satisfies a general kinetic equation of the form
\begin{eqnarray}
  \frac{\partial f_a }{\partial t}~
  ~+~
  {\bf v}_a\,\!\cdot\!
  \frac{\partial f_a}{\partial {\bf x}}
  ~+~
  {\bf F}_a \!\cdot\!
  \frac{\partial f_a}{\partial {\bf p}}
  = 
  \left( \frac{\partial f_a }{\partial t}\right)_c
  \equiv
 {\sum}_b K_{ab}^\nu[f]  
  \ ,
  \label{be_col}
\end{eqnarray}
where the velocity is given by ${\bf v}_a = {\bf p}/m_a$, and ${\bf F}_a$ 
is the total force acting on $a$ at ${\bf x}$, {\em e.g.} ${\bf F}_a = e_a 
{\bf E({\bf x})}$ in the case of an external electric field. The scattering 
rate $(\partial f_a/\partial t)_c$ 
is a generic expression that accounts for the effects of scattering or collisions.
It is calculated in kinetic theory text books under various conditions, the most 
relevant being for the Boltzmann kernel $B_{ab}[f]$ and the Lenard-Balescu 
kernel $L_{ab}[f]$. For now, we will keep the form of the kernel generic and
simply write $K_{ab}^\nu[f]$. 
For stopping power calculations and other Coulomb energy exchange
processes, we will set the external force to zero, so the distribution
function $f_a$ satisfies 
\begin{eqnarray}
  \frac{\partial f_a}{\partial t} 
  + 
  {\bf v}_a \! \cdot \frac{\partial f_a}{\partial {\bf x}}
  &=& 
  {\sum}_b K_{ab}^\nu[f]  
   \ .
\label{kequgen_Fzero}
\end{eqnarray}
The kinetic energy density of plasma species $a$ is defined by
\begin{eqnarray}
  {\cal E}_a
  = 
  \int\!\frac{d^\nu p_a}{(2\pi\hbar)^\nu} \,
  \frac{p_a^2 }{2 m_a}~ f_a({\bf p}_a,t) \ ,
\label{Ea}
\end{eqnarray}
where $f_a$ is the corresponding distribution function. 
The stopping power is related to the rate of energy loss by
\begin{eqnarray}
  \frac{d{\cal E}_a}{dx} 
  &=& 
  \frac{1}{v_a}\,\frac{d{\cal E}_a}{dt}
  =
 \frac{1}{v_a} \int\!\frac{d^\nu p_a}{(2\pi\hbar)^\nu} \,
  \frac{p_a^2 }{2 m_a}~ \frac{\partial f_a({\bf p}_a,t)}{\partial t} 
  \ .
\label{dEdx_def}
\end{eqnarray}
Using the kinetic equation (\ref{kequgen_Fzero}), the divergence 
over ${\bf x}$ integrates to zero, and we find
\begin{eqnarray}
  \frac{d {\cal E}_a}{dt} 
  = 
  \int\!\frac{d^\nu p_a}{(2\pi\hbar)^\nu} \,
  \frac{p_a^2 }{2 m_a}~
  \frac{\partial f_a({\bf p}_a,t)}{\partial t}
  = 
  {\sum}_b\int \frac{d^\nu p_a}{(2\pi\hbar)^\nu} \,
  \frac{p_a^2 }{2 m_a}~ K_{ab}^\nu[f] \ .
\label{dedtbenu}
\end{eqnarray}
We can therefore identify the rate of change in the kinetic-energy
density of species $a$ resulting from its Coulomb interactions with
species $b$ by 
\begin{eqnarray}
  \frac{d {\cal E}_{ab}}{dt} 
  = 
  \int \frac{d^\nu p_a}{(2\pi\hbar)^\nu} \,
  \frac{p_a^2 }{2 m_a}~ K_{ab}^\nu[f] \ .
\label{dedtbenuab}
\end{eqnarray}

\subsection{\label{sec:reduction} Dimensional Reduction of BBGKY}
 
As we have seen, moving to an arbitrary dimension $\nu$ acts as a regulator, 
rendering the kinetic equations, both the Boltzmann equation (BE) and the 
Lenard-Balescu equation (LBE), finite in  their respective dimensional regimes.  
Dimensional regularization, however,  does far more than this.  A more subtle 
advantage of working in a general dimension is that it acts as a ``physics sieve'', 
in that it selects the proper scattering kernel to leading order (LO) in $g$ in 
the dimension at hand:

\vbox{
\begin{eqnarray}
  \text{BBGKY in}~\nu>3~\Rightarrow~
  \frac{\partial f_a}{\partial t} + 
  {\bf v}_a \cdot  \frac{\partial f_a}{\partial {\bf x}}
  &=& 
  {\sum}_b B_{ab}[f]   
 ~~~\,\text{to LO in } g 
 \ ,
\label{BEnu}
\end{eqnarray}
where $B_{ab}$ is the $\nu$-dimensional Boltzmann scattering kernel, and
\begin{eqnarray}
  \text{BBGKY in}~\nu<3~\Rightarrow~
  \frac{\partial f_a}{\partial t} + 
  {\bf v}_a \cdot \frac{\partial f_a}{\partial {\bf x}}
  &=& 
  {\sum}_b L_{ab}[f]   
 ~~~\text{to LO in } g \ ,
\label{LBEnu}
\end{eqnarray}
} % vbox

\vskip0.2cm
\noindent
where $L_{ab}[f]$ is the $\nu$-dimensional scattering kernel for the 
Lenard-Balescu equation.  Proving this statement, which I call the
{\em dimensional reduction theorem},  is the main purpose of these 
notes. Figure~\ref{fig_bbgky_breakdown} serves as a useful pictorial 
representation of the theorem. 

%%
%\pagebreak
\subsubsection{\label{sec:BEnu} The Boltzmann Kernel}

In this subsection I will review the Boltzmann scattering kernel
in some detail. In formal work I will write the Boltzmann equation 
in schematic form as
\begin{eqnarray}
  \frac{\partial f_a}{\partial t} + 
  {\bf v}_a \! \cdot \! \frac{\partial f_a}{\partial {\bf x}}
  &=& 
  {\sum}_b B_{ab}[f]   ~~~~:\, \nu> 3 
  \ ,
\label{BEsimpnu}
\end{eqnarray}
or in calculations I will use the form 
\begin{eqnarray}
  B_{ab}[f]
  &=&
  \int \frac{d^\nu p_b}{(2\pi\hbar)^\nu}\,
  \vert{\bf v}_a - {\bf v}_b \vert\, d\sigma_{ab}\,
  \bigg\{
  f_a({\bf p}_a^\prime) f_b({\bf p}_b^\prime)
  -
  f_a({\bf p}_a) f_b({\bf p}_b)
  \bigg\} 
  \ .
\label{BEfeSigma}
\end{eqnarray}
BPS included the quantum effects of two-body Coulomb scattering 
by replacing the classical cross section by the corresponding quantum 
transition amplitude $T(ab \to a^\prime b^\prime) \equiv T_{a^\prime 
b^\prime;\, ab}(W, q^2)$, where $W$ is the center-of-mass energy and
$q^2$ is the square of the momentum exchange during the collision.
The cross section $d\sigma_{ab}$ and the square of the scattering 
amplitude $\vert T_{a^\prime b^\prime; a b}(W,q^2) \vert^2$ are related by
\begin{eqnarray}
  \vert{\bf v}_a - {\bf v}_b \vert\,
  d\sigma_{ab}
 &\equiv& 
 \int
  \frac{d^\nu p_a^\prime}{(2\pi\hbar)^\nu}\,
  \frac{d^\nu p_b^\prime}{(2\pi\hbar)^\nu}\,
  \big\vert T_{a^\prime b^\prime;\,ab}(W, q^2) \big\vert^2\, 
   (2\pi\hbar)^\nu\,\delta^\nu\!\Big( {\bf p}_a^\prime + {\bf p}_b^\prime - 
  {\bf p}_a - {\bf p}_b \Big) {\scriptstyle \times}
  \nonumber \\ && \hskip4.4cm
  (2\pi\hbar) \delta\Big(E_a^\prime + E_b^\prime - E_a - E_b\Big) 
   \ ,
\label{dSigma}
\end{eqnarray}
and the Boltzmann equation can then be written
\begin{eqnarray}
\nonumber
  B_{ab}[f]
  &=&
  \int \frac{d^\nu p_a^\prime}{(2\pi\hbar)^\nu}\,
  \frac{d^\nu p_b^\prime}{(2\pi\hbar)^\nu}\,
  \frac{d^\nu p_b}{(2\pi\hbar)^\nu}\,
  \big\vert T_{a^\prime b^\prime;\,ab}(W, q^2)\big\vert^2\, 
  \bigg\{
  f_a({\bf p}_a^\prime) f_b({\bf p}_b^\prime)
  -
  f_a({\bf p}_a) f_b({\bf p}_b)
  \bigg\} 
\\[5pt] && \hskip0.5cm 
  (2\pi\hbar)^\nu\,\delta^\nu\!\Big( {\bf p}_a^\prime + {\bf p}_b^\prime - 
  {\bf p}_a - {\bf p}_b \Big)\,
  (2\pi\hbar) \delta\Big(E_a^\prime + E_b^\prime - E_a - E_b\Big) \ .
\label{BEfe}
\end{eqnarray}
The latter expression is more useful for formal manipulations, even in
the classical regime, where one can define a classical ``transition amplitude''
$T_{a^\prime b^\prime;\,ab}$ from (\ref{dSigma}) by using the classical 
Rutherford cross section for $d\sigma_{ab}$. Surprisingly, the classical 
amplitude is identical to quantum Born amplitude.  When $\nu>3$,
expression (\ref{BEsimpnu}) allows us to write the rate of change of
the energy density resulting from the now finite Boltzmann kernel as
\begin{eqnarray}
  \frac{d{\cal E}_{ab}^\smGT}{dt}
  =
  \int \!\frac{d^{\,\nu}p_a}{(2\pi\hbar)^\nu}~\frac{p_a^2}{2 m_a}~
  B_{ab}[f] ~~~~ : ~~ \nu>3 \ .
\label{dedtgtthree}
\end{eqnarray}
I have used a ``greater than'' superscript to remind us that we should
calculate (\ref{dedtgtthree}) in dimensions greater than three.  

As we have discussed, in dimensions greater than three the derivation 
of the Boltzmann equation for Coulomb scattering is rigorous and finite. 
This is because the short distance physics of the
Coulomb potential is dominant in dimensions $\nu>3$, and the Boltzmann
equation is designed to capture short distance scattering physics. Furthermore,
the long distance physics, where the Boltzmann equation breaks down in
three dimensions, falls off faster than $1/r$ at large distances, thereby 
rendering the scattering finite for $\nu > 3$. 
It should not be a surprise that a simple scaling argument shows why 
$B_{ab}[f]$ is finite for $\nu>3$. Write $\epsilon = \nu - 3 > 0$, and note
 that the amplitude scales as 
$\vert T \vert^2 \sim 1/q^2$, for momentum transfer ${\bf q}$. Finally,
a $\delta$-function in $q$ contributes a power $q^{-1}$, so that
\begin{eqnarray}
\nu > 3 :
  B_{ab}
  &\sim&
  \int dq \, q^{\nu-1} \cdot \frac{1}{q^2} \cdot \frac{1}{q}
  \sim
  \int dq \, q^{\nu -4}
  \sim
    \int dq \, q^{-1 + \epsilon }
  \\[5pt]
 & \sim&
  q^{\epsilon} \to 0 ~~~\text{as}~~ q \to 0 
  \ .
\end{eqnarray}
The momentum transfer is related to the corresponding wavenumber 
by $q = \hbar k$, and we see that small values of $q$ correspond to 
large distances. This means that the Boltzmann equation does not 
possess an IR divergence for $\nu > 3$. 

\pagebreak
\subsubsection{\label{sec:LBEnu} The Lenard-Balescu Kernel}

I usually write the Lenard-Balescu equation in schematic form, 
\begin{eqnarray}
  \frac{\partial f_a}{\partial t} + 
  {\bf v}_a \! \cdot \! \frac{\partial f_a}{\partial {\bf x}}
  &=& 
  {\sum}_b L_{ab}[f] ~~~~ : \nu < 3 
  \ ,
\label{LBEsimpnu}
\end{eqnarray}
although for calculations, we will use the explicit form\,\footnote{
\footnoteskip
  Note that Eq.~(3.57) for $L_{ab}[f]$ in Ref.~\cite{bps2} contains 
  a spurious integration over the momentum ${\bf p}_a$. Fortunately,
  this typo was innocuous and did not affect the results that followed. 
} % end footnote
\begin{eqnarray}
  L_{ab}[f]
  &=&
  - \frac{\partial}{\partial {\bf p}_a} \cdot {\bf J}({\bf p}_a)
  \\[8pt] 
  {\bf J}({\bf p}_a)
  &=&
  \int\! 
  \frac{d^\nu p_b}{(2\pi\hbar)^\nu}\,\frac{d^\nu k}{(2\pi)^\nu}
  \, 
  {\bf k}   \,
  \bigg\vert \frac{e_a\, e_b}{k^2\, \epsilon({\bf k},{\bf k} \cdot {\bf v}_a)}
  \bigg\vert^2 
  \pi\,\delta({\bf k} \cdot {\bf v}_a - {\bf k} \cdot  {\bf v}_b)
  \nonumber\\[5pt]
  && \hskip4.5cm 
  \bigg[
  {\bf k}\!\cdot\! \frac{\partial}{\partial {\bf p}_a}
  -
  {\bf k}\!\cdot\! \frac{\partial}{\partial {\bf p}_b}
  \bigg]  
  f_a({\bf p}_a) \,  f_b({\bf p}_b) \ ,
\label{dedteigtAA}
\end{eqnarray} 
where ${\bf v}_a={\bf p}_a/m_a$ is really an integration variable.
The dielectric function $\epsilon$ is given by  (\ref{epsilonmulti}), 
which we repeat here for convenience with a change in summation 
index, 
\begin{eqnarray}
  \epsilon({\bf k},\omega) 
  = 
  1 + {\sum}_c \, \frac{e_c^2}{k^2} \int \frac{d^\nu p_c}{(2\pi\hbar)^\nu}\, 
  \frac{1}{\omega - {\bf k} \!\cdot\!  {\bf v}_c+ i \eta}\, {\bf k} \cdot 
   \frac{\partial f_c({\bf p}_c) }{\partial {\bf p}_c} \,,
\label{epsilon}
\end{eqnarray}
and the prescription $ \eta \to 0^+ $ is implicit, defining the
correct retarded time response.
Therefore, when $\nu<3$, the rate (\ref{LBEnu})  
allows us to express
\begin{eqnarray}
  \frac{d{\cal E}_{ab}^\smLT}{dt}
  =
  \int \!\frac{d^{\,\nu}p_a}{(2\pi\hbar)^\nu}~\frac{p_a^2}{2 m_a}~
  L_{ab}[f] ~~~~ : ~~ \nu<3 \ .
\label{dedtltthree}
\end{eqnarray}
I have used a ``less than'' superscript to remind us that we should
calculate (\ref{dedtltthree}) in dimensions less than three, where
it is finite and well defined. 

In dimensions less than three one finds a complementary situation to
the Boltzmann equation, namely, the derivation of the Lenard-Balescu
equation is rigorous and finite when $\nu < 3$. This is
because the long distance physics of the Coulomb potential is dominant
in dimensions $\nu<3$, and the Lenard-Balescu equation is designed to
capture such long distance physics. Furthermore, the Coulomb potential 
falls off faster than $1/r$ at large distances, where the LBE breaks down
in three dimensions, and this renders the kernel finite in $\nu<3$.  
A scaling argument shows why $L_{ab}[f]$ is finite for $\nu<3$. 
Since $\epsilon = \nu - 3 < 0$, we will work with the quantity
$\vert \epsilon\vert > 0$.  From (\ref{dedteigtAA}),  the kernel 
contains an obvious linear term $k$, and a factor $k^{-4}$ arising
from the Fourier transform of the potential. Note that 
$\epsilon({\bf k}, {\bf k} \cdot {\bf v}) \to 1$ for large values of $k$,
so the dielectric function does not change the $k \to \infty$ scaling
behavior. 
The $\delta$-function gives a factor $k^{-1}$, and the
${\bf p}$-derivative terms provide a compensating factor of $k$,
so that
\begin{eqnarray}
\nu < 3 :
  L_{ab}
  &\sim&
  \int dk \, k^{\nu-1} \cdot k \cdot \frac{1}{k^4} \cdot \frac{1}{k}
  \cdot k
  \sim
  \int dk \, k^{\nu-4}
  \sim
  \int dk \, k^{-1 -  \vert \epsilon \vert}
  \\[5pt]
 & \sim&
  k^{-\vert \epsilon \vert} \to 0 ~~~\text{as}~~ k \to \infty
  \ .
\end{eqnarray}
Since large values of $k$ corresponds to small distances,
this means that the LBE is finite at small distances for $\nu < 3$. 

\subsection{Completing the Picture: The Rate and Analytic Continuation}

As a matter of completeness, let us finish the calculation of the rate
$d{\cal E}_{ab}/dt$. Recall that we have calculated the rates  the rates 
$d{\cal E}_{ab}^\smGT /dt$ and $d{\cal E}_{ab}^\smLT /dt$ to leading
order (LO) in $g$, and they contain simple poles $1/(\nu-3)$, and they
take the general  form
\begin{eqnarray}
  \frac{d{\cal E}_{ab}^\smGT}{dt}
  &=& 
  H(\nu)\,\frac{g^2}{\nu-3} 
  +
  {\cal O}(\nu-3) 
  \hskip1.15cm  \text{is LO in $g$ when}~ \nu > 3 
\label{dedtonecala}
\\[8pt]
  \frac{d{\cal E}_{ab}^\smLT}{dt}
  &=&
  G(\nu)\, \frac{g^{\nu-1}}{3 - \nu} 
  +
  {\cal O}(\nu-3) 
  ~~~~~~~~\,
  \text{is LO in $g$ when}~ \nu < 3 
  \ ,
  \label{dedtonecalb}
\end{eqnarray}
where $H(\nu)$ and $G(\nu)$ are coefficients that depend upon $\nu$.
The heavy lifting for a real process is in calculating the functions $H(\nu)$
and $G(\nu)$ using the exact expressions for $B_{ab}$ and $L_{ab}$. 
Once these calculations have been completed, in order to compare the 
rates  (\ref{dedtonecala}) and  (\ref{dedtonecalb}), we must then analytically 
continue to a {\em common} value of the dimension $\nu$ (and then take 
the limit $\nu \to 3$). Analytically continuing the spatial dimension makes 
sense because we can view the quantities $d {\cal E }^\smGT_{ab}/dt$ and 
$d {\cal E }^\smLT_{ab}/dt$ as
functions of a complex parameter $\nu$, even though they were only
calculated for positive integer values of $\nu$. This is analogous to
analytically continuing the factorial function on the positive
integers to the Gamma function on the complex plane.  For
definiteness, I will analytically continue  (\ref{dedtonecalb}) to
$\nu>3$, in which case $g^{\nu-1} = g^{2 + (\nu-3)}$ becomes subleading 
relative to the $g^2$ dependence of  (\ref{dedtonecala}), so that
\begin{eqnarray}
  \frac{d{\cal E}_{ab}^\smLT}{dt}
  &=&
  -G(\nu)\, \frac{g^{2+(\nu-3)}}{\nu-3} 
  +
  {\cal O}(\nu-3) 
  ~~~\text{is NLO in $g$ when } \nu > 3 \ .
\label{NLOgterm}
\end{eqnarray}
This is illustrated in Fig.~\ref{fig:NLOdedt}. To finish calculating the rates, 
we need to work consistently to linear order in the small parameter
$\epsilon = \nu-3$; therefore, we should expand $H(\nu)$ and $G(\nu)$
to first order in $\epsilon$, 
\begin{eqnarray}
  H(\nu) 
  &=&
  -A + \epsilon \,H_1 + {\cal O}(\epsilon^2)
\label{Hexp}
\\[5pt]
  G(\nu) 
  &=&
  -A + \epsilon \,G_1 + {\cal O}(\epsilon^2) \ .
\label{Gexp} 
\end{eqnarray}
It is crucially important here that $H(\nu)$ and $G(\nu)$ give the
same value at $\nu=3$, a term that I have called $-A$ in (\ref{Hexp})
and (\ref{Gexp}), otherwise the divergent poles will not cancel. 
\begin{figure}[t!]
\begin{picture}(120,120)(-50,0)
\put(-101,60){\vector(1,0){235}}
\put(139,58){$\nu$}

\put(0,56){\line(0,1){8}}
\put(-2.3,57.3){$\bullet$}
\put(-2.0,49){${\scriptstyle 3}$}

\put(-60,66){$d{\cal E}_{ab}^\smLT/dt$}
\put(-100,44){$g^{\nu-1}=g^{2-(3 - \nu)}$}
\put(-60,30){$\Downarrow$}
\put(-80,15){$g^{2-\vert \nu-3 \vert}$}
\put(-120,0){LO: large when $g \ll 1$}

\put(30,66){$d{\cal E}_{ab}^\smLT/dt$}
\put(55,44){$g^{\nu-1}=g^{2+(\nu-3)}$}
\put(95,30){$\Downarrow$}
\put(75,15){$g^{2+\vert \nu-3 \vert}$}
\put(20,0){NLO: small when $g \ll 1$}

\put(0,75){\oval(45,45)[t]}
\put(22,72){\vector(0,-1){1}}
\put(-40,115){analytically continue}
\put(-42,105){around the $\nu\!=\!3$ pole}

\end{picture}
\caption{\captionskip The analytic continuation of $d{\cal E }_{
ab}^\smLT/dt$ from $\nu<3$ to the region $\nu>3$ in the complex
$\nu$-plane. The same expression can be used for $d{\cal E }_{
ab}^\smLT/dt$ throughout the complex plane since the pole at
$\nu=3$ can easily be avoided. The quantity $d{\cal E }_{ab}^\smLT
/ dt \sim g^{2+(\nu-3)}$ is leading order in $g$ for $\nu <
3$. However, upon analytically continuing to $\nu>3$ we find that
$d{\cal E}_{ab}^\smLT/dt\sim g^{2+\vert \nu-3\vert}$ is
next-to-leading order in $g$ relative to $d{\cal E}_{ab}^\smGT/dt
\sim g^2$.  }
\label{fig:NLOdedt}
\end{figure}
Finally, upon writing $g^\epsilon = \exp\{\epsilon \ln g\}$ in
(\ref{NLOgterm}), and expanding to first order in $\epsilon$, we find
\begin{eqnarray}
  \frac{g^\epsilon}{\epsilon}
  = 
  \frac{1}{\epsilon} 
  +
  \ln g + {\cal O}(\epsilon) \ .
\label{gexp}
\end{eqnarray}
This is where the nonanalyticity in $g$ arises, {\em i.e.} the $\ln g$
term, and we can now express the rates as
%%%
\begin{eqnarray}
  \frac{d{\cal E}_{ab}^\smGT}{dt}
  &=& 
  -
  \frac{A}{\nu-3} \, g^2
  +
  H_1\,g^2
  +
  {\cal O}(\nu-3;g^3)
  \hskip2.7cm ~\nu>3 
\label{LOcalcA}
\\[5pt]
  \frac{d{\cal E}_{ab}^\smLT}{dt}
  &=& 
    \phantom{-}
  \frac{A}{\nu-3} \, g^2 
  - 
  G_1 \,g^2
  - 
  A\, g^2 \ln g 
  +
  {\cal O}(\nu-3;g^3) 
  \hskip0.75cm ~\nu>3 
  \ .
\label{LOcalcB}
\end{eqnarray}
These expressions hold in the common dimension $\nu > 3$, and
to obtain the leading and next-to-leading order result in three
dimensions, we add and take the limit:
\begin{eqnarray}
  \frac{d{\cal E}_{ab}}{dt}
  =
  \lim_{\nu \to 3^+}
  \left[
  \frac{d{\cal E}_{ab}^\smGT}{dt}
  +
  \frac{d{\cal E}_{ab}^\smLT}{dt}
  \right] + {\cal O}(g^3) 
  =
  - A g^2 \ln g + B g^2 + {\cal O}(g^3) \ ,
\label{dedtorderg3}
\end{eqnarray}
with $B=H_1-G_1$.  This gives the energy exchange rate from
Coulomb interactions between plasma species, accurate to leading 
order and next-to-leading order in $g$, in terms of the
Coulomb logarithm, 
\begin{eqnarray}
  \frac{d{\cal E}_{ab}}{dt}
  =
  - A g^2 \ln C \,g + {\cal O}(g^3)
  \ , 
\label{dedtorderg4}
\end{eqnarray}
where $\ln C = -B/A$. The quantity $L= \ln C g$ is known as the
{\em Coulomb logarithm}.

\pagebreak
\section{The BBGKY Hierarchy in Arbitrary Dimensions}
\label{sec_bbgky_arbitrary}

Now that we have reviewed statistical mechanics and Coulomb physics  
in \hbox{$\nu$-dimensions}, we can start addressing the main claim 
of these notes, namely, that to leading order in the plasma coupling 
$g$,  the BBGKY hierarchy reduces to the Boltzmann equation in $\nu>3$, 
and to the Lenard-Balescu equation in $\nu<3$. To prove this, we 
first derive the BBGKY hierarchy in a general number of dimensions, 
primarily to establish notation, and because it will be the starting
point in our derivation. We then develop perturbation theory in 
powers of the coupling $g$, and we calculate the BBGKY equations 
to order $g^2$. To solve these equations, we must make some 
approximations, and we discuss how the two regimes $\nu < 3$ 
and $\nu > 3$ affect the validity of the approximations. More
precisely, we will show that complementary 2-point correlations
should be dropped in these respective dimensional regimes. 
This will serve as a starting point for Section~\ref{sec_be_fr_bbgky}  
on the Boltzmann equation and 
Section~\ref{sec_LBE} on the Lenard-Balescu equation.  The 
derivation of the BBGKY hierarchy presented here was adapted 
from a three dimensional argument given in Clemmow and 
Dougherty\,\cite{cd}, which will be our primary reference 
throughout this section.

\subsection{Liouville's Theorem and Ensemble Averages}

For simplicity, we consider a plasma with a single species of particle. We
can add a charge neutralizing background if desired; for example, one
might consider an electron plasma with fixed ions as the background. 
Multiple plasma species can (and soon will) be added. There are several 
ways of representing the state of many-particle systems such as plasmas. 
The first  is through the use of a phase-spaced called $\mu$, which is the  
$2\nu$-dimensional phase space $({\bf x}, {\bf p})$ of a single particle. 
The state of a plasma
with $N$ particles is given by specifying the \hbox{$\nu$-dimensional} 
positions ${\bf x}_i$ and momenta ${\bf p}_i$ for every particle 
\hbox{$i=1, 2, \cdots, N$} in the plasma. Each particle is represented by  
a  point $({\bf x}_i, {\bf p}_i)$ in $\mu$-space, and
the system looks like a swarm of $N$ particles, each interacting with all of 
the particles of the system. The $\nu$-dimensional spatial slice of $\mu$-space 
is the part of phase space that we observe with our eyes in the laboratory. 
The single particle distribution function $f_1({\bf x}, {\bf p})$ lives in
$\mu$-space, and specifies the number of particles within a phase
space element $d^\nu x\, d^\nu p$. A second way of representing
the state of a multi-particle system is through the $2\nu N$-dimensional 
phase  space defined by $({\bf x}_1, {\bf p}_1, \cdots, {\bf x}_\smN,  
{\bf p}_\smN)$. This larger phase space is called $\Gamma$-space, and by 
design, the entire system is represented by a single point in $\Gamma$, 
rather than the swarm of points in $\mu$. To simplify notation, we denote 
the $2\nu$-dimensional phase space for particle $i$ by $X_i = ({\bf x}_i, {\bf p}_i )$, 
so that points in $\Gamma$-space  are specified by coordinate values 
$(X_1, \cdots, X_\smN)$. Once the initial condition of the $N$-body system 
is specified in $\Gamma$-space, that is to say, once the locations and velocities 
of all $N$ particles are specified at some initial time $t=0$, then the subsequent 
evolution at any future time $t$ is uniquely determined. The system therefore 
traces out a path in $\Gamma$-space as it evolves in time. There are only two 
possible types of paths: either the path is periodic or it never intersects itself. 
This is related to the ergodic properties of the system.  

Let us now consider a large ensemble of systems in $\Gamma$-space, each
evolving in time, tracing out a unique path for every member of the ensemble.
We can think of building the ensemble by initially populating $\Gamma$-space 
uniformly among all possible initial conditions. None of the
systems in $\Gamma$-space interact, and every system evolves along its
own private non-intersecting trajectory in this $2\nu N$-dimensional space.
The world-lines in $\Gamma$ look like a tangle of non-intersecting spaghetti,
with each strand oriented forward in time, never looping back on itself. Since 
we have populated $\Gamma$-space with all possible initial 
configurations, then by the ergodic principle, ensemble averages in 
$\Gamma$-space give time averaged quantities as measured by experiment.
For an ensemble of systems in $\Gamma$-space, the ensemble density is 
defined by
\begin{eqnarray}
\nonumber
  \rho(X_1, \cdots, X_\smN, t) \, dX_1 \cdots dX_\smN
  &\equiv&
  \text{probability that a system selected from the ensemble}
  \\[-8pt] &&
  \text{lies within}~ dX_i ~\text{of}~ X_i 
   ~\text{for}~ i=1, \cdots, N,  ~\text{at time}~ t,
  \label{rho_def}
\end{eqnarray}
where the measures are defined by
\begin{eqnarray}
  dX_i
  = 
  \frac{d^\nu x_i \, d^\nu p_i}{(2\pi\hbar)^\nu}
  \ .
  \label{eq_dX_def}
\end{eqnarray}
The density is of course normalized to unity, 
\begin{eqnarray}
  \int \! dX_1 \cdots dX_\smN \,\rho = 1
  \ ,
  \label{rho_norm}
\end{eqnarray}
and $\rho$ is symmetric in each of its arguments $X_i$ (again, this relates 
to the ergodic mixing properties of the system).

Since the individual systems in $\Gamma$ are non-interacting, the density
satisfies the conservation equation
\begin{eqnarray}
  \frac{\partial\rho}{\partial t} + \sum_{i=1}^N
  {\bm \nabla}_{\smX_i} \cdot 
  \Big( \rho \dot X_i  \Big)
  = 0 
  \ .
  \label{rho_conserve_X}
\end{eqnarray}
It is more convenient to break the variables $X_i=({\bf x}_i, {\bf p}_i)$ into their 
space and velocity components, thereby giving
\begin{eqnarray}
  \frac{\partial\rho}{\partial t} 
  + 
  \sum_{i=1}^N\frac{\partial}{\partial {\bf x}_i} \cdot \Big(\rho  \dot {\bf x}_i \Big)
    + 
  \sum_{i=1}^N\frac{\partial}{\partial {\bf p}_i} \cdot \Big( \rho  \dot {\bf p}_i \Big)
  = 0 
  \ ,
  \label{rho_conserve_xv}
\end{eqnarray}
which we write in the form
\begin{eqnarray}
  && \frac{\partial\rho}{\partial t} 
  + 
  \sum_{i=1}^N \dot {\bf x}_i  \cdot \frac{\partial \rho}{\partial {\bf x}_i} 
  + 
  \sum_{i=1}^N \dot {\bf p}_i  \cdot \frac{\partial \rho}{\partial {\bf p}_i} 
  +
    \sum_{i=1}^N \rho \left(
    \frac{\partial}{\partial {\bf x}_i} \cdot \dot {\bf x}_i \,  
    + 
   \frac{\partial}{\partial {\bf p}_i} \cdot \dot {\bf p}_i 
  \right)
  = 0 
  \ .
  \label{rho_break}
\end{eqnarray}
Using Hamilton's equations of motion, 
\begin{eqnarray}
  \dot {\bf x}_i  &=& \phantom{-}\frac{\partial H}{\partial {\bf p}_i}
  \\[8pt]
    \dot {\bf p}_i  &=& - \frac{\partial H}{\partial {\bf x}_i}
  \ ,
\end{eqnarray}
where $H$ is the Hamiltonian, 
we see that the last term in Eq.~(\ref{rho_break}) vanishes,
\begin{eqnarray}
  \left(
    \frac{\partial}{\partial {\bf x}_i} \cdot \dot {\bf x}_i \,  
    + 
   \frac{\partial}{\partial {\bf p}_i} \cdot \dot {\bf p}_i 
   \right)
  = 
  \sum_{\ell=1}^\nu
  \left(
   \frac{\partial^2 H}{\partial x_i^\ell \,\partial  p_i^\ell}\,  
    -
   \frac{\partial^2 H}{\partial p_i^\ell \, \partial x_i^\ell}
  \right)
  = 0
  \ .
\end{eqnarray}
The density in $\Gamma$-space therefore satisfies Liouville's equation
\begin{eqnarray}
  && \frac{\partial\rho}{\partial t} 
  + 
  \sum_{i=1}^N {\bf v}_i  \cdot \frac{\partial \rho}{\partial {\bf x}_i} 
  + 
  \sum_{i=1}^N {\bf F}_i  \cdot \frac{\partial \rho}{\partial {\bf p}_i} 
  = 0 
  \ ,
  \label{rho_l}
\end{eqnarray}
where we have substituted ${\bf v}_i= \dot{\bf x}_i$ and  ${\bf F}_i = 
\dot{\bf p}_i$. The force ${\bf F}_i$
includes the effects from all the other particles, and therefore ${\bf F}_i$
depends upon all of the coordinates $X_1, \cdots, X_\smN$. 
Note that (\ref{rho_l}) takes the same form as the 
collisionless Boltzmann equation. This is because for non-interacting 
particles, the function $f({\bf x}, {\bf p},t)$ acts like the ensemble 
density $\rho(X_1, \cdots, X_N)$.

\subsection{The Hierarchy of Distribution Functions}

We now show that there is a hierarchy of distributions functions that
measure successively higher-order correlations in the plasma. 
Note that we can  define the average value of a general quantity
$Q = Q(X_1, \cdots, X_\smN)$, where we have not allowed for a 
possible explicit time dependence in $Q$ (although we could),
by 
\begin{eqnarray}
 \langle Q(X_1, \cdots, X_\smN, t) \rangle = \int dX_1 \cdots dX_\smN\,
  \rho(X_1, \cdots, X_\smN, t)   Q(X_1, \cdots, X_\smN) 
  \ .
\end{eqnarray}
Note that the time dependence of the average is due to that in
$\rho(t)$. Let us now define the one-particle function by
\begin{eqnarray}
  F(X) = \sum_{i=1}^N \delta(X - X_i)
  \ ,
\end{eqnarray}
where the particles are located  at $X_1, \cdots, X_\smN$.  Writing 
$X = ({\bf x}, {\bf p})$, we can recover the single-particle distribution 
by  performing the ensemble average
\begin{eqnarray}
  f_1({\bf x}, {\bf p}, t) = \langle F(X) \rangle
   \ .
\end{eqnarray}
To see this, we explicitly perform the ensemble average,
\begin{eqnarray}
 f_1({\bf x}, {\bf p}, t)
  &=&
  \int dX_1 dX_2 \cdots dX_\smN \, \rho(X_1, X_2, \cdots, X_\smN,t)
   \sum_{i=1}^N \delta(X - X_i)
   \\[5pt]
   &=&
    N \int dX_2 \cdots dX_\smN \, \rho(X, X_2, \cdots, X_\smN,t)
  \ .
\end{eqnarray}
The factor of $N$ occurs because $\rho$ is symmetric in its arguments,
and therefore each term in the sum over $\delta$-functions is identical.
Using the normalization   (\ref{rho_norm}) we see that
\begin{eqnarray}
\int \frac{d^\nu x \, d^\nu p}{(2\pi\hbar)^\nu}
\, f_1({\bf x}, {\bf p}, t)
  &=&
    N \int dX dX_2 \cdots dX_\smN \, \rho(X, X_2, \cdots, X_\smN,t)
    = N
  \ ,
\end{eqnarray}
as required. Therefore the single-particle distribution $f_1$ is 
automatically normalized correctly. This motivates the definition 
of the $s$-particle correlation function
\begin{eqnarray}
  f_s(X_1, \cdots, X_s) 
  =
  \frac{N!}{(N-s)!} \int dX_{s+1}\cdots dX_\smN \,
  \rho(X_1, \cdots, X_s, X_{s+1}, \cdots, X_\smN)
  \ ,
  \label{def_fs}
\end{eqnarray}
\vskip0.2cm
\noindent
which has the normalization
\begin{eqnarray}
  \int dX_1 \cdots dX_s \,f_s(X_1, \cdots, X_s) 
  =
  \frac{N!}{(N-s)!} 
  \ .
\end{eqnarray}
We have now defined a hierarchy of distribution functions $f_1, f_2, \cdots,
f_s, \cdots, f_{\smN-1},  f_\smN$, where $f_\smN \equiv \rho$. In the next 
section we will find a set of coupled kinetic equations for each $f_s$ by 
integrating over successive sub-spaces of the Liouville's equation.

\subsection{The BBGKY Hierarchy of Kinetic Equations}

Let us express the forces ${\bf F}_i$ in Liouville's equation (\ref{rho_l}) 
in terms of external forces ${\bf F}_i^{(0)}$ and the Coulomb interaction 
force ${\bf F}_i^{(j)}$, 
\begin{eqnarray}
  {\bf F}_i = {\bf F}_i^{(0)} + \sum_{j=1}^N {\bf F}_i^{(j)}
  =
  \sum_{j=0}^N {\bf F}_i^{(j)}
  \ ,
  \label{eq_Fi_sum}
\end{eqnarray}
where the Coulomb force on $i$ from $j$ takes the form
\begin{eqnarray}
  {\bf F}_i^{(j)}
  = 
    e^2 \, \frac{\Gamma(\nu/2)}{2 \pi^{\nu/2}}\,
    \frac{{\bf x}_i - {\bf x}_j}{\vert{\bf x}_i - {\bf x}_j\vert^\nu}
  \ .
\label{Fij_def}
\end{eqnarray}
We must exclude the $j=i$ term from the sum in (\ref{eq_Fi_sum});
or equivalently we can include the value $j=i$ in the sum by formally
setting ${\bf F}_i^{(i)}=0$. We now show that  the distribution functions 
$f_s$ satisfy the following coupled set of kinetic equations called the 
BBGKY hierarchy, 
\begin{eqnarray}
  &&
  \frac{\partial f_s}{\partial t} 
  +
   \sum_{i=1}^s {\bf v}_i  \cdot \frac{\partial f_s}{\partial {\bf x}_i} 
  +
   \sum_{i=1}^s \sum_{j=0}^s {\bf F}_i^{(j)}  \cdot \frac{\partial f_s}{\partial {\bf p}_i} 
  =
  -\sum_{i=1}^s \int dX_{s+1} \,
  {\bf F}_i^{(s+1)} \cdot \frac{\partial f_{s+1}}{\partial {\bf p}_i} 
  \ ,
  \label{eq_bbgky}
\end{eqnarray}

\vskip0.2cm
\noindent
where $s=1, \cdots N-1$.  The BBGKY hierarchy is the the ``$F = m a$'' 
of kinetic theory. These equations are completed by Liouville's equation
(\ref{rho_l}) for $x=N$, 
\begin{eqnarray}
  && \frac{\partial f_\smN}{\partial t} 
  + 
  \sum_{i=1}^N {\bf v}_i  \cdot \frac{\partial f_\smN}{\partial {\bf x}_i} 
  + 
  \sum_{i=1}^N \sum_{j=0}^N {\bf F}_i^{(j)}  \cdot 
  \frac{\partial f_\smN}{\partial {\bf p}_i} 
  = 0 
  \ .
  \label{rho_l_gzero}
\end{eqnarray}
We thus have a complete set of $N$ equations in $N$ variables $f_1, f_2, 
\cdots, f_s,  \cdots, f_\smN$, and a unique solution will exist. It should be 
emphasized that the BBGKY hierarchy is time reversal invariant. It is only 
upon {\em closing} the equations at some level, usually $s=1$ or $s=2$, 
that we introduce time non-invariance. In other words, it is closing the 
hierarchy of kinetic equations that introduces the arrow of time. Except 
under the most contrived of conditions, we cannot hope to find 
an exact solution, or even a numerical solution, as $N$ is macroscopically 
large. However, the BBGKY hierarchy is still an extremely useful piece
of theoretical machinery, particularly in more formal arguments, and 
provides for a deeper understanding of kinetic theory. 

To prove (\ref{eq_bbgky}), let us  integrate (\ref{rho_l}) over 
$dX_{s+1}\cdots dX_\smN$, and multiply by the normalization factor of $f_s$, 
thereby giving the exact equation
\begin{eqnarray}
  \frac{N!}{(N-s)!}
  \int dX_{s+1} \cdots dX_\smN 
  \left(
   \frac{\partial\rho}{\partial t} 
  + 
  \sum_{i=1}^N {\bf v}_i  \cdot \frac{\partial \rho}{\partial {\bf x}_i} 
  + 
  \sum_{i=1}^N \sum_{j=0}^N {\bf F}_i^{(j)}  \cdot \frac{\partial \rho}{\partial {\bf p}_i} 
  \right)
  = 
  0
  \ .
  \label{eq_three_terms}
\end{eqnarray}
The first two terms of (\ref{eq_three_terms}) are rather trivial
to evaluate, and they correspond to the first two terms of (\ref{eq_bbgky}), 
\begin{eqnarray}
  {\rm term1}
  &\!=\!&
   \frac{N!}{(N-s)!} \int dX_{s+1} \cdots dX_\smN \,  
   \frac{\partial\rho}{\partial t} 
   =
  \frac{\partial}{\partial t} \, \frac{N!}{(N-s)!}   \int dX_{s+1} \cdots dX_\smN \,
  \rho
  =
  \frac{\partial f_s}{\partial t}
  \label{rho_termone}
  \\[10pt]
  {\rm term2}
  &\!=\!&
  \frac{N!}{(N-s)!}  \sum_{i=1}^N  \int dX_{s+1} \cdots dX_\smN 
  {\bf v}_i  \cdot \frac{\partial \rho}{\partial {\bf x}_i} 
  =
   \sum_{i=1}^s {\bf v}_i  \cdot \frac{\partial f_s}{\partial {\bf x}_i} 
   \ .
   \label{eq_termtwo}
\end{eqnarray}
In expression (\ref{eq_termtwo}), note that the sum over $i$ has 
been truncated from $N$ to $s$. This is because the terms $i=s+1, 
\cdots, N$ vanish 
by the use of  divergence theorem, and the fact that $\rho$ vanishes 
on the distant surface at infinity. To see that such terms explicitly
vanish, let $i \ge s+1$, and consider the integral
\begin{eqnarray}
   \int dX_{s+1} \cdots dX_\smN 
  {\bf v}_i  \cdot \frac{\partial \rho}{\partial {\bf x}_i} 
  &=&
  \int dX_{s+1} \cdots dX_\smN \,
  \frac{\partial}{\partial {\bf x}_i} \cdot \Big(  \rho {\bf v}_i  \Big)
  \\[5pt]
  &=&
  \int dX_{s+1} \cdots dX_{i-1}\, dX_{i+1} \cdots dX_\smN 
  \int \frac{d^3p_i}{(2\pi\hbar)^\nu} \int_V d^\nu x_i
  \frac{\partial}{\partial {\bf x}_i} \cdot \Big(\rho {\bf v}_i   \Big)
  \nonumber
  \\[8pt]
  &=&
  \int dX_{s+1} \cdots dX_{i-1}\, dX_{i+1} \cdots dX_\smN 
  \int \frac{d^3p_i}{(2\pi\hbar)^\nu}
  \oint_{\partial V} d{\bf S}_i \cdot \rho{\bf v}_i   
  ~=~ 0 
  \ ,
  \nonumber
\end{eqnarray}
which vanishes because $\rho$ vanishes on the surface at infinity,
the boundary $\partial V$. 
Note that we have enclosed the $\nu$-space space  ${\bf x}_i$ in 
a very large but finite volume $V$ (the volume will be taken to infinity 
in the limit). The boundary of $V$, denoted $\partial V$, is often 
called the surface at infinity. This is all standard, but it might be 
useful for the novice to have seen such a calculation all the way
through.  

At this point in the derivation, the equation for $f_s$ is
\begin{eqnarray}
  \frac{\partial f_s}{\partial t} 
  +
   \sum_{i=1}^s {\bf v}_i  \cdot \frac{\partial f_s}{\partial {\bf x}_i} 
  +
  \frac{N!}{(N-s)!}  \sum_{i=1}^N \sum_{j=0}^N  
  \int dX_{s+1} \cdots dX_\smN \,
  {\bf F}_i^{(j)}  \cdot \frac{\partial \rho}{\partial {\bf p}_i} 
  = 0 
  \ .
\label{eq_bbgky_inter}
\end{eqnarray}
We must now consider the last term in (\ref{eq_bbgky_inter}),
which we decompose about the $i=s$ contribution,
\begin{eqnarray}
  \sum_{j=0}^N{\bf F}_i^{(j)}
  = 
   \sum_{j=0}^s {\bf F}_i^{(j)} 
  +
  \sum_{j=s+1}^N {\bf F}_i^{(j)} 
  \ .
  \label{eq_Fi_break}
\end{eqnarray}
The first sum in  (\ref{eq_Fi_break}) is handled as before, 
and we can write (\ref{eq_bbgky_inter}) in the form
\begin{eqnarray}
  &&\frac{\partial f_s}{\partial t} 
  +
   \sum_{i=1}^s {\bf v}_i  \cdot \frac{\partial f_s}{\partial {\bf x}_i} 
  +
   \sum_{i=1}^s \sum_{j=0}^s {\bf F}_i^{(j)}  \cdot \frac{\partial f_s}{\partial {\bf p}_i} 
   +
  \nonumber\\[5pt]
  &&
  \frac{N!}{(N-s)!}  \sum_{i=1}^N \sum_{j=s+1}^N  \int dX_{s+1} \cdots dX_\smN \,
  {\bf F}_i^{(j)}  \cdot \frac{\partial \rho}{\partial {\bf p}_i} 
  = 0 \ . 
  \label{eq_bbgky_almost}
\end{eqnarray}
The final step in the calculation is to address the last term in 
(\ref{eq_bbgky_almost}). Recall that the distribution function $\rho$ 
is symmetric in its arguments $X_1, \cdots, X_N$. This means that 
every term of the $j$-sum in the last term of (\ref{eq_bbgky_almost}) 
is identical. Therefore, let us represent the sum by arbitrarily choosing 
the first term $j=s+1$, and multiplying by $(N-s)$ to account for the 
remaining terms in the sum.  This allows us to express the  last term
in  (\ref{eq_bbgky_almost}) as
\begin{eqnarray}
   \frac{N! \, (N-s)}{(N-s)!} \sum_{i=1}^s \int dX_{s+1} \cdots dX_\smN \,
  {\bf F}_i^{(s+1)}  \cdot \frac{\partial \rho}{\partial {\bf p}_i} 
  \ ,
\label{eq_jterm}
\end{eqnarray}
where we have, for the usual reasons, truncated the $i$-sum at $i=s$.
We can express  (\ref{eq_jterm})~as
\begin{eqnarray}
 && 
   \sum_{i=1}^s \int dX_{s+1} \, {\bf F}_i^{(s+1)} \cdot \frac{\partial }{\partial {\bf p}_i} \,
  \frac{N!}{(N-s-1)!} \int dX_{s+2} \cdots dX_\smN \, \rho
  \\[5pt]
  &&
  = \sum_{i=1}^s   \int dX_{s+1} \,
  {\bf F}_i^{(s+1)} \cdot \frac{\partial f_{s+1}}{\partial {\bf p}_i} 
  \ ,
  \label{eq_last_term_bbgky}
\end{eqnarray}
and substituting  (\ref{eq_last_term_bbgky}) back into  (\ref{eq_bbgky_almost}).
This establishes the BBGKY hierarchy (\ref{eq_bbgky}), which is repeated
again for convenience, 
\begin{eqnarray}
  &&
  \frac{\partial f_s}{\partial t} 
  +
   \sum_{i=1}^s {\bf v}_i  \cdot \frac{\partial f_s}{\partial {\bf x}_i} 
  +
   \sum_{i=1}^s \sum_{j=0}^s {\bf F}_i^{(j)}  \cdot \frac{\partial f_s}{\partial {\bf p}_i} 
  =
  -\sum_{i=1}^s \int dX_{s+1} \,
  {\bf F}_i^{(s+1)} \cdot \frac{\partial f_{s+1}}{\partial {\bf p}_i} 
  \ ,
  \label{eq_bbgky_again}
\end{eqnarray}
for $s=1, \cdots, N-1$. The system is closed with Liouville's equation,
\begin{eqnarray}
  &&
  \frac{\partial f_\smN}{\partial t} 
  +
   \sum_{i=1}^N {\bf v}_i  \cdot \frac{\partial f_\smN}{\partial {\bf x}_i} 
  +
   \sum_{i=1}^N \sum_{j=0}^N {\bf F}_i^{(j)}  \cdot \frac{\partial f_\smN}{\partial {\bf p}_i} 
  =
  0
 \ .
  \label{eq_L_again}
\end{eqnarray}
\pagebreak
\subsection{Dimensionless Variables}

We have now developed the BBGKY  hierarchy  (\ref{eq_bbgky_again}) 
and (\ref{eq_L_again}) in the quite general setting of a non-equilibrium 
but {\em single-component} plasma. The plasma could be generalized 
to have multiple components,
but at the expense of increasing the complexity of the counting arguments
and the simplicity of the formalism.  In these notes,  it turns out to be quite 
easy to generalize the results of a single-component calculation 
to that of a multi-component plasma. For the sake of simplicity, we continue
with a single-component plasma, whose constituents have  charge $e$ and
mass $m$.  In equilibrium, the plasma is characterized by temperature $T$ 
and number density $n$. We measure $T$ in energy units, while $n$ is the 
number of particles per unit hypervolume. The Debye wavenumber 
$\kappa$, and  the plasma frequency $\omega_p$, are given by
\begin{eqnarray}
  \kappa^2 &=& \frac{e^2\,n}{T}
  \label{eq_kappa_def}
  \\[8pt]
  \omega_p^2 &=& \frac{e^2\,n}{m}
  \label{eq_omega_def}
  \ .
\end{eqnarray}
By dimensional analysis, these expressions hold in any spatial dimension 
$\nu$, and we can therefore use $\kappa$ and $\omega_p$ as defined 
by (\ref{eq_kappa_def}) and (\ref{eq_omega_def}) in any dimension under
consideration, as the electric charge absorbs any dimensional factors 
involving $\nu$. Let us generalize the equilibrium system by imposing 
a small non-equilibrium background on the equilibrium plasma. This new
quasi-equilibrium system is still described by the BBGKY hierarchy, and 
it is quite informative to express the BBGKY kinetic equations in terms
of dimensionless variables. This is  possible because the background 
equilibrium plasma provides natural length and time scales. We shall 
see that the coupling constant $g$ emerges quite naturally, 
and that a consistent perturbation theory in powers of $g$ can be 
developed. 
 
We first express the basic kinematic variables in dimensionless
form. We do this with the following scale transformation, where 
the over-bar denotes the dimensionless form of the corresponding 
variable, 
\begin{eqnarray}
   {\bf x} &=&  \bar {\bf x}/\kappa
   \hskip2.0cm
    t =  \bar t / \omega_p
   \label{eq_length_time}
   \\[5pt]
   {\bf v} &=& (\omega_p/\kappa)\, \bar{\bf v}  
  \hskip1.15cm
   {\bf p} =  (m \omega_p/\kappa)\,\bar{\bf p} = (T \kappa/\omega_p)\, 
   \bar{\bf p} 
   \\[5pt]
      dX &=& \left(\frac{m \omega_p}{\hbar \kappa^2}\right)^\nu \! 
      d\bar X
   \hskip0.50cm 
   {\bf F}^{(0)} = \kappa T \, \bar{\bf F}^{(0)}
   \\[5pt]
  f_s(X_1, \cdots, X_s, t) 
  &=&  
  \left(\frac{\hbar \kappa^2}{m \omega_p}\right)^{\nu s} \,
  \bar f_s(\bar X_1, \cdots, \bar X_s,  t)
  \ .
 \end{eqnarray}
Motivated by the scaling $\kappa T$ for the external force
${\bf F}^{(0)}$, 
we are immediately led to express the Coulomb force as
\begin{eqnarray}
  {\bf F}_i^{(j)}
  &=&
  e^2 \,\frac{\Gamma(\nu/2)}{2 \pi^{\nu/2}}\, \frac{{\bf x}_i - {\bf x}_j}
  {\vert {\bf x}_i - {\bf x}_j \vert^{\nu}}
%\\[8pt]
%  &=&
%  e^2 \,\frac{\Gamma(\nu/2)}{2 \pi^{\nu/2}}\, \kappa^{\nu-1}\,
\\[5pt]
%  \bar {\bf F}_i^{(j)}
  &=&
  g \,  \kappa T\,  \bar {\bf F}_i^{(j)}
  \ ,
\label{EnuAltaF_singlecomp}
\end{eqnarray}
where the dimensionless Coulomb force is defined by
\begin{eqnarray}
  \bar{\bf F}_i^{(j)} 
  &=&
  \frac{\bar {\bf x}_i - \bar {\bf x}_j  }
  {\vert \bar {\bf x}_i - \bar {\bf x}_j \vert^{\nu}} 
  \ ,
 \label{eq_F_def}
\end{eqnarray}
and the remaining factors combine to form the plasma 
coupling constant, 
\begin{eqnarray}
  g 
  &=&  
  \frac{\Gamma(\nu/2)}{2 \pi^{\nu/2}} \, 
  \frac{e^2 \kappa^{\nu-2}}{T}
  \ .
\label{eq_g_def}
 \end{eqnarray}
We see that the expansion parameter $g$ simply falls out of 
the algebra. Finally,  the BBGKY hierarchy (\ref{eq_bbgky_again}) 
and (\ref{eq_L_again}) can be expressed in the dimensionless 
form
\begin{eqnarray}
  &&
  \frac{\partial \bar f_s}{\partial \bar t} 
  +
   \sum_{i=1}^s \bar{\bf v}_i  \cdot \frac{\partial \bar f_s}{\partial \bar{\bf x}_i} 
  +
   \sum_{i=1}^s  \bar{\bf F}_i^{(0)}  \cdot \frac{\partial \bar f_s}{\partial \bar{\bf p}_i} 
   +
   g \sum_{i=1}^s \sum_{j=1}^s \bar{\bf F}_i^{(j)}  \cdot 
   \frac{\partial \bar f_s}{\partial \bar{\bf p}_i} 
  =
  \nonumber\\[5pt] 
  && \hskip5.3cm
  - g \sum_{i=1}^s  \int d \bar X_{s+1} \,
  \bar {\bf F}_i^{(s+1)} \cdot \frac{\partial \bar f_{s+1}}{\partial \bar {\bf p}_i} 
  \ ,
\label{eq_bbgky_dimless_final}
\end{eqnarray}
for $i=1, \cdots, N-1$, in the square bracket along with the $i=N$ equation
\begin{eqnarray}
  &&
  \frac{\partial \bar f_\smN}{\partial \bar t} 
  +
   \sum_{i=1}^N \bar{\bf v}_i  \cdot \frac{\partial \bar f_\smN}{\partial \bar{\bf x}_i} 
  +
   \sum_{i=1}^N  \bar{\bf F}_i^{(0)}  \cdot \frac{\partial \bar f_\smN}{\partial \bar{\bf p}_i} 
   +
   g \sum_{i=1}^N \sum_{j=1}^N \bar{\bf F}_i^{(j)}  \cdot 
   \frac{\partial \bar f_\smN}{\partial \bar{\bf p}_i} 
  =
 0
  \ .
\label{eq_liouville_dimless_final}
\end{eqnarray}
It should be emphasized again that the coupling constant $g$ as 
defined by (\ref{eq_g_def}) emerges quite naturally, and when 
$\nu=3$,  the coupling takes the usual form $g = e^2\kappa/
4\pi T$ (in rationalized cgs units).

In the next section, we will develop a  method that permits  us to 
solve the BBGKY equations perturbatively as an expansion in powers 
of $g$. Before doing this, it is instructive to work through the algebra 
establishing (\ref{eq_bbgky_dimless_final}) and 
(\ref{eq_liouville_dimless_final}). We start by measuring space in 
units of inverse $\kappa$ and time in units of inverse $\omega_p$,
\begin{eqnarray}
   {\bf x} &=&  \bar {\bf x}/\kappa
\label{eq_x_bar_two}
  \\[5pt]
   t &=&  \bar t / \omega_p
\label{eq_t_bar_two}
   \ ,
 \end{eqnarray}
where the bared quantities are dimensionless. Note that 
\begin{eqnarray}
  \omega_p/\kappa 
  = 
  \sqrt{T/m} = v_{\rm th}
\label{eq_vth_def}
 \end{eqnarray}
is the thermal velocity of the plasma, which we use to
form a dimensionless velocity
\begin{eqnarray}
  {\bf v} &=& (\omega_p/\kappa)\, \bar{\bf v}  
  \ .
\label{eq_v_bar} 
 \end{eqnarray}
Since $\omega_p/\kappa$ has units of velocity, we see that
$m\omega_p/\kappa$ has units of momentum. The relation for
the thermal velocity (\ref{eq_vth_def}) implies that $\omega_p/
\kappa = \kappa T/\omega_p$, and we can thus scale the 
momentum in two separate but equivalent ways
\begin{eqnarray}
  {\bf p} 
  &=& 
  (m \omega_p/\kappa)\, \bar{\bf p} 
  =
  (\kappa T/\omega_p)\, \bar{\bf p} 
  \ .
\label{eq_p_bar} 
\end{eqnarray}
Both forms of the momentum scaling will be used interchangeably.
Since the temperature $T$ has energy units, the quantity $\kappa T$
has units of force, and we define the dimensionless external force
as
\begin{eqnarray}
  {\bf F}^{(0)}
  =
  \kappa T \, \bar{\bf F}^{(0)}
  \ .
\label{eq_f0_bar} 
\end{eqnarray}
This motivates expressing the Coulomb force by
\begin{eqnarray}
  {\bf F}_i^{(j)}
  &=&
  e^2 \,
  \frac{\Gamma(\nu/2)}{2 \pi^{\nu/2}}\,
  \frac{{\bf x}_i - {\bf x}_j}{\vert {\bf x}_i - {\bf x}_j \vert^{\nu}}
  =
  e^2 \,\frac{\Gamma(\nu/2)}{2 \pi^{\nu/2}}\, 
  \frac{\kappa^{\nu-2}}{T}\,
  \cdot
  \kappa T 
  \cdot 
  \frac{\hat{\bf x}_i - \hat{\bf x}_j}{\vert \hat{\bf x}_i - \hat{\bf x}_j 
  \vert^{\nu}}
\\[5pt]
%  \bar {\bf F}_i^{(j)}
  &=&
  g \,  \kappa T\,  \bar {\bf F}_i^{(j)}
  \ .
\end{eqnarray}
The next quantity that we consider is the phase space measure, 
and it transforms as 
\begin{eqnarray}
  dX 
  = 
  \frac{d^\nu x \,d^\nu p}{(2\pi\hbar)^\nu}
  =
   \left(\frac{m \omega_p}{\hbar \kappa^2}\right)^\nu\, 
   \frac{d^\nu \bar x \,d^\nu \bar p}{(2\pi)^\nu}
   =
   \left(\frac{m \omega_p}{\hbar \kappa^2}\right)^\nu\, d\bar X
  \ .
  \label{eq_dX_scale}
\end{eqnarray}
The final quantity to consider is the distribution function itself, 
which transforms by a constant factor
\begin{eqnarray}
  f_s(X_1, \cdots, X_s, t) = {\cal N}_s \,\bar f_s(\bar X_1, \cdots, 
  \bar X_s, \bar t)
   \ .
 \end{eqnarray}
We can find ${\cal N}_s$ by the requirement that
\begin{eqnarray}
  \int d\bar X_1 \cdots d \bar X_s \, \bar f_s(\bar X_1, \cdots, \bar X_s) 
  &=&
  \int dX_1 \cdots dX_s \,f_s(X_1, \cdots, X_s) 
  \\[5pt]
  &=&
  \left(\frac{m \omega_p}{\hbar \kappa^2}\right)^{\nu s}\,
  {\cal N}_s
  \int d\bar X_1 \cdots d \bar X_s \,\bar f_s(X_1, \cdots, X_s) 
  \ ,
\end{eqnarray}
which implies
\begin{eqnarray}
  {\cal N}_s 
  &=&
  \left(\frac{\hbar \kappa^2}{m \omega_p}\right)^{\nu s}
  \ .
  \label{eq_Ns_def}
\end{eqnarray}
We will later require the ratio 
\begin{eqnarray}
  \frac{{\cal N}_{s+1}}{{\cal N}_s} 
  = 
  \left(\frac{\hbar \kappa^2}{m \omega_p}\right)^\nu 
  =\,
  {\cal N}_1
 \ ,
  \label{eq_Ns_ratio}
\end{eqnarray}
but for now  we will express our results in terms of ${\cal N}_s$. 
Upon changing to dimensionless variables, the first two terms of 
the dimensional BBGKY hierarchy (\ref{eq_bbgky_again}) become
\begin{eqnarray}
  \text{term1}  
  &\equiv&
    \frac{\partial f_s}{\partial t} 
    = 
    \omega_p\, {\cal N}_s \, \frac{\partial \bar f_s}{\partial \bar t} 
  \\[5pt]
    \text{term2} 
  &\equiv&
   \sum_{i=1}^s {\bf v}_i  \cdot \frac{\partial f_s}{\partial {\bf x}_i} 
   =
  \sum_{i=1}^s  
  \left(\frac{\omega_p}{\kappa}\, \bar {\bf v}_i \right)
 \! \cdot \!
  \left(\kappa \, {\cal N}_s\, \frac{\partial \bar f_s}{\partial \bar {\bf x}_i} 
  \right)
   =
   \omega_p \, {\cal N}_s\, \sum_{i=1}^s \bar {\bf v}_i  \cdot
  \frac{\partial \bar f_s}{\partial \bar {\bf x}_i} 
  \ .
\end{eqnarray}
The third term of (\ref{eq_bbgky_again}) can be decomposed 
into an external force and an internal Coulomb contribution,
\begin{eqnarray}
  && \hskip-2.0cm
  \text{term3a} 
  \equiv
   \sum_{i=1}^s  {\bf F}_i^{(0)}  \cdot \frac{\partial f_s}{\partial {\bf p}_i} 
   =
  \sum_{i=1}^s  
   \Big(
   \kappa T {\bf F}_i^{(0)}  
   \Big)
   \cdot 
   \left(
   \frac{\omega_p\, {\cal N}_s}{\kappa T}\,
   \frac{\partial \bar f_s}{\partial {\bf \bar p}_i}
  \right)
   =
   \omega_p\, {\cal N}_s
   \sum_{i=1}^s  \bar{\bf F}_i^{(0)}  \cdot \frac{\partial \bar f_s}{\partial {\bf \bar p}_i}
   \\[8pt]
   && \hskip-2.0cm
   \text{term3b} 
  \equiv
   \sum_{i=1}^s \sum_{j=1}^s {\bf F}_i^{(j)}  \cdot \frac{\partial f_s}{\partial {\bf p}_i} 
   =
     \omega_p \, {\cal N}_s \, g
  \sum_{i=1}^s \sum_{j=1}^s \bar{\bf F}_i^{(j)}  \cdot 
  \frac{\partial \bar f_s}{\partial \bar{\bf p}_i} 
  \ .
\end{eqnarray}
The BBGKY equations now become
\begin{eqnarray}
  &&
  \frac{\partial \bar f_s}{\partial \bar t} 
  +
   \sum_{i=1}^s \bar{\bf v}_i  \cdot \frac{\partial \bar f_s}{\partial \bar{\bf x}_i} 
  +
   \sum_{i=1}^s  \bar{\bf F}_i^{(0)}  \cdot \frac{\partial \bar f_s}{\partial \bar{\bf p}_i} 
   +
   g \sum_{i=1}^s \sum_{j=1}^s \bar{\bf F}_i^{(j)}  \cdot 
   \frac{\partial \bar f_s}{\partial \bar{\bf p}_i} 
  +
  \nonumber\\[5pt] 
  && \hskip4cm
  \frac{1}{\omega_p \,{\cal N}_s}
  \sum_{i=1}^s \int dX_{s+1} \,
  {\bf F}_i^{(s+1)} \cdot \frac{\partial f_{s+1}}{\partial {\bf p}_i} 
  = 0 \ .
  \label{eq_bbgky_dimless}
\end{eqnarray}
Finally,  the last term in (\ref{eq_bbgky_dimless}) involving $f_{s+1}$
can be written
\begin{eqnarray}
  {\rm term4}
  &\equiv&
  \frac{1}{\omega_p \, {\cal N}_s}
  \sum_{i=1}^s \int dX_{s+1} \,
  {\bf F}_i^{(s+1)} \cdot \frac{\partial f_{s+1}}{\partial {\bf p}_i} 
  \\[5pt]
  &=&
  \frac{1}{\omega_p} \, \frac{{\cal N}_{s+1}}{ {\cal N}_s}
  \cdot 
  \left( \frac{m \omega_p}{\hbar \kappa^2} \right)^\nu
  \cdot
   g \, \kappa T
  \cdot
  \frac{\omega_p}{T \kappa} \, 
  \sum_{i=1}^s  \int d \bar X_{s+1} \,
   \bar {\bf F}_i^{(s+1)} \cdot \frac{\partial \bar f_{s+1}}{\partial \bar {\bf p}_i} 
  \\[5pt]
  &=&
  g \sum_{i=1}^s  \int d \bar X_{s+1} \,
   \bar {\bf F}_i^{(s+1)} \cdot \frac{\partial \bar f_{s+1}}{\partial \bar {\bf p}_i} 
  \ ,
\end{eqnarray}
where we have used (\ref{eq_Ns_ratio}) for ${\cal N}_{s+1}/{\cal N}_s
= {\cal N}_1$. We have now established (\ref{eq_bbgky_dimless_final})
and (\ref{eq_liouville_dimless_final}), which we reproduce below for
convenience,
\begin{eqnarray}
  &&
  \frac{\partial \bar f_s}{\partial \bar t} 
  +
   \sum_{i=1}^s \bar{\bf v}_i  \cdot \frac{\partial \bar f_s}{\partial \bar{\bf x}_i} 
  +
   \sum_{i=1}^s  \bar{\bf F}_i^{(0)}  \cdot \frac{\partial \bar f_s}{\partial \bar{\bf p}_i} 
   +
   g \sum_{i=1}^s \sum_{j=1}^s \bar{\bf F}_i^{(j)}  \cdot 
   \frac{\partial \bar f_s}{\partial \bar{\bf p}_i} 
  =
  \nonumber\\[5pt] 
  && \hskip5.3cm
  - g \sum_{i=1}^s  \int d \bar X_{s+1} \,
  \bar {\bf F}_i^{(s+1)} \cdot \frac{\partial \bar f_{s+1}}{\partial \bar {\bf p}_i} 
  \ ,
\label{eq_bbgky_dimless_final_x}
\end{eqnarray}
for $s=1, \cdots, N-1$, and 
\begin{eqnarray}
  &&
  \frac{\partial \bar f_\smN}{\partial \bar t} 
  +
   \sum_{i=1}^N \bar{\bf v}_i  \cdot \frac{\partial \bar f_\smN}{\partial \bar{\bf x}_i} 
  +
   \sum_{i=1}^N  \bar{\bf F}_i^{(0)}  \cdot \frac{\partial \bar f_\smN}{\partial \bar{\bf p}_i} 
   +
   g \sum_{i=1}^N \sum_{j=1}^N \bar{\bf F}_i^{(j)}  \cdot 
   \frac{\partial \bar f_\smN}{\partial \bar{\bf p}_i} 
  =
 0
  \ .
\label{eq_liouville_dimless_final_x}
\end{eqnarray}

\pagebreak
\subsection{Perturbation Theory}
\label{sec_pert_theory}

As expressed in the form (\ref{eq_bbgky_dimless_final_x}) and
(\ref{eq_liouville_dimless_final_x}), it is unclear 
how to solve the BBGKY hierarchy perturbatively in powers of $g$. This 
is because the relation between the distribution functions $\bar f_s$ and 
the coupling constant $g$ is not straightforward. The proper procedure
is to expand in powers of the  so called {\em reduced} distribution functions
$\bar h_s= \bar h_s(\bar X_1, \cdots, \bar X_s)$. We define the reduced 
distribution $\bar h_s$ by subtracting all possible lower order correlations 
from $\bar f_s$, a procedure that will be made more precise in just a 
moment. Consequently, the distribution $\bar h_s$ is also called  the
{\em correlation function}, as it encodes the full complement of $s$-body 
correlations. Perturbation theory is then constructed by expanding 
in powers of~$\bar h_s$. We start this recursive procedure by first 
constructing the 2-point correlation function~$\bar h_2$. To do this, 
let us briefly return  to dimensional variables, and define 
\begin{eqnarray}
  h_2( X_1, X_2) =  f_2( X_1,  X_2)  
  - 
   f_1( X_1)  f_1( X_2)
  \ .
\label{eq_h2_def} 
\end{eqnarray}
It is clear that $h_2( X_1, X_2)$ captures the 2-body correlations, 
as the uncorrelated piece $f_1(X_1) f_1(X_2)$ has been subtracted 
from the full 2-body distribution $f_2(X_1, X_2)$:~the remainder 
can only be the correlations. We will {\em assume} that $h_2$ is 
of order $g$, and more generally that $g h_s \propto g^s$. In
dimensionless coordinates, we can therefore express the 
\hbox{2-point} function by the expansion  
\begin{eqnarray}
  \bar f_2(\bar X_1, \bar X_2)
  &=&
  \bar f_1(\bar X_1)   \bar f_1(\bar X_2)   
  + 
  g \bar h_2(\bar X_1, \bar X_2)
  \ .
\label{eq_f2_expand_h2}
\end{eqnarray}
We will justify this perturbative assumption by proving that we 
can expand (\ref{eq_bbgky_dimless_final_x}) and 
(\ref{eq_liouville_dimless_final_x}) to second order in~$g$ 
(in principle we could work to any desired order in~$g$). In 
a similar manner, the reduced \hbox{3-point} function $\bar h_3$ 
is defined by the expansion
\begin{eqnarray}
  \bar f_3(\bar X_1, \bar X_2, \bar X_3)
  &=&
  \bar f_1(\bar X_1)   \bar f_1(\bar X_2)   \bar f_1(\bar X_3) 
  +
  g \Big[
 \bar h_2( \bar X_1,  \bar X_2)  \bar  f_1( \bar X_3)  
  + 
  \bar h_2( \bar X_2,  \bar X_3) \bar f_1( \bar X_1)  
  +
\nonumber \\[5pt] && \hskip3.3cm
 \bar h_2( \bar X_3,  \bar X_1)   \bar f_1( \bar X_2) 
  \Big]
  + 
  g^2\,  \bar h_3( \bar X_1,  \bar X_2,  \bar X_3)
  \ .
\label{eq_f3_def_h3}
\end{eqnarray}
We have removed the following lower order correlations from $\bar f_3$: 
(i) a completely uncorrelated piece consisting of the 
product of three 1-point functions $\bar f_1 {\scriptscriptstyle\times} 
\bar f_1 {\scriptscriptstyle\times}  \bar f_1$,  and (ii) three \hbox{2-point} 
correlations involving $\bar h_2 {\scriptscriptstyle\times} \bar f_1$, 
evaluated on the cyclic permutations of $X_1$, $X_2$, and $X_3$, and
(iii) the \hbox{3-point} correlation $\bar h_3$. Note that $\bar h_3$ is of 
order $g^2$, or in dimensional form, $g h_3 \propto g^3$. To the order 
$g^2$ in which we are working, the $\bar h_3$ term must therefore 
be dropped from (\ref{eq_f3_def_h3}) for consistency. Although we will 
not do so in these notes, one may press onward and calculate the order 
$g^3$ terms. To do this, we would keep the $g^2 \bar h_3$ contribution
 to $\bar f_3$. We 
would also need to construct the \hbox{4-point} correlation function 
$\bar h_4$ by subtracting off the lower order correlations  from 
$\bar f_4$, which schematically takes the form 
\begin{eqnarray}
  g^3\,  \bar h_4( \bar X_1,  \bar X_2,  \bar X_3, \bar X_3)
  &=&
  \bar f_4(\bar X_1, \bar X_2, \bar X_3, \bar X_4)
  -
  \bar f_1(\bar X_1)   \bar f_1(\bar X_2)   \bar f_1(\bar X_3) 
  \bar f_1(\bar X_4) 
  -
\label{eq_f4_def_h4}
\\[8pt] && \hskip-3.0cm
  g^2 \Big[
   \bar h_3(\bar X_1,  \bar X_2,  \bar X_3) \bar  f_1( \bar X_4)   
  + \cdots 
  \Big]
    - g^2 \Big[
  \bar h_2( \bar X_1,  \bar X_2)  \bar h_2( \bar X_3,  \bar X_4)  + 
  \cdots~~
  \Big]
\nonumber\\[5pt] && \hskip3.05cm
  -g\Big[
  \bar h_2( \bar X_1,  \bar X_2)  \bar f_1( \bar X_3) \bar f_1( \bar X_4)  
  + \cdots \Big]
  \ .
\nonumber
\end{eqnarray}
As usual, we subtract off the completely uncorrelated piece, 
this time consisting of the product of four 1-point functions. 
At this order, there are several more combinations of lower-order 
correlations that must be removed. For example, there are 
terms like $\bar h_3 {\scriptscriptstyle\times} \bar f_1$, in 
addition to pair-wise  2-point contributions  like $\bar h_2 
{\scriptscriptstyle\times}  \bar h_2$. Finally, there are
contributions of the form $\bar h_2 {\scriptscriptstyle\times} 
\bar f_1 \bar f_1$.  The next higher-order contribution would 
have even more combinations of lower-order correlations, and 
we see that higher-order calculations become quite involved 
very rapidly.

We now show that one can work consistently to order $g^2$, dropping 
terms of order $g^3$ and higher.  We must prove that the $s=1$ and 
$s=2$ equations contain terms of order $g^2$ or lower, and that the 
$s \ge 3$ equations contain terms of order of $g^3$ and higher. For 
consistency, we must therefore work only with the $s=1$ equation, 
and a {\em truncated} version of the $s=2$ equation (as we have seen, 
we must also drop the $h_3$-contribution in the $\bar f_3$ term, as 
this contribution is of order~$g^3$). Writing the factors of $g$ explicitly, 
we will show that the $s=1$ equation takes the form
\begin{eqnarray}
  \left[
  \frac{\partial }{\partial t} + V_\smA + g V_\smB[\bar f_1]
  \right] \bar f_1
  =
  g^2 K[\, \bar h_2]
\label{sone_V_def}
\ ,
\end{eqnarray}
where $K[h_2]$ is a homogeneous integration kernel, while $V_\smA$ 
and $V_\smB[\bar f_1]$ are differential and integro-differential 
operators on $\bar X_1$ space, respectively. Note that the operator 
$V_\smB[\bar f_1]$ contains a functional dependence on $\bar f_1$.  
The truncated $s=2$ equation can be expressed in the form
\begin{eqnarray}
  \left[
  \frac{\partial }{\partial t} 
  +
   V_\smC
  + 
   g  V_\smD[\bar f_1]
   \right] g \bar h_2
  =
 gS[\bar f_1]   + {\cal O}(g^3)
 \ ,
\label{stwo_V_def}
\end{eqnarray}
where $S[\bar f_1]$ is a source term depending upon $\bar f_1$, 
while $V_\smC$ is a differential operator, and $V_\smD[\bar f_1]$ 
is an integro-differential operator. Both operators act on 
$\bar X_1$-$\bar X_2$ space, of which $\bar h_2$ is a function. The 
precise form of the source term, the operators, and the kernel are 
not important to this perturbative argument, although we shall 
calculate these quantities explicitly in the next paragraph. The 
point here is that both (\ref{sone_V_def}) and (\ref{stwo_V_def}) 
are of order~$g^2$, and that higher-$s$ equations are of order~$g^3$ 
and higher.  Since the kernel $K[\bar h_2]$ is homogeneous, note 
that the $g$-dependence on the right-hand-side 
of (\ref{sone_V_def}) may be recast in the more suggestive form 
$g K[g \bar h_2]$, so that (\ref{sone_V_def}) and (\ref{stwo_V_def}) 
is a system of coupled integro-differential equations for $\bar f_1$ 
and $g \bar h_2$. These equations are accurate to order $g^2$,  
with error of order $g^3$. With a lot of work, one can show that 
the $s=3$ equation is of order $g^3$, and consistency demands 
that we neglect it as well (and all higher order equations). This 
justifies the assumption that $\bar h_2$ is of order $g$, and that
we are indeed working consistently with an accuracy of order 
$g^2$, and an absolute error of order $g^3$.

%%
%\pagebreak
\subsubsection{General Number of Spatial Dimensions $\nu$}

Let us now verify equations (\ref{sone_V_def}) and (\ref{stwo_V_def}). 
We shall drop the bar from the dimensionless quantities for ease of 
notation, and the $s=1$ equation of (\ref{eq_bbgky_dimless_final_x}) 
becomes
\begin{eqnarray}
  \left(\frac{\partial}{\partial t} 
  +
 {\bf v}_1 \cdot \frac{\partial}{\partial {\bf x}_1}
  +
  {\bf F}_1^{(0)} \cdot  \frac{\partial}{\partial {\bf p}_1}
  \right) f_1(X_1) 
  =
  - g \! \int dX_2 \,  {\bf F}_1^{(2)} \cdot 
  \frac{\partial}{\partial {\bf p}_1}\, f_2(X_1,X_2)
  \ .
\label{bbgkyOne}
\end{eqnarray}

\vskip0.1cm
\noindent
When working with the Boltzmann equation in $\nu>3$, this form 
will be particularly useful. For the perturbative analysis, however, 
it is better to expand $f_2$ (and $f_3$ in the $s=2$ equation) in 
terms of the 2-point correlation $h_2$. For convenience we repeat 
here the expansions (\ref{eq_f2_expand_h2}) and(\ref{eq_f3_def_h3}), 
but in an annotated form, 

\begin{eqnarray}
  f_2
  &=&
  \underbrace{
  f_1( X_1)    f_1( X_2)   
  }_{\text{uncorrelated}}
  ~+~ 
  g 
  \underbrace{~
   h_2( X_1,  X_2)
   ~}_{\text{1-2 correlation}}
  \ .
\label{eq_f2_expand_h2_braces}
\\[20pt]
  f_3 &=& 
  \underbrace{
  f_1( X_1)    f_1( X_2)    f_1( X_3) 
  }_{\text{uncorrelated}}
  ~+
\label{eq_f3_expand_h2_braces}
\\[5pt]
  &&
  g 
  \Big[
  \underbrace{
  h_2( X_1,   X_2)  f_1(  X_3)   
  }_{\text{1-2 correlation}}
  ~+~ 
  \underbrace{
  h_2( X_2,   X_3) f_1(  X_1)   
  +
   h_2( X_1,   X_3) f_1(  X_2)  
  }_{\text{2-3 and 1-3 correlations}}
  \Big]
  + ~ \text{higher-order}
  \ .
\nonumber
\end{eqnarray}

\vskip0.2cm
\noindent
Using (\ref{eq_f2_expand_h2_braces}) in (\ref{bbgkyOne}) 
gives the coupled integro-differential equation

\begin{eqnarray}
 && \hskip-1.0cm
  \frac{\partial f_1(X_1) }{\partial t} 
  +
 {\bf v}_1 \cdot \frac{\partial f_1(X_1) }{\partial {\bf x}_1}
  +
  {\bf F}_1^{(0)} \cdot  \frac{\partial f_1(X_1) }{\partial {\bf p}_1}
  +
  g \! \int dX_3 \,  f_1(X_3) \, {\bf F}_1^{(3)} \cdot 
  \frac{\partial f_1(X_1)}{\partial {\bf p}_1}\, 
  =
\nonumber\\[8pt] && \hskip7.5cm
 - g^2 \! \int dX_3 \,  {\bf F}_1^{(3)} \cdot 
  \frac{\partial h_2(X_1,X_3)}{\partial {\bf p}_1}\, 
  \ .
\label{bbgkyOne_intdif}
\end{eqnarray}

\noindent
We have replaced the integration variable $X_2$ in (\ref{bbgkyOne}) 
by $X_3$ to avoid conflicts with the variable $X_2$ when we turn to 
the $s=2$ equation. We can recast the above equation in a more 
compact form by defining the {\em self-consistent electric field} at 
${\bf x}_1$ by
\begin{eqnarray}
  {\bf F}_1[f_1]
  =
  \int dX_3 \, f_1(X_3) {\bf F}_1^{(3)}
  =
  \int dX_3 \, f_1(X_3) {\bf F}({\bf x}_1- {\bf x}_3)
  \ ,
\label{eq_Fi_self}
\end{eqnarray}

\noindent
so that (\ref{bbgkyOne_intdif}) becomes
\begin{eqnarray}
  \Bigg(\frac{\partial}{\partial t} 
  +
  \underbrace{
 {\bf v}_1 \cdot \frac{\partial}{\partial {\bf x}_1}
  +
  {\bf F}_1^{(0)} \cdot  \frac{\partial}{\partial {\bf p}_1}
  ~}_{V_\smA}
  +
  \underbrace{~
  g {\bf F}_1[f_1] \cdot  \frac{\partial}{\partial {\bf p}_1}
 ~ }_{V_\smB[f_1]}
  \Bigg) f_1(X_1) 
  =
  - g^2 \hskip-0.2cm
  \underbrace{~
  \int dX_3 \,  {\bf F}_1^{(3)} \cdot 
   \frac{\partial h_2(X_1, X_3)}{\partial {\bf p}_1}
   }_{K[h_2]}
   \ .
\nonumber\\[-8pt]
\label{bbgkyOne_h2}
\end{eqnarray}
We have identified the quantities $V_\smA$,  $V_\smB[ f_1]$, 
and $K[h_2]$ in (\ref{sone_V_def}) by the under-braces. Note 
that there is a factor of $g$ for every Coulomb interaction 
${\bf F}_1^{(3)}$, and a factor of $g$ for the correlation 
$\bar h_2$. In dimensionless units, there is no difference 
between the electric force and the electric field, as the 
factors of electric charge have been collected in the coupling 
constant~$g$. 

Let us now turn to the the $s=2$ equation of the BBGKY 
hierarchy (\ref{eq_bbgky_dimless_final_x}), which we
write in the form
\begin{eqnarray}
  &&
  \left[
  \frac{\partial }{\partial t} 
  +
   \sum_{i=1}^2 
   \left(
   {\bf v}_i  \cdot \frac{\partial }{\partial {\bf x}_i} 
  +
  {\bf F}_i^{(0)}  \cdot \frac{\partial}{\partial {\bf p}_i} 
  \right)
  \right]f_2
  +
 g {\bf F}_1^{(2)}  \cdot 
 \left[
 \frac{\partial }{\partial {\bf p}_1} 
  -
  \frac{\partial }{\partial {\bf p}_2} 
  \right] f_2
 =
\nonumber\\[5pt] && \hskip8.0cm
 - g \sum_{i=1}^2 \!  \int \!  dX_3 \,
  {\bf F}_i^{(3)} \cdot \frac{\partial f_3}{\partial {\bf p}_i} 
  \ . 
\label{bbgkyTwo}
\end{eqnarray}
Note that we have expanded the 1-2 scattering term as
\begin{eqnarray}
  \sum_{i=1}^2 \sum_{j=1}^2
  g {\bf F}_i^{(j)} \cdot \frac{\partial f_2 }{\partial {\bf p}_i} 
  =
  g {\bf F}_1^{(2)}  \cdot 
 \left[
 \frac{\partial }{\partial {\bf p}_1} 
  -
  \frac{\partial }{\partial {\bf p}_2} 
  \right] f_2
\end{eqnarray}
by using Newton's third law ${\bf F}_1^{(2)} = -{\bf F}_2^{(1)}$. 
Expression (\ref{bbgkyTwo}) can be recast into an equation for 
$h_2$ by expanding $f_2$ and $f_3$ in terms of the 2-point 
correlation $h_2$.  We will do this in stages, emphasizing the 
role played by the spatial dimension $\nu$ at each step, showing
how the physics changes depending upon whether $\nu<3$ or 
$\nu > 3$.  Using 
(\ref{eq_f2_expand_h2_braces}) for $f_2$ in the 1-2 scattering
term allows us to express (\ref{bbgkyTwo}) as

\begin{eqnarray}
  &&
  \left[
  \frac{\partial }{\partial t} 
  +
   \sum_{i=1}^2 
   \left(
   {\bf v}_i  \cdot \frac{\partial }{\partial {\bf x}_i} 
  +
  {\bf F}_i^{(0)}  \cdot \frac{\partial}{\partial {\bf p}_i} 
  \right)
  \right]f_2
  +
 g {\bf F}_1^{(2)}  \cdot 
 \left[
 \frac{\partial }{\partial {\bf p}_1} 
  -
  \frac{\partial }{\partial {\bf p}_2} 
  \right] g h_2
  +
\nonumber\\[5pt] && \hskip5.5cm
 g \sum_{i=1}^2 \!  \int \!  dX_3 \,
  {\bf F}_i^{(3)} \cdot \frac{\partial f_3}{\partial {\bf p}_i} 
  =
  g S[f_1]
  \ ,
\label{bbgkyTwo_pert_a}
\end{eqnarray}
where the source term is defined by
\begin{eqnarray}
  S[f_1]
  &=&  
  -
 {\bf F}_1^{(2)}  \cdot 
 \left[
 \frac{\partial }{\partial {\bf p}_1} 
  -
  \frac{\partial }{\partial {\bf p}_2} 
  \right]f_1(X_1) f_1(X_2)
  \ .
\label{eq_Sf1_def}
\end{eqnarray}
It is instructive to contrast equation (\ref{bbgkyTwo}) with 
(\ref{bbgkyTwo_pert_a}). The latter form is more amenable 
to the perturbative analysis we are performing. It will also 
be used in deriving the Lenard-Balescu equation for $\nu<3$, 
where the Coulomb forces become long-range and 2-body 
scattering becomes soft. In this regime, we can drop the 
correlation term 
\begin{eqnarray}
g {\bf F}_1^{(2)}  \cdot 
 \left[
 \frac{\partial }{\partial {\bf p}_1} 
  -
  \frac{\partial }{\partial {\bf p}_2} 
  \right] g h_2
\end{eqnarray}
from (\ref{bbgkyTwo_pert_a}). In contrast, this term must be kept 
when $\nu > 3$. This is because the scatter becomes short-range, 
and momentum exchange can be become quite large. In this case, 
it is best {\em not} to make the substitution for $f_2$ in the 1-2 
scattering 
term, and to use (\ref{bbgkyTwo}) instead. This is justified, however, 
only after perturbation theory has been established. We will have 
more say about this in the next section. For now, we retain all terms 
for completeness. 

Let us return to the  general perturbative argument. I will present 
the detailed algebraic manipulations, since this calculation provides a 
template for proving that the $s=3$ kinetic equation is indeed higher order. 
Upon expanding the remaining $f_2$-term in (\ref{bbgkyTwo_pert_a}), 
the $s=1$ equation can now be written as

\begin{eqnarray}
\label{bbgkyTwo_f2_sub}
  && \hskip0.0cm
  \left[
  \frac{\partial }{\partial t} 
  +
   \sum_{i=1}^2 
   \left(
   {\bf v}_i  \cdot \frac{\partial }{\partial {\bf x}_i} 
  +
  {\bf F}_i^{(0)}  \cdot \frac{\partial}{\partial {\bf p}_i} 
  \right)
  \right]g h_2 
  +
  g{\bf F}_1^{(2)}  \cdot 
 \left[
 \frac{\partial }{\partial {\bf p}_1} 
  -
  \frac{\partial }{\partial {\bf p}_2} 
  \right] g h_2
  ~+ 
\\[11pt] && \hskip3.2cm
  \bigg[
  \frac{\partial }{\partial t}
  +
  \sum_{i=1}^2  \left(
  {\bf v}_i \cdot \frac{\partial}{\partial {\bf x}_i}
  +
  {\bf F}_i^{(0)} \cdot \frac{\partial }{\partial {\bf p}_i}
  \right)
  \bigg]f_1(X_1) f_1(X_2) 
  ~+
\nonumber\\[5pt] && \hskip7.5cm
 g \sum_{i=1}^2 \!  \int \!  dX_3 \,
  {\bf F}_i^{(3)} \cdot \frac{\partial f_3}{\partial {\bf p}_i} 
  =
  g S[f_1]
  \ .
\nonumber
\end{eqnarray}
The contribution 
from the uncorrelated piece of $f_2$ is written in the second line
of (\ref{bbgkyTwo_f2_sub}), which breaks up into two 
collections of terms, one proportional to $f_1(X_2)$ and the other 
proportional to $f_1(X_1)$:

\begin{eqnarray}
  &&
  \bigg[
  \frac{\partial f_1(X_1)}{\partial t}
  +
  {\bf v}_1 \cdot \frac{\partial f_1(X_1)}{\partial {\bf x}_1}
  +
  {\bf F}_1^{(0)} \cdot \frac{\partial f_1(X_1)}{\partial {\bf p}_1}
  \bigg] f_1(X_2) +
\\[8pt] && \hskip0.0cm
  \bigg[
  \frac{\partial f_1(X_2)}{\partial t}
  +
  {\bf v}_2 \cdot \frac{\partial f_1(X_2)}{\partial {\bf x}_2}
  +
  {\bf F}_2^{(0)} \cdot \frac{\partial f_1(X_2)}{\partial {\bf p}_2}
  \bigg] f_1(X_1) 
  \ .
\nonumber
\end{eqnarray}
Our strategy will be to expand $f_3$ in (\ref{bbgkyTwo_f2_sub}) 
using (\ref{eq_f3_expand_h2_braces}), and then to collect terms
that reproduce the $s=1$ equation (\ref{bbgkyOne_h2}) within 
the square brackets. This equation will be evaluated at $X_1$ 
and $X_2$ in each square bracket, respectively, but they will
otherwise vanish. This leaves only an equation involving 
$g h_2$ on the left-hand-side (which is explicitly of order~$g^2$). 
To perform this calculation, we express  the $f_3$ scattering 
term by
\begin{eqnarray}
  && \hskip-1.0cm
  g \sum_{i=1}^2 \!  \int \!  dX_3 \,
  {\bf F}_i^{(3)} \cdot \frac{\partial f_3}{\partial {\bf p}_i} 
  =
\label{eq_f3_int}
  \\ &&
  g \sum_{i=1}^2  \!  \int dX_3 \,
  {\bf F}_i^{(3)} \cdot \frac{\partial}{\partial {\bf p}_i} \,
  \Big[
  f_1(  X_1)  \, g h_2(  X_2,   X_3) 
  + 
  f_1(  X_2) \,  g h_2(  X_1,   X_3) 
  \Big]
  +
  \nonumber\\[5pt] &&
  g \sum_{i=1}^2 {\bf F}_i[f_1] \cdot \frac{\partial}{\partial {\bf p}_i}\,
  f_1(X_1)  f_1(X_2) 
  +
  g \sum_{i=1}^2 {\bf F}_i[f_1] \cdot \frac{\partial}{\partial {\bf p}_i}\,
  g h_2(X_1, X_2)
  \ ,
\nonumber
\end{eqnarray}
The second line of (\ref{eq_f3_int}) can be traced to the 2-3 and
1-3 correlations in (\ref{eq_f3_expand_h2_braces}), while the
terms in the third line come from the uncorrelated piece of
$f_3$ and the 1-2 correlation, respectively. We have generalized 
to definition of the self-consistent field to any position ${\bf x}_i$, 
\begin{eqnarray}
  {\bf F}_i[f_1]
  =
  \int dX_3 \, f_1(X_3) {\bf F}_i^{(3)}
  =
  \int dX_3 \, f_1(X_3) {\bf F}({\bf x}_i- {\bf x}_3)
  \ .
\label{eq_Fi_self_geni}
\end{eqnarray}
We now express equation (\ref{bbgkyTwo_f2_sub}) in
the form

\begin{eqnarray}
  && \hskip-1.0cm
  \left[
  \frac{\partial }{\partial t} 
  +
   \sum_{i=1}^2 
   \left(
   {\bf v}_i  \cdot \frac{\partial }{\partial {\bf x}_i} 
  +
  {\bf F}_i^{(0)}  \cdot \frac{\partial}{\partial {\bf p}_i} 
  +
  g {\bf F}_i[f_1]  \cdot \frac{\partial}{\partial {\bf p}_i} 
  \right)
  \right]g h_2 
  +
  g{\bf F}_1^{(2)}  \cdot 
 \left[
 \frac{\partial }{\partial {\bf p}_1} 
  -
  \frac{\partial }{\partial {\bf p}_2} 
  \right] g h_2
  + 
\nonumber\\[5pt] && \hskip0.0cm
  \sum_{i=1}^2  \!  \int dX_3 \,
  g{\bf F}_i^{(3)} \cdot \frac{\partial}{\partial {\bf p}_i} \,
  \Big[
  f_1(  X_1)  \, g h_2(  X_2,   X_3) 
  + 
  f_1(  X_2) \,  g h_2(  X_1,   X_3) 
  \Big]
  +
 \nonumber\\[11pt] && \hskip0.0cm
  \bigg[
  \frac{\partial f_1(X_1)}{\partial t}
  +
  {\bf v}_1 \cdot \frac{\partial f_1(X_1)}{\partial {\bf x}_1}
  +
  {\bf F}_1^{(0)} \cdot \frac{\partial f_1(X_1)}{\partial {\bf p}_1}
  + 
   g{\bf F}_1[f_1]\cdot \frac{\partial f_1(X_1)}{\partial {\bf p}_1}
  \bigg] f_1(X_2) +
\label{bbgkyTwo_pert_tmp}
\\[5pt] && \hskip0.0cm
  \bigg[
  \frac{\partial f_1(X_2)}{\partial t}
  +
  {\bf v}_2 \cdot \frac{\partial f_1(X_2)}{\partial {\bf x}_2}
  +
  {\bf F}_2^{(0)} \cdot \frac{\partial f_1(X_2)}{\partial {\bf p}_2}
  +
  g{\bf F}_2[f_1]\cdot \frac{\partial f_1(X_2)}{\partial {\bf p}_2}
  \bigg] f_1(X_1) 
%\nonumber\\[3pt] && \hskip8.0cm
%  g{\bf F}_1^{(2)}  \cdot 
% \left[
% \frac{\partial }{\partial {\bf p}_1} 
%  -
%  \frac{\partial }{\partial {\bf p}_2} 
%  \right] g h_2
  =
  g S[f_1]
  \ .
\nonumber
\end{eqnarray}
Note that there are terms directly proportional to $f_1(X_2)$, 
and others proportional to $f_1(X_1)$, which have been grouped 
together in the square brackets. As mentioned above, each of 
the square brackets will turn out to vanish upon using the $s=1$ 
equation at $X_1$ and $X_2$, respectively. Note that $i=1$ term 
of the sum in the second line (from the 2-3 and 1-3 correlations)
takes the form

\begin{eqnarray}
  g^2\! \int dX_3 \,
  {\bf F}_1^{(3)} \cdot \frac{\partial f_1(  X_1)}{\partial {\bf p}_1} \,
  h_2(  X_2,   X_3) 
  + 
  \underbrace{
  g^2 \! \int dX_3 \,
  {\bf F}_1^{(3)} \cdot \frac{\partial h_2(  X_1,   X_3)}{\partial {\bf p}_1} 
  }_{\text{kernel for $s=1$ equation at $X_1$}}
  {\scriptstyle\times} f_1(  X_2)   
  \ ,
\label{eq_si1_term}
\end{eqnarray}
and the $i=2$ term is
\begin{eqnarray}
  g^2 \int dX_3 \,
  {\bf F}_2^{(3)} \cdot \frac{\partial  f_1(  X_2)}{\partial {\bf p}_2} \,
  h_2(  X_1,   X_3) 
  +
  \underbrace{
  g^2 \int dX_3 \,
  {\bf F}_2^{(3)} \cdot \frac{\partial h_2(  X_2,   X_3) }{\partial {\bf p}_2} 
  }_{\text{kernel for $s=1$ equation at $X_2$}}
 {\scriptstyle\times} f_1(  X_1)   
 \ .
\label{eq_si2_term}
\end{eqnarray}
The second term in (\ref{eq_si1_term}) marked by an under-brace 
is the kernel of the $s=1$ equation (\ref{bbgkyOne_h2}), evaluated
at the default position $X_1$. When combined with the terms in 
the first square bracket, those proportional to $f_1(X_2)$, we find 
the $s=1$ equation evaluated at $X_1$, and these terms vanish. 
Note that (\ref{bbgkyOne_h2}) is evaluated at the phase space 
position $X_1$, and since $X_1$ is just a {\em free variable} (in 
the formal mathematical sense), we can make the replacement 
$X_1 \to X_2$ in (\ref{bbgkyOne_h2}). Thus the $s=1$ equation 
can also be evaluated at $X_2$. We see that the second term in 
(\ref{eq_si2_term}) contains the kernel of the $s=1$ equation at 
$X_2$, and the second square bracket also vanishes. The truncated 
$s=2$ equation therefore becomes

\begin{eqnarray}
  &&
  \left[
  \frac{\partial }{\partial t} 
  +
   \sum_{i=1}^2 
   \left(
   {\bf v}_i  \cdot \frac{\partial }{\partial {\bf x}_i} 
  +
  {\bf F}_i^{(0)}  \cdot \frac{\partial}{\partial {\bf p}_i} 
  +
  g\,{\bf F}_i[f_1]  \cdot \frac{\partial}{\partial {\bf p}_i} 
  \right) \right]g h_2 
  \,+
\label{eq_bbgky_pert_final}
\\[8pt] && \hskip-1.0cm
  g \!\int dX_3 \,
  {\bf F}_1^{(3)} \cdot \frac{\partial f_1(  X_1)}{\partial {\bf p}_1} \,
  g h_2(  X_2,   X_3) 
  +
  g \! \int dX_3 \,
  {\bf F}_2^{(3)} \cdot \frac{\partial  f_1(  X_2)}{\partial {\bf p}_2} \,
  g h_2(  X_1,   X_3)
  +
\nonumber\\[8pt] && \hskip5.5cm
  g \, {\bf F}_1^{(2)}  \cdot 
 \left[
 \frac{\partial }{\partial {\bf p}_1} 
  -
  \frac{\partial }{\partial {\bf p}_2} 
  \right] g h_2
  =
  g S[f_1]
 \ ,
\nonumber
\end{eqnarray}

\vskip0.2cm
\noindent
which is in the form given by (\ref{stwo_V_def}). Also note that 
the absolute error incurred by dropping the \hbox{$h_3$-contribution} 
from $g f_3$ is of order $g^3$. 

To fully complete the argument, we 
must show that the $s=3$ equation, expressed here for 
completeness, 
\begin{eqnarray}
  && \hskip-0.25cm
  \Bigg(\frac{\partial}{\partial t} 
  +
  \sum_{i=1}^3\left[
  {\bf v}_i \cdot  \frac{\partial}{\partial {\bf x}_i}
  +
  {\bf F}_i^{(0)}  \cdot \frac{\partial}{\partial {\bf p}_i} 
  +
  g \sum_{j=1}^3 {\bf F}_i^{(j)} \cdot \frac{\partial}{\partial {\bf p}_i} 
  \right]
  \Bigg) f_3(X_1,X_2, X_3)
\label{bbgkyThree}
\\[5pt] && \hskip5.0cm
  =
  - g   \int dX_4 \,
  \sum_{i=1}^3 {\bf F}_i^{(4)} \cdot \frac{\partial}{\partial {\bf p}_i}
  f_4(X_1,X_2,X_3, X_4)
  \ ,
\nonumber
\end{eqnarray}

\vskip0.2cm
\noindent
is of order $g^3$ or higher. This is performed in complete analogy
to the $s=2$ case just presented.  We first express (\ref{bbgkyThree}) 
in terms of $g h_3$, up to order order $g^3$. The definition of $h_3$ 
and $h_4$ ensure that the accuracy of (\ref{bbgkyThree}) is of order 
order $g^3$. There will be terms analogous to the square brackets in 
(\ref{bbgkyTwo_pert_tmp}), proportional to factors of $f_1$, but these
terms will vanish by using the lower-order equations for $f_1$ and
$g h_2$. The final result will be of order $g^3$, and must therefore
be dropped for consistency.

%%
%\pagebreak
\subsubsection{Coulomb Physics in $\nu<3$ and $\nu>3$}

We have expanded the BBGKY hierarchy to order $g^2$ in a general 
number of spatial dimensions $\nu$, with little regard to the behavior 
of the Coulomb physics as a function $\nu$. The equations have been 
quite general, but we must make some approximations to proceed, 
and the validity of the approximations depends upon whether the
scattering is hard or soft, that is to say, upon whether $\nu>3$ or 
$\nu<3$, respectively. We have already addressed the 1-2 correlation 
and how it must be kept in $\nu>3$, and how can it can, in part, be 
dropped in~$\nu < 3$. It will turn out that complementary collections 
of 2-body correlations dominate in dimensions $\nu<3$ compared 
with $\nu>3$, and vice verse. In $\nu=3$, both collection of terms 
contribute equally to the transport equations, but 
they have the misfortune of diverging logarithmically in at long- and 
short-distances. Recall from Section~\ref{sec_Coulomb_Potential}, 
that the electric electric field at position ${\bf x}$ from a point charge 
$e$ at the origin is
\begin{eqnarray}
  {\bf E}({\bf x}) 
  = 
  e\,\frac{\Gamma(\nu/2)}{2 \pi^{\nu/2}}\, \frac{\hat{\bf x} }{r^{\nu-1}}
  \ ,
\label{Enu_again}
\end{eqnarray}
where $\hat{\bf x}$ is the unit vector at the origin, and $r = \vert {\bf x} 
\vert$ is the distance to ${\bf x}$. It is often more convenient to work 
with the Coulomb potential
\begin{eqnarray}
  \phi({\bf x})
  = 
  e \, \frac{\Gamma(\nu/2-1)}{4 \pi^{\nu/2}} \,
  \frac{1}{r^{\nu-2}} 
  \ .
\label{Vnu_again}
\end{eqnarray}

\vskip0.2cm
\noindent
These two expressions are just equations (\ref{Enu}) and  (\ref{Vnu}),
and they produce a  qualitative difference in the Coulomb field for 
$\nu>3$ and $\nu<3$. The reader is encouraged
to revisit Fig.~\ref{fig_coulomb} for details. Short-distance ultraviolet 
(UV) physics is dominant in dimensions $\nu>3$, and conversely, 
long-distance infrared (IR) physics dominates when $\nu<3$. The 
dimension $\nu=3$ is a critical case, in which the UV and IR physics 
are equally dominant. For $\nu<3$, the Coulomb potential diverges 
less severely that $1/r$ as $r \to 0$. In this regime,  the Lenard-Balescu 
scattering kernel does not suffer a UV divergence. 
%This is the Lenard-Balescu regime.  
In like manner, for $\nu > 3$, the Coulomb force converges to zero 
faster than $1/r$ as $r \to \infty$. In this regime,  and the Boltzmann
scattering kernel does not suffer an IR divergence. 
%This is the Boltzmann regime. 

The primary results from the previous section are the $s=1$ equation 
(\ref{bbgkyOne_h2}), repeated here for convenience, 
\begin{eqnarray}
&& \hskip-1.5cm 
  \Bigg(\frac{\partial}{\partial t} 
  +
 {\bf v}_1 \cdot \frac{\partial}{\partial {\bf x}_1}
  +
  {\bf F}_1^{(0)} \cdot  \frac{\partial}{\partial {\bf p}_1}
  +
  g {\bf F}_1[f_1] \cdot  \frac{\partial}{\partial {\bf p}_1}
  \Bigg) f_1(X_1) 
  =
  - g \!
  \int dX_3 \,  {\bf F}_1^{(3)} \cdot 
   \frac{\partial g h_2(X_1, X_3)}{\partial {\bf p}_1}
   \ ,
\nonumber\\[-8pt]
\label{bbgkyOne_h2_repeat}
\end{eqnarray}
and the truncated $s=2$ equation (\ref{eq_bbgky_pert_final}), 
\begin{eqnarray}
  && \hskip-1.0cm 
  \left[
  \frac{\partial }{\partial t} 
  +
   \sum_{i=1}^2 
   \left(
   {\bf v}_i  \cdot \frac{\partial }{\partial {\bf x}_i} 
  +
  {\bf F}_i^{(0)}  \cdot \frac{\partial}{\partial {\bf p}_i} 
  +
  g\,{\bf F}_i[f_1]  \cdot \frac{\partial}{\partial {\bf p}_i} 
  \right) \right]g h_2 
  \,+
\label{eq_bbgky_pert_repeat}
\\[8pt] && \hskip1.5cm
  g\sum_{i=1}^2 {\bf F}_i[g h_2] \cdot 
  \frac{\partial f_1}{\partial {\bf p}_i} 
  +
%\nonumber\\[8pt] && \hskip-1.0cm
%  g \!\int dX_3 \,
%  {\bf F}_1^{(3)} \cdot \frac{\partial f_1(  X_1)}{\partial {\bf p}_1} \,
%  g h_2(  X_2,   X_3) 
%  +
%  g \! \int dX_3 \,
%  {\bf F}_2^{(3)} \cdot \frac{\partial  f_1(  X_2)}{\partial {\bf p}_2} \,
%  g h_2(  X_1,   X_3)
%  +
  g \, {\bf F}_1^{(2)}  \cdot 
 \left[
 \frac{\partial }{\partial {\bf p}_1} 
  -
  \frac{\partial }{\partial {\bf p}_2} 
  \right] g h_2
  =
  g S[f_1]
 \ .
\nonumber
\end{eqnarray}
We have written (\ref{eq_bbgky_pert_repeat}) in a compact 
form, using the self consistent field ${\bf F}_i[f_1]$ defined 
in (\ref{eq_Fi_self_geni}), and the self-consistent field induced 
by~$h_2$, 
\begin{eqnarray}
  {\bf F}_i[h_2]
  &=&
  \int dX_3 \, h_2(X_3, X_j) \, {\bf F}_i^{(3)}
  \ .
\label{eq_h2i_self_iagain}
\end{eqnarray}
Here, $j=2$ when $i=1$, and $j=1$ when $i=2$.  Note that long-distance 
2-body Coulomb scattering is soft in dimensions $\nu<3$, and the momentum 
exchange is small. We can therefore drop the 
1-2 correlation term
\begin{eqnarray}
  g {\bf F}_1^{(2)} \cdot
  \left[
  \frac{\partial }{\partial {\bf p}_1} 
  -
  \frac{\partial }{\partial {\bf p}_2} 
  \right] g h_2(X_1, X_2)
  \ , 
\label{eq_pdiff_h2}
\end{eqnarray}
which we assume is formally of order $g^3$. This assumption does 
not mean that the momentum difference in the collision is being
neglected, {\em i.e.} we are not dropping the term
\begin{eqnarray}
  -g {\bf F}_1^{(2)} \cdot
  \left[
  \frac{\partial }{\partial {\bf p}_1} 
   -
  \frac{\partial }{\partial {\bf p}_2} 
  \right] f_1(X_1) f_1(X_2)
  =
  g S[f_1]
  \ .
\end{eqnarray}
In $\nu < 3$, we can therefore neglect (\ref{eq_pdiff_h2}) from 
(\ref{eq_bbgky_pert_repeat}), giving 
\begin{eqnarray}
  && \hskip-1.0cm
  \left[
  \frac{\partial }{\partial t} 
  +
   \sum_{i=1}^2 
   \left(
   {\bf v}_i  \cdot \frac{\partial }{\partial {\bf x}_i} 
  +
  {\bf F}_i^{(0)}  \cdot \frac{\partial}{\partial {\bf p}_i} 
 +
  g {\bf F}_i[f_1] \cdot \frac{\partial}{\partial {\bf p}_i} 
  \right) \right]g h_2(X_1, X_2)
  \,+
\label{bbgkyTwo_h2_LB}
\\[11pt] && \hskip-1.0cm
  \int dX_3 \, g h_2(  X_3,   X_2) \,
  g{\bf F}_1^{(3)} \cdot \frac{\partial f_1(X_1)}{\partial {\bf p}_1} \,
  +
  \int dX_3 \,g h_2(  X_3,   X_1)\,
  g {\bf F}_2^{(3)} \cdot \frac{\partial  f_1(X_2)}{\partial {\bf p}_2} \,
  =
  g S[f_1]
 \ ,
\nonumber
\end{eqnarray}
where we have expanded the ${\bf F}_i[h_2]$ term for clarity.
The equation for $h_2$ can be recast in the form
\begin{eqnarray}
  \frac{\partial h_2}{\partial t} + V_1 h_2 + V_2 h_2 = S[f_1]
  \ ,
\label{eq_hV1V2_new}
\end{eqnarray}
with source
\begin{eqnarray}
  S[f_1]
  &=&  
  -
 {\bf F}_1^{(2)}  \cdot 
 \left[
 \frac{\partial }{\partial {\bf p}_1} 
  -
  \frac{\partial }{\partial {\bf p}_2} 
  \right]f_1(X_1) f_1(X_2)
  \ .
\label{eq_Sf1_mom}
\end{eqnarray}
Here,  $V_1$ is a linear integro-differential operator defined in 
$X_1$-space by
\begin{eqnarray}
  V_1 h_2(X_1, X_2)
  &=&
  {\bf v}_1  \cdot \frac{\partial h_2}{\partial {\bf x}_1} 
  +
  {\bf F}_1^{(0)}\cdot \frac{\partial h_2}{\partial {\bf p}_1}  
  + 
 g{\bf F}_1 \cdot \frac{\partial h_2}{\partial {\bf p}_1}  
  +  
  g\!\int dX_3 \,  h_2(X_3,X_2) \,
  {\bf F}_1^{(3)} \cdot  
  \frac{\partial f_1(X_1)}{\partial {\bf p}_1} 
   \ ,
   \nonumber\\[3pt]
 \label{eq_VOne_a}
\end{eqnarray}
and $V_2$ is the corresponding operator in $X_2$-space, 
\begin{eqnarray}
  V_2 h_2(X_1, X_2)
  &=&
  {\bf v}_2  \cdot \frac{\partial h_2}{\partial {\bf x}_2} 
  +
  {\bf F}_2^{(0)}\cdot \frac{\partial h_2}{\partial {\bf p}_2}  
  +  
  g{\bf F}_2 \cdot \frac{\partial h_2}{\partial {\bf p}_2}  
  +  
  g\!\int dX_3 \,  h_2(X_3, X_1) \, 
  {\bf F}_2^{(3)} \cdot 
  \frac{\partial f_1(X_2)}{\partial {\bf p}_2} 
   \ .
   \nonumber\\[3pt]
  \label{eq_VTwo_a}
\end{eqnarray}
These expressions simplify marginally in the case of a uniform 
plasma, and this will be our starting point in Section~\ref{sec_LBE} 
on the Lenard-Balescu equation. 

In dimensions $\nu >3$, the behavior of the Coulomb field
is quite different. The potential becomes short-range, and 
2-body scattering is not always soft. This means that
we cannot drop the 2-body momentum exchange term in
(\ref{eq_pdiff_h2}). This term arose from the expansion
of $f_2$, so it is convenient {\em not} to make this expansion,
and to express the $s=1$ equation as
\begin{eqnarray}
  \left(\frac{\partial}{\partial t} 
  +
 {\bf v}_1 \cdot \frac{\partial}{\partial {\bf x}_1}
  +
  {\bf F}_1^{(0)} \cdot  \frac{\partial}{\partial {\bf p}_1}
  \right) f_1
  =
  - g \! \int dX_2 \,  {\bf F}_1^{(2)} \cdot 
  \frac{\partial f_2}{\partial {\bf p}_1}
  \ .
\label{bbgkyOne_Boltzmann}
\end{eqnarray}
We can, however, drop the short-range contributions
to the scattering in $f_3$, and use the truncated  $s=2$ equation 
\begin{eqnarray}
  &&
  \left[
  \frac{\partial }{\partial t} 
  +
   \sum_{i=1}^2 
   \left(
   {\bf v}_i  \cdot \frac{\partial }{\partial {\bf x}_i} 
  +
  {\bf F}_i^{(0)}  \cdot \frac{\partial}{\partial {\bf p}_i} 
  +
  {\bf F}_i[f_1]  \cdot \frac{\partial}{\partial {\bf p}_i} 
  \right)
  \right]f_2
  +
 g {\bf F}_1^{(2)}  \cdot 
 \left[
 \frac{\partial }{\partial {\bf p}_1} 
  -
  \frac{\partial }{\partial {\bf p}_2} 
  \right] f_2
 = 0
 \ .
 \nonumber\\
\label{bbgkyTwo_repeat}
\end{eqnarray}
This will be our starting point for Section~\ref{sec_be_fr_bbgky} 
on the Boltzmann equation. Note that the term $f_2$ implicitly 
contains a functional dependence on the 2-point correlation 
$h_2$. 

%\vskip3.0cm

%\pagebreak
\subsubsection{The Uniform Plasma}

In the calculations that follow, we make one more significant
assumption, namely, we take the plasma to be spatially 
{\em uniform} in that the 1-point function $f_1 = f({\bf p}_1)$
depends only upon the momentum. By Galilean invariance,
the 2-point function can only be of the form $h_2 = 
h({\bf x}_1 - {\bf x}_2, {\bf p}_1, {\bf p}_2)$. This plasma
conforms to the experimental situations involved in inertial
confinement fusion (ICF), the testing ground of charged particle
stopping power. I did not wish to introduce this assumption
sooner, as I wanted to prove the validity of perturbation theory
for the BBGKY hierarchy in a more general setting. The kinetic
equations simplify somewhat, in that the self-consistent forces
${\bf F}_i[f]$ vanish. 
%in (\ref{eq_Fi_self_geni}) vanish.
% and ${\bf F}_i[h_2]$ in (\ref{eq_h2i_self_iagain}) vanish. 
Keeping the external force
${\bf F}^{(0)}$ for generality, in $\nu < 3$, the $s=1$ equation 
becomes
%\vskip2.0cm

%%
\begin{eqnarray}
&& \hskip-1.5cm 
  \Bigg(\frac{\partial}{\partial t} 
  +
  {\bf F}_1^{(0)} \cdot  \frac{\partial}{\partial {\bf p}_1}
  \Bigg) f({\bf p}_1,t) 
  =
  - g^2 \!
  \int dX_3 \,  {\bf F}_1^{(3)} \cdot 
  \frac{\partial h(X_1, X_3,t)}{\partial {\bf p}_1}
  \ .
\label{bbgkyOne_h2_uniform}
\end{eqnarray}

\vskip0.2cm
\noindent
We shall change spatial integration variables from ${\bf x}_3$ 
to ${\bf x}={\bf x}_1 - {\bf x}_3$. Setting ${\bf p}={\bf p}_3$ 
and $X=({\bf x}, {\bf p})$, we can express 
(\ref{bbgkyOne_h2_uniform}) as

\begin{eqnarray}
&& \hskip-1.5cm 
  \Bigg(\frac{\partial}{\partial t} 
  +
  {\bf F}_1^{(0)} \cdot  \frac{\partial}{\partial {\bf p}_1}
  \Bigg) f({\bf p}_1,t) 
  =
  - g \!
  \int dX \,  {\bf F}({\bf x}) \cdot 
   \frac{\partial}{\partial {\bf p}_1}\, g h_2({\bf x}, {\bf p}_1, {\bf p},t)
   \ .
\label{bbgkyOne_h2_uniform_a}
\end{eqnarray}

\noindent
In a similar manner, the truncated $s=2$ equation in
$\nu<3$ becomes
\begin{eqnarray}
  \frac{\partial h_2}{\partial t} + V_1 h_2 + V_2 h_2 = S[f_1]
  \ ,
\label{eq_hV1V2_uniform}
\end{eqnarray}
with source 
\begin{eqnarray}
  S[f_1]
  &=&  
  -
 {\bf F}_1^{(2)}  \cdot 
 \left[
 \frac{\partial }{\partial {\bf p}_1} 
  -
  \frac{\partial }{\partial {\bf p}_2} 
  \right]f({\bf p}_1) f_1({\bf p}_2)
  \ .
\label{eq_Sf1_mom_x}
\end{eqnarray}
%%

%\hfill
%\vskip2.0cm
\noindent
Here,  $V_1$ is a linear integro-differential operator defined in 
$X_1$-space by
\begin{eqnarray}
  V_1 h_2(X_1, X_2)
  &=&
  {\bf v}_1  \cdot \frac{\partial h_2}{\partial {\bf x}_1} 
  +
  {\bf F}_1^{(0)}\cdot \frac{\partial h_2}{\partial {\bf p}_1}  
  +  
  g\!\int dX_3 \,  h_2(X_3,X_2) \,
  {\bf F}_1^{(3)} \cdot  
  \frac{\partial f_1({\bf p}_1)}{\partial {\bf p}_1} 
   \ ,
   \nonumber\\
 \label{eq_VOne_uniform}
\end{eqnarray}
and $V_2$ is the corresponding operator in $X_2$-space, 
\begin{eqnarray}
  V_2 h_2(X_1, X_2)
  &=&
  {\bf v}_2  \cdot \frac{\partial h_2}{\partial {\bf x}_2} 
  +
  {\bf F}_2^{(0)}\cdot \frac{\partial h_2}{\partial {\bf p}_2}  
  +  
  g\!\int dX_3 \,  h_2(X_3, X_1) \, 
  {\bf F}_2^{(3)} \cdot  
  \frac{\partial f_1({\bf p}_2)}{\partial {\bf p}_2} 
   \ .
   \nonumber\\
  \label{eq_VTwo_uniform}
\end{eqnarray}
This will be our starting point when $\nu<3$ of Section~\ref{sec_LBE}
on the Lenard-Balescu equation. 
%%%%
%\pagebreak
%\noindent
In contrast, when $\nu>3$
in the Boltzmann analysis in Sec~\ref{sec_be_fr_bbgky}, 
we shall start with

\begin{eqnarray}
  \left(\frac{\partial}{\partial t} 
%  +
%  {\bf v}_1 \cdot \frac{\partial}{\partial {\bf x}_1}
  +
  {\bf F}_1^{(0)} \cdot  \frac{\partial}{\partial {\bf p}_1}
  \right) f_1
  =
  - g \! \int dX_2 \,  {\bf F}_1^{(2)} \cdot 
  \frac{\partial f_2}{\partial {\bf p}_1}
\label{bbgkyOne_uniform}
\end{eqnarray}
and
\begin{eqnarray}
  &&
  \left[
  \frac{\partial }{\partial t} 
  +
   \sum_{i=1}^2 
   \left(
   {\bf v}_i  \cdot \frac{\partial }{\partial {\bf x}_i} 
  +
  {\bf F}_i^{(0)}  \cdot \frac{\partial}{\partial {\bf p}_i} 
  \right)
  \right]f_2
  +
 g {\bf F}_1^{(2)}  \cdot 
 \left[
 \frac{\partial }{\partial {\bf p}_1} 
  -
  \frac{\partial }{\partial {\bf p}_2} 
  \right] f_2
  =
  0
  \ . 
\label{bbgkyTwo_uniform}
\end{eqnarray}
The latter equation is just the $s=2$ equation in which $f_3 \to 0$.

%%%%%%%%%%%%%%%%%%%%%%
\pagebreak
\section{The Boltzmann Equation from BBGKY in $\bm{\nu > 3}$}
\label{sec_be_fr_bbgky}

In this section will prove that in spatial dimensions $\nu>3$, the 
Boltzmann equation (BE) follows from the BBGKY hierarchy to 
leading order in the plasma coupling $g$. For simplicity we work 
with a single-component plasma, for which the Boltzmann equation 
takes the form 
\begin{eqnarray}
  \frac{\partial f}{\partial t} + 
  {\bf v}_1  \cdot  \frac{\partial f}{\partial {\bf x}_1}
  &=& 
 B[f]   
  \ ,
\label{BEsimpnu_single}
\end{eqnarray}
with scattering kernel
\begin{eqnarray}
  B[f]
  &=&
  \int \frac{d^\nu p_2}{(2\pi\hbar)^\nu} \,
  \vert{\bf v}_1 - {\bf v}_2 \vert\, d\sigma_{12}\,
  \bigg\{
  f({\bf p}_1^\prime) f({\bf p}_2^\prime)
  -
  f({\bf p}_1) f({\bf p}_2)
  \bigg\} 
  \ ,
\label{BEfeSigma_single}
\end{eqnarray}
where ${\bf p}_1 = m  {\bf v}_1$.
The $\nu$-dimensional cross section $d\sigma_{12}$ is defined in 
Appendix~\ref{sec_cross_section}. The argument of this section is 
based on that of Huang from Section~3.5 of Ref.~\cite{huang}.  
Huang's argument in fact breaks down for the Coulomb force in 
$\nu=3$ spatial dimensions (the actual case of physical interest)
because of a long-distance infra-red (IR) divergence. However, 
the argument 
goes through unscathed when generalized to arbitrary spatial 
dimensions $\nu > 3$. This is because the Coulomb force goes 
like $1/r^{\nu-1}$, which falls off faster than $1/r$ at large $r$ 
for $\nu>3$,  thereby rendering finite any potential IR divergence. 
Since the short distance physics of the BE is correct, the scattering 
kernel does not, as we expect, suffer a short-distance ultra-violet
(UV) divergence. 

In Section~\ref{sec_pert_theory}, we showed that for short-range
interactions, in particular the Coulomb force in dimensions $\nu>3$, 
the BBGKY  hierarchy to order $g^2$ can be expressed as\,\footnote{
\footnoteskip
We have restored a slowly varying spatial dependence to $f_1$. 
} % end footnote
\begin{eqnarray}
  &&
  \left(\frac{\partial}{\partial t} 
  +
 {\bf v}_1 \!\cdot\! \frac{\partial}{\partial {\bf x}_1}
  \right) f_1(X_1) 
  =
   - \!\int dX_2  \,  {\bf F}_1^{(2)} \!\cdot\! 
   \frac{\partial}{\partial {\bf p}_1}\, f_2(X_1,X_2)
\label{bbgkytwoa}
\\[10pt]
  &&
  \Bigg(\frac{\partial}{\partial t} 
  +
  {\bf v}_1 \!\cdot\!  \frac{\partial}{\partial {\bf x}_1}
  +
  {\bf v}_2 \!\cdot\!  \frac{\partial}{\partial {\bf x}_2}
  +
  {\bf F}_1^{(2)}\!\cdot\! \left[ 
  \frac{\partial}{\partial {\bf p}_1} -  \frac{\partial}{\partial {\bf p}_2}
  \right] \Bigg) f_2(X_1,X_2) 
  = 0
  \ ,
\label{bbgkytwoc}
\end{eqnarray}
where $f_1$ is the single-particle distribution function, and $f_2$
is the \hbox{2-point} distribution. Expression (\ref{bbgkytwoa}) is 
the first BBGKY equation, while (\ref{bbgkytwoc}) is the second 
BBGKY equation, except that the \hbox{3-point} function has 
been dropped from the right-hand-side of the full $s=2$ equation
(\ref{bbgkyTwo}). To continue,  let us express  (\ref{bbgkytwoc}) 
in center-of-mass coordinates.\footnote{
\footnoteskip
See Appendix~\ref{sec_CM} for 
the details of this coordinate transformation.
} % end footnote
Since our final goal 
is to apply this formalism to a multi-species plasma, in this calculation 
let us temporarily suppose particle-$1$ has mass $m_1$ and 
particle-$2$ has mass $m_2$. We will denote the total mass by 
$M=m_1 + m_2$, and the reduced mass by $m_{12} = m_1 m_2/M$. 
We then define  the total and relative momentum, and the 
center-of-mass and relative position by
\begin{eqnarray}
    {\bf x} &=& {\bf x}_1 - {\bf x}_2
  \hskip2.15cm 
  {\bf p} = m_{12} \big( {\bf v}_1 - {\bf v}_2\big)
\\[5pt]
  {\bf R} &=& \frac{m_1{\bf x}_1 + m_2{\bf x}_2}{M}
  \hskip1cm 
  {\bf P} = m_1  {\bf v}_1 + m_2 {\bf v}_2
  \ .
\end{eqnarray}
Recalling that ${\bf P} = 0$ in the center-of-mass frame,
in Appendix~\ref{sec_cross_section} we show that
\begin{eqnarray}
  {\bf v}_1 \!\cdot\!  \frac{\partial}{\partial {\bf x}_1}
  +
  {\bf v}_2 \!\cdot\!  \frac{\partial}{\partial {\bf x}_2}
  &=&
  \big({\bf v}_1 - {\bf v}_2 \big) \cdot \frac{\partial}{\partial {\bf x}}
\label{eq_conv}
\\[10pt]
 \frac{ \partial}{\partial {\bf p}_1} -  \frac{ \partial}{\partial {\bf p}_2}
 &=&
 \frac{ \partial}{\partial {\bf p}}
 \ ,
\label{eq_prel_deriv}
\end{eqnarray}
where ${\bf p}_1 = m_1 {\bf v}_1$ and ${\bf p}_2 = m_2 {\bf v}_2$. To
find the BE, we must  consider the asymptotic time limit $t \to \infty$.
This is because, as per  Bogoliubov's hypothesis, the \hbox{2-point} 
correlation $h_2$ comes into equilibrium much sooner than the single 
particle distribution $f_1$. We may consequently set the time derivative 
in (\ref{bbgkytwoc}) to zero, giving the static equation
\begin{eqnarray}
 && \Bigg(
  \big({\bf v}_1 - {\bf v}_2 \big) \cdot \frac{\partial}{\partial {\bf x}}
  +
  {\bf F}_1^{(2)}({\bf x})\cdot  \frac{\partial}{\partial {\bf p}} 
  \Bigg) f_2= 0 
  \ ,
\label{ftwoequilib}
\end{eqnarray}
where we have used (\ref{eq_conv}) and (\ref{eq_prel_deriv}) to write 
(\ref{bbgkytwoc}) in terms of relative coordinates. Let us now express 
the first BBGKY equation (\ref{bbgkytwoa}) in terms of a scattering 
kernel, 
\begin{eqnarray}
 \label{foneequilib}
  &&\left(\frac{\partial}{\partial t} 
  +
  {\bf v}_1\, \cdot \frac{\partial}{\partial {\bf x}_1}
  \right) f_1
  =
  B[f]
  \ ,
\end{eqnarray}
where the kernel is defined by
\begin{eqnarray}
  B[f]
  &\equiv& 
  - \int d X_2 ~ {\bf F}_1^{(2)} \cdot \frac{\partial f_2}{\partial {\bf p}_1}
\label{eq_B_def}
\\[8pt]
  &=&
  - \int d X_2 ~ {\bf F}({\bf x}_1 - {\bf x}_2) \cdot 
  \left[
  \frac{\partial}{\partial {\bf p}_1} -   \frac{\partial}{\partial {\bf p}_2}
  \right] f_2
  =
  -\int d X_2 ~ {\bf F}_1^{(2)}({\bf x})
  \!\cdot\! \frac{\partial f_2}{\partial {\bf p}}
  \ .
\label{eq_B_center}
\end{eqnarray}
We have added zero in the form of the total derivative  
$\partial/\partial{\bf p}_2$ in (\ref{eq_B_center}), and we have
used (\ref{eq_prel_deriv})  to express the resulting difference
in momentum derivatives in terms of the derivative of the relative 
momentum. We can now 
use (\ref{ftwoequilib}) to  write the scattering kernel in the form
\begin{eqnarray}
  B[f]
  &=&
 \int  \frac{d^\nu p_2}{(2\pi\hbar)^\nu} \,
 \int d^\nu x_2 \, 
 ({\bf v}_1 - {\bf v}_2) \cdot \frac{\partial f_2}{\partial {\bf x}} 
  \ .
  \label{eq_B_vrel}
\end{eqnarray}
It is understood that (\ref{eq_B_vrel}) is to be evaluated in the limit
$t \to \infty$, or rather, at asymptotic times compared to the time 
scale of the \hbox{2-point} correlations $h_2$.

\begin{figure}[t!]
\includegraphics[scale=0.45]{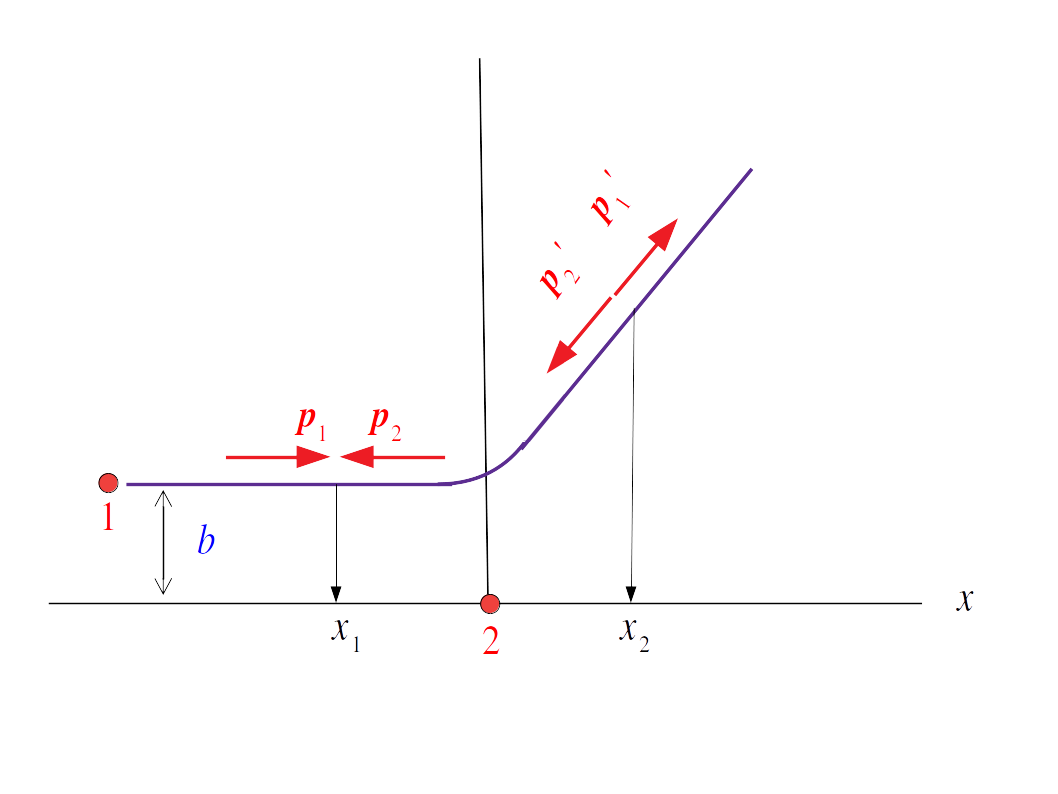} 
\vskip-1.5cm 
\caption{\footnoteskip  
  Two-body scattering for a short-range force. Particle-1 has
  velocity ${\bf v}_1$ and particle-2 has velocity ${\bf v}_2$,
  although for simplicity particle-2 is pictured at rest. 
 The particle velocities are directed towards one another,
  with the beam-line defining the $x$-axis. Therefore,
  ${\bf v}_1 - {\bf v}_2 = \vert {\bf v}_1 - {\bf v}_2\vert \hat{\bf x}$.
  The cross section is given in terms of the impact parameter 
  $b$ by $d\sigma_{12} = d\Omega_{\nu-2} b^{\nu-2}\, db$, 
  and therefore the volume element about particle-2 can be 
  written $d^\nu x_2 = d\sigma_{12}\, dx$. 
}
\label{fig_scattering_angle}
\end{figure}

We shall  now express (\ref{eq_B_vrel}) in terms of the $\nu$-dimensional
cross section $d\sigma_{12}$. See Appendix~\ref{sec_cross_section} of 
these notes for a detailed treatment of the cross section in a general 
number of dimension. As illustrated in Fig.~\ref{fig_scattering_angle},
let the beam-line of the 1+2 collision define the $x$-axis, so that 
${\bf v}_1 - {\bf v}_2= \vert {\bf v}_1 - {\bf v}_2 \vert \,\hat{\bf x}$. In 
two-body scattering, the velocity vectors ${\bf v}_1$ and ${\bf v}_2$ 
are directed toward one another along the beam-line, but they are 
offset (in a normal direction to~$x$) by a distance $b$ called the impact 
parameter. Using expression (\ref{Cnu}), the $\nu$-dimensional 
volume element in cylindrical coordinates about particle-$2$ can 
be expressed as

\vbox{
\begin{eqnarray}
  d^\nu x_2 
  = 
  d\Omega_{\nu-2}\, b^{\nu-2} db \, dx 
  \ .
\end{eqnarray}
Section~\ref{sec_cross_section} of these notes proves that  
$\nu$-dimensional differential scattering cross section takes 
the form  
\begin{eqnarray}
  d\sigma_{12} 
  = 
  d\Omega_{\nu-2}\, b^{\nu-2} db
  \ ,
\end{eqnarray}
} % end vbox

\vskip0.2cm
\noindent
and therefore the spatial volume element can be written
\begin{eqnarray}
d^\nu x_2 =  d\sigma_{12} \, dx
  \ .
\end{eqnarray}
Since the Coulomb force in $\nu > 3$ is short range with a characteristic 
distance scale $r_0 \sim \kappa^{-1}$, we can choose points $x_1$ and 
$x_2$ on either side of $x$ such that the force virtually vanishes for 
$x < x_1$ and $x > x_2$, although the force cannot be neglected for 
$x_1 < x < x_2$.  In other words, 
we can choose the points $x_1$ and $x_2$ right after and right before 
the collision of interest. This is illustrated in Fig.~\ref{fig_scattering_angle}.
%
%\vskip-2.5cm 
We can therefore write   (\ref{eq_B_vrel}) in the form
\begin{eqnarray}
  B[f]
  &=&
  \int \frac{d^\nu p_2}{(2\pi\hbar)^\nu} \, \int d\sigma_{12}\,
  \int_{x_1}^{x_2} \!\! dx  ~
  \vert {\bf v}_1 - {\bf v}_2 \vert \,\frac{\partial f_2}{\partial x}
\\[8pt]
  &=&
  \int  \frac{d^\nu p_2}{(2\pi\hbar)^\nu} \, 
  \vert{\bf v}_1 - {\bf v}_2 \vert \, d\sigma_{12}\,
  \Big[f_2(x_2) - f_2(x_1)\Big] \ .
\end{eqnarray}
Recall that the 2-point function $f_2$ is the product of two
factors of $f_1$ and a correlation function $h_2$. Since the
Coulomb force is short range in $\nu>3$, the function $h_2$ 
vanishes at $x_1$ and $x_2$, and we have 
\begin{eqnarray}
  f_2 
  =
   f_1 {\scriptstyle\times} f_1 + h_2 \to f_1 {\scriptstyle\times} f_1 
   \ ,
\end{eqnarray}
so that
\begin{eqnarray}
  f_2(x_1) &=& f_1({\bf p}_1) f_1({\bf p}_2)
\\
  f_2(x_2) &=& f_1({\bf p}_1^\prime) f_1({\bf p}_2^\prime) 
  \ .
\end{eqnarray}
Here, ${\bf p}_1=m_1{\bf v}_1$ and ${\bf p}_2=m_2{\bf v}_2$ are
the momenta before the collision, and ${\bf p}_1^\prime$ and 
${\bf p}_2^\prime$ are the momenta after the collision. In standard
derivations of the Boltzmann equation, the assumption of {\em 
molecular chaos} is invoked at this juncture. This principle states 
that the momenta before and after a collision are uncorrelated, 
and we see that the short-range nature of  the Coulomb force in 
$\nu>3$ justifies this assumption. We finally arrive at the Boltzmann 
equation 
\begin{eqnarray}
  B[f]
  &=&
  \int \frac{d^\nu p_2}{(2\pi\hbar)^\nu} \! \int \! d\Omega \, 
  \vert{\bf v}_1 - {\bf v}_2 \vert \, 
  \frac{d\sigma_{12}}{d\Omega} \,
  \Big[f_1({\bf p}_1^\prime) f_1({\bf p}_2^\prime) - 
  f_1({\bf p}_1) f_1({\bf p}_2)\Big] 
  \ .
\end{eqnarray}
In terms of a {\em quantum} transition amplitude $T$, we can 
express the Boltzmann scattering kernel in the form
\begin{eqnarray}
\nonumber
  B[f]
  &=&
  \int \frac{d^\nu p_1^\prime}{(2\pi\hbar)^\nu}\,
  \frac{d^\nu p_2^\prime}{(2\pi\hbar)^\nu}\,
  \frac{d^\nu p_2}{(2\pi\hbar)^\nu}\,\big\vert T_{1^\prime 2^\prime;\,12} 
  \big\vert^2\, 
  \bigg\{
  f_1({\bf p}_1^\prime) f_1({\bf p}_2^\prime)
  -
  f_1({\bf p}_1) f_1({\bf p}_2)
  \bigg\} 
\\[5pt] && \hskip1.3cm 
  (2\pi\hbar)^\nu\,\delta^\nu\!\Big( {\bf p}_1^\prime + {\bf p}_2^\prime - 
  {\bf p}_1 - {\bf p}_2 \Big)\,
  (2\pi\hbar) \delta\Big(E_1^\prime + E_2^\prime - E_1 - E_2\Big) \ .
\end{eqnarray}
The time non-invariance of the BE happens in two places in this argument: 
(i) using the $s=2$ equation at asymptotic times, and (ii) the molecular chaos
assumption.  The generalization to multi-species is easy. Two-point functions 
are still uncorrelated at $x_1$ and $x_2$, so that
\begin{eqnarray}
  f_2(x_1) &=& f_a({\bf p}_a) f_b({\bf p}_b)
\\
  f_2(x_2) &=& f_a({\bf p}_a^\prime) f_b({\bf p}_b^\prime) 
  \ ,
\end{eqnarray}
and the Boltzmann scattering kernel becomes
\begin{eqnarray}
\nonumber
  B_{ab}[f]
  &=&
  \int \frac{d^\nu p_a^\prime}{(2\pi\hbar)^\nu}\,
  \frac{d^\nu p_b^\prime}{(2\pi\hbar)^\nu}\,
  \frac{d^\nu p_b}{(2\pi\hbar)^\nu}\,\big\vert T_{a^\prime b^\prime;\,a b} 
  \big\vert^2\, 
  \bigg\{
  f_a({\bf p}_a^\prime) f_b({\bf p}_b^\prime)
  -
  f_a({\bf p}_a) f_b({\bf p}_b)
  \bigg\} 
\\[5pt] && \hskip1.3cm 
  (2\pi\hbar)^\nu\,\delta^\nu\!\Big( {\bf p}_a^\prime + {\bf p}_b^\prime - 
  {\bf p}_a - {\bf p}_b \Big)\,
  (2\pi\hbar) \delta\Big(E_a^\prime + E_b^\prime - E_a - E_b\Big) \ .
\end{eqnarray}
%%

%%%%%%%%%%%%%%%%%%%
\pagebreak
\section{The Lenard-Balescu Equation from BBGKY in $\bm{\nu < 3}$}
\label{sec_LBE}

We now derive the Lenard-Balescu equation (LBE) from the BBGKY 
hierarchy in spatial dimensions $\nu < 3$. We rely on Chapter~12 of
Clemmow and Dougherty\,\cite{cd} as our primary source in this section, 
since it is so clearly written and easily generalizes to multiple dimensions. 
The calculation is very long, but quite informative. The calculation of
Ref.~\cite{cd} actually breaks down in $\nu=3$ spatial dimensions 
because of a short-distance ultra-violet (UV) divergence. However, all 
quantities become finite when  $\nu<3$, and the calculation can proceed 
as presented. This is because the Coulomb force falls off like $1/r^{\nu-1}$, 
and this renders the UV divergence finite in $\nu < 3$. Since the long 
distance physics of the LBE is correct, the kernel does not suffer a
long-distance infra-red (IR) divergence.  In a single-component plasma, 
the LBE takes the form
%\,\footnote{
%\footnoteskip
%  As with the Boltzmann equation, we have restored a slowly 
%  varying spatial dependence to the single-particle distribution 
%  $f$. 
%} % end footnote
%%
\begin{eqnarray}
  \frac{\partial f}{\partial t} + 
  {\bf v}_1 \cdot \frac{\partial f}{\partial {\bf x}_1}
  &=& 
  L[f]
  \ ,
\label{LBEsimpnu_single}
\end{eqnarray}
with scattering kernel
\begin{eqnarray}
  L[f]
  &=&
  - \frac{\partial}{\partial {\bf p}} \cdot {\bf J}({\bf p})
  \\[8pt] 
  {\bf J}({\bf p})
  &=&
  \int\! 
  \frac{d^\nu p_2}{(2\pi\hbar)^\nu}\,\frac{d^\nu k}{(2\pi)^\nu}
  \, 
  {\bf k}   \,
  \bigg\vert \frac{e^2}{k^2\, \epsilon({\bf k},{\bf k} \cdot {\bf v}_1)}
  \bigg\vert^2 
  \pi\,\delta({\bf k} \cdot {\bf v}_1 - {\bf k} \cdot  {\bf v}_2)
  \nonumber
  %\\[5pt]&& \hskip4.5cm 
  \bigg[
  {\bf k}\!\cdot\! \frac{\partial}{\partial {\bf p}_1}
  -
  {\bf k}\!\cdot\! \frac{\partial}{\partial {\bf p}_2}
  \bigg]  
  f({\bf p}_1)   f({\bf p}_2) 
  \ .
\label{dedteigtAA}
\end{eqnarray} 
As always, we  take ${\bf v}_i={\bf p}_i/m$ for $i=1,2$. For a single
component plasma, the dielectric function $\epsilon$ is given by 
\begin{eqnarray}
  \epsilon({\bf k},\omega) 
  = 
  1 + \frac{e^2}{k^2} \! \int \frac{d^\nu p}{(2\pi\hbar)^\nu}\, 
  \frac{1}{\omega - {\bf k} \!\cdot\!  {\bf v}+ i \eta}\, {\bf k} \cdot 
   \frac{\partial f({\bf p}) }{\partial {\bf p}} \,,
\label{epsilon_single}
\end{eqnarray}
and the prescription $ \eta \to 0^+ $ is implicit, defining the
correct retarded time response.

\subsection{Formal Solution to the Perturbative Equations}

We now explicitly assume the plasma to be uniform, in the 
sense that the 1-point function is constant in space, being a
function only of momentum, 
\begin{eqnarray}
  f_1(X, t) = f({\bf p}, t)
   \ ,
\label{eq_f_uniform}
\end{eqnarray}
with $X = ({\bf x}, {\bf p})$. 
Galilean invariance then constrains the 2-point function
to take the form 
\begin{eqnarray}
   h_2(X_1, X_2, t) 
   &=& 
   h({\bf x}_1- {\bf x}_2,{\bf p}_1, {\bf p}_2, t) 
   \ .
\label{eq_h_uniform}
\end{eqnarray}
The time dependence $t$ will often be left implicit, and we will 
employ a slight abuse of notation by writing $h(X_1, X_2)$. The so 
called self-consistent fields ${\bf F}_i[f]$ defined in (\ref{eq_Fi_self_geni})
vanish under the condition 
of uniformity, and the kinetic equations take slightly simpler forms.
In Section~\ref{sec_pert_theory}, we showed that for long-range
interactions, in particular for the Coulomb force in $\nu<3$, the
the coupled system integro-differential equations is
\begin{eqnarray}
&& \hskip-1.5cm 
  \frac{\partial}{\partial t} \, f({\bf p}_1,t) 
  =
  - g \!
  \int dX \,  {\bf F}({\bf x}) \cdot 
   \frac{\partial}{\partial {\bf p}_1}\, g h_2({\bf x}, {\bf p}_1, {\bf p},t)
   \ .
\label{bbgkyOne_h2_uniform_b}
\end{eqnarray}
and 
\begin{eqnarray}
  \frac{\partial h}{\partial t} + V_1 h + V_2 h = S[f]
  \ ,
 \label{eq_hV1V2_LBE}
\end{eqnarray}
where the source term is
\begin{eqnarray}
  S[f]
  &=&  
  -
 {\bf F}_1^{(2)}  \cdot 
 \left[
 \frac{\partial }{\partial {\bf p}_1} 
  -
  \frac{\partial }{\partial {\bf p}_2} 
  \right]f({\bf p}_1) f({\bf p}_2)
  \ .
\label{eq_Sf1_mom_again}
\end{eqnarray}
Note that ${\bf F}_1^{(2)} = e {\bf E}({\bf x}_1 - {\bf x}_2)$ is the
Coulomb force at ${\bf x}_1$ from a point charge at ${\bf x}_2$.
These equations are accurate to order $g^2$ in the plasma
coupling. The quantity  $V_1$ is an integro-differential operator 
defined in $X_1$-space by 
\begin{eqnarray}
  V_1\, h(X_1, X_2)
  &=&
  {\bf v}_1  \cdot \frac{\partial h(X_1, X_2)}{\partial {\bf x}_1} 
  +
  \int dX_3 \, \, h(X_3, X_2) \,
  {\bf F}_1^{(3)} \cdot    \frac{\partial f({\bf p}_1)}{\partial {\bf p}_1} 
   \ ,
 \label{eq_VOne_a}
\end{eqnarray}
and $V_2$ is the corresponding operator in $X_2$-space, 
\begin{eqnarray}
  V_2 \, h(X_1, X_2)
  &=&
  {\bf v}_2  \cdot \frac{\partial h(X_1, X_2)}{\partial {\bf x}_2} 
  +  
  \int dX_3 \, h(X_3, X_1)\,
  {\bf F}_2^{(3)} \cdot \frac{\partial f({\bf p}_2)}{\partial {\bf p}_2} 
   \ ,
  \label{eq_VTwo_a}
\end{eqnarray}
with ${\bf F}_i^{(3)}$ (for $i=1,2$) being the Coulomb force at 
${\bf x}_i$ from ${\bf x}_3$. When the correlation function $h$ 
is written without arguments, it is assumed to be $h(X_1, X_2)$. 
The variable $X_2$ in~(\ref{eq_VOne_a}) \hbox{``\! just goes 
along for the ride''},  and we may regard $V_1$ as an operator 
in $X_1$-space acting on a functions $h(X_1)$. Similarly, the variable 
$X_1$ is  {\em free} in the operator $V_2$. The operators $V_1$ and 
$V_2$ therefore commute when acting on functions $h(X_1, X_2)$
of two variables, and $V_1$ and $V_2$ may consequently be treated 
as numbers when solving the differential equation (\ref{eq_hV1V2_LBE}) 
for $h(t)$. In deriving the Lenard-Balescu equation, we require 
the asymptotic time limit $t \to \infty$ of $h(t)$. This is because 
of Bogoliubov's hypothesis, which states that $h({\bf x}, {\bf p},t)$ 
quickly relaxes to its asymptotic value $h({\bf x}, {\bf p}, \infty)$, 
relative to $f({\bf p}, t)$. We may therefore treat $t$ as a parameter 
in the source term $S$, and the operators $V_1$, and $V_2$. 

Let us now find a formal solution for $h(t)$. We leave the phase-space 
variables ${\bf x}$ and ${\bf p}$ implicit. We will employ the method 
of Laplace transforms, where the Laplace transform and its inverse are 
related by
\begin{eqnarray}
  \tilde h(p) 
  &=& 
  \int_0^\infty dt \, e^{-p t}\,  h(t)
\label{eq_g_lxformA}
  \\[5pt]
  h(t) 
  &=& 
  \frac{1}{2\pi i}\int_C dp \, e^{pt} \,\tilde h(p)
  \ ,
\label{eq_g_lxformB}
\end{eqnarray}
where the contour $C$  runs parallel to the imaginary axis with all 
poles of $\tilde h(p)$ lying to the left of $C$. The analytic structure 
of $\tilde h(p)$ in the complex $p$-plane determines the function 
$h(t)$ for all values of $t$ greater than zero. Let us multiply 
(\ref{eq_hV1V2_LBE}) by $e^{-pt}$ and integrate over $t$, giving 
the equation
\begin{eqnarray}
  \int_0^\infty dt\,e^{-pt} 
  \left(\frac{\partial h}{\partial t} + V_1 h + V_2 h\right) 
  = 
  \int_0^\infty dt \, e^{-pt} S
  =
  p^{-1} S
  \ .
\label{eq_hV1V2_laplace}
\end{eqnarray}
Upon integrating by parts,
\begin{eqnarray}
   \int_0^\infty dt\,e^{-pt} \, \frac{\partial h}{\partial t}
   =
  e^{-pt} \, h(t)\bigg\vert_0^\infty +  \int_0^\infty dt\, p \,e^{-pt}\,  h
  =
  -h(0) + p \,\tilde h(p)
  \ ,
\end{eqnarray}
we can express this as
\begin{eqnarray}
  (p + V_1 + V_2) \tilde h(p) = p^{-1} S + h(0)
  \ .
\label{eq_pV1V2_htilde}
\end{eqnarray}
Solving for the Laplace transform $\tilde h$ gives the formal
solution
\begin{eqnarray}
   \tilde h(p) = (p + V_1 + V_2)^{-1} \left( p^{-1} S + h(0)\right)
   \ .
   \label{eq_solh}
\end{eqnarray}

We can find the asymptotic value $h(\infty)$ from (\ref{eq_solh}) in
the following manner. From (\ref{eq_solh}) we see that 
$\tilde h(p)$ has a pole at $p=0$, in addition to the other poles lying 
in the left half-plane with ${\rm Re}\,p < 0$. As $t \to \infty$, the dominant 
contribution to the integral (\ref{eq_g_lxformA}) comes from the 
$p=0$ pole.  This means we can replace the 
contour $C$ by a circular contour $C_r$ of radius $r$ about the origin,
and we can evaluate $h(\infty)$ by integrating around $C_r$ and taking 
the limit $r \to 0^+$. Points on $C_r$ are given by $p = r e^{i\theta}$.
Therefore $dp = i p \, d\theta$, and we can change variables from $p$ 
to $\theta$. Since we are interested in the $r \to 0^+$ limit, we can
replace factors of $p$ in the integrand by factors of $r$ (there is no
$\theta$-dependence at the origin), giving
\begin{eqnarray}
  h(\infty)
  &=&
   \lim_{t \to \infty}
   \lim_{r \to 0^+}
   \frac{1}{2\pi i}\oint_{C_r} dp \, e^{p t} \, \tilde h(p)
   =
   \lim_{t \to \infty}
  \lim_{r \to 0^+}
  \frac{1}{2\pi}\int_0^{2\pi} d\theta \, r \, e^{r t} \, \tilde h(r)
  \\[8pt]
  &=&
   \lim_{r \to 0^+} \frac{1}{2\pi} \cdot 2\pi \cdot  r \cdot 
   1 \cdot \tilde h(r)
   \ ,
   \nonumber\\
   \label{eq_h_inf_why}
\end{eqnarray}
where the factor of unity comes from $e^{rt} \to 1$ as $r \to 0^+$
(note that $t$ is fixed while the $r$-limit is taken). Consequently, 
we find the elegant and compact result
\begin{eqnarray}
   h(\infty)
  =
   \lim_{p \to 0^+} p \, \tilde h(p)   
   \ ,
   \label{eq_h_inf}
\end{eqnarray}
where we have changed variables from $r$ back to $p$ in the limit. 
Upon using  (\ref{eq_solh}) for the Laplace transform $\tilde h(p)$, 
we can write the asymptotic form as
\begin{eqnarray}
  h(\infty) = \lim_{p \to 0^+} (p + V_1 + V_2)^{-1} S
   \ .
\label{eq_hinf_a}
\end{eqnarray}
Note that the initial condition $\tilde h(0)$ does not appear in the 
asymptotic form. 

Recall that the Laplace transform of $e^{-at}$ is $(p + a)^{-1}$, 
and we can therefore write
\begin{eqnarray}
 (p + V_1 + V_2)^{-1} 
 &=&
  \int_0^\infty dt \, e^{-(p + V_1 + V_2)t}
  \\[5pt]
  &=&
  \frac{1}{(2\pi i)^2}\int_0^\infty dt \, e^{-pt}\int_{C_1} dp_1 \,
  \frac{ e^{p_1 t}}{p_1 + V_1}
  \int_{C_2} dp_2 \, \frac{e^{p_2 t}}{p_2 + V_2}
  \ ,
\end{eqnarray}
where we have expressed $e^{-V_1 t}$ and $e^{-V_2 t}$ as 
inverse Laplace transforms defined by contours $C_1$ and 
$C_2$, respectively.  These two contours are suitable inverse 
Laplace transform contours parallel to the imaginary axis, with 
all poles lying to their left. Upon performing the 
\hbox{$t$-integral}, we find
\begin{eqnarray}
 (p + V_1 + V_2)^{-1} 
 &=&
    \frac{1}{(2\pi i)^2}\int_{C_1} dp_1 \int_{C_2} dp_2 \,\frac{1}{p - p_1 - p_2} \,
    \frac{ 1}{p_1 + V_1} \,  \frac{1}{p_2 + V_2}
  \ ,
    \label{eq_hinf_b}
\end{eqnarray}
where ${\rm Re}\,p > {\rm Re}(p_1 + p_2)$ for the $t$-integral 
convergence at large $t$. The asymptotic form of $h$ can now 
be written
\begin{eqnarray}
 h(\infty)
  &=&
    \lim_{p \to 0^+}
    \frac{1}{(2\pi i)^2}\int_{C_1} d p_1 \int_{C_2} d p_2 \,
    \frac{1}{p - p_1 - p_2} \,\frac{ 1}{p_1 + V_1}\,\frac{1}{p_2 + V_2}\, S[f]
    \  .
  \label{eq_h_inf}
\end{eqnarray}
The source $S[f]$ is defined by (\ref{eq_Sf1_mom}). The problem 
now reduces to an exercise in complex analysis, albeit a rather involved 
exercise.  The next step involves calculating the action of the operators 
$(p + V_2)^{-1}$ and $(p + V_1)^{-1}$ on the source $S$.

\subsection{Preliminary Example}

As a prelude to finding the inverse operators above, let us 
consider a simpler problem in which $h$ is a function of only one 
phase-space variable $X$ (rather than $X_1$ and $X_2$), so that
$h(X,t)= h({\bf x}, {\bf p},t)$. Suppose now that $h$ satisfies the 
simplified equation
\begin{eqnarray}
  \frac{\partial h}{\partial t} + V h  = 0
  \ ,
 \label{eq_hV_LBE_simple}
\end{eqnarray}
where the operator is defined by
\begin{eqnarray}
  V  h({\bf x}, {\bf p})
  &=&
  {\bf v}  \cdot \frac{\partial h}{\partial {\bf x}} 
  +  
  \int dX_3 \, h({\bf x}_3, {\bf p}_3)\,
  {\bf F}_{{\bf x}}^{(3)} \cdot \frac{\partial f({\bf p})}{\partial {\bf p}} 
  \ .
\label{eq_VTwo_simple}
 \end{eqnarray}
As usual, ${\bf F}_{{\bf x}}^{(3)} = e {\bf E}({\bf x} - {\bf x}_3)$ is the 
Coulomb force at  ${\bf x}$ from a point charge at ${\bf x}_3$, and
the integration variable is $X_3=({\bf x}_3, {\bf p}_3)$. We shall express
 the operator $V$ in the more suggestive form
\begin{eqnarray}
  V h 
  &=&
   {\bf v}  \cdot \frac{\partial h}{\partial {\bf x}} 
  +  
  e {\bf E}[h] \cdot \frac{\partial f({\bf p})}{\partial {\bf p}} 
  \ ,
\label{eq_Vh_Eh}
\end{eqnarray}
where we define the electric field functional by
\begin{eqnarray}
  {\bf E}[h]({\bf x})
  =
  \int dX_3 \, h({\bf x}_3, {\bf p}_3)\, {\bf E}({\bf x} - {\bf x}_3)
  \ .
\label{eq_Eh_def}
\end{eqnarray}
The quantity ${\bf E}[h]$ is analogous to the self-consistent electric 
field ${\bf E}[f]$, although it does not vanish in a uniform plasma. To 
solve (\ref{eq_hV_LBE_simple}) for $h$, let us take the spatial Fourier 
transform and the temporal Laplace transform of (\ref{eq_hV_LBE_simple}),
\begin{eqnarray}
  (p + V)\, \tilde h({\bf k},  {\bf p}, p) 
  =  
  \tilde h({\bf k}, {\bf p}, 0)
  \ ,
\label{eq_pV_ex}
\end{eqnarray}
which has the  formal solution
\begin{eqnarray}
   \tilde h({\bf k},  {\bf p}, p) = (p + V)^{-1} \, \tilde h({\bf k},  {\bf p}, 0)
   \ .
 \label{eq_solh_ex}
\end{eqnarray}
The tilde 
over a function is used to denote both the Fourier and Laplace 
transforms. The transform of relevance should be clear from 
context (and from the presence of the variable ${\bf x}$ {\em vs.} 
${\bf k}$ or $t$ {\em vs.} $p$).  In other words,  we are using a 
mixed notation in which  $\tilde h({\bf k}, {\bf p}, p)$ is the spatial 
Fourier transform and the temporal Laplace transform of $h({\bf x}, 
{\bf p}, t)$, while $\tilde h({\bf k}, {\bf p}, 0)$ is the spatial Fourier 
transform of  $h({\bf x}, {\bf p}, t=0)$.  The momentum variable 
``just goes along for the ride,'' so we will keep it implicit. 

To find an expression for $(p + V)^{-1}$ that we can use in a 
calculation,  let us repeat the steps leading to the formal expression 
(\ref{eq_solh_ex}), except that now we shall employ the explicit 
form (\ref{eq_Vh_Eh}) for $V$. Note that we are using the
Fourier conventions give by (\ref{f1ft_convention}) and   
(\ref{f1ft_inv_convention}), or equivalently by (\ref{VxVtilde})
 and (\ref{VtildeVx}).  % 
Clemmow and Dougherty\,\cite{cd} use a convention with the 
opposite sign of ${\bf k}$ and different factors of $2\pi$, so care  
must be taken when comparing the results from these notes to 
Ref.~\cite{cd}. Note that the spatial integral in (\ref{eq_Eh_def}) 
is a convolution of $h({\bf x})$ and the Coulomb 
field ${\bf E}({\bf x})$ of a point charge. We can therefore use the 
convolution theorem when taking the spatial Fourier transform
of (\ref{eq_Vh_Eh}), giving
\begin{eqnarray}
  p \tilde h + i {\bf k}\cdot{\bf v} \,\tilde h 
  +
  e \tilde{\bf E }[h] \cdot  \frac{\partial f}{\partial {\bf p}} 
  = 
  \tilde h({\bf k}, {\bf p}, 0)
  \label{eq_phaeqnow_single}
  \ .
\end{eqnarray}
As stated above, the spatial Fourier transform $\tilde {\bf E}[h]$ 
of the self-induced field ${\bf E}[h]$ is performed by applying the 
convolution theorem, so that (being explicit with the arguments)
\begin{eqnarray}
 \tilde {\bf E}[h]({\bf k},p)
  &=& 
   \int\frac{d^\nu p_3}{(2\pi\hbar)^\nu} \
  \tilde h({\bf k}, {\bf p}_3, p)  \,  \tilde{\bf E}({\bf k})
  \ ,
  \label{eq_Ekp_def_single}
\end{eqnarray}
where $\tilde{\bf E}({\bf k})$ is the Fourier transform of the static 
Coulomb field, 
\begin{eqnarray}
  \tilde {\bf E}({\bf k}) 
  = 
  - i {\bf k}\,\tilde \phi({\bf k}) 
  =  
  i {\bf k}\, \frac{e}{k^2}
  \ .
\label{eq_E_point_ft}
\end{eqnarray}
For notational simplicity, we drop the functional dependence on 
$h$ from (\ref{eq_Ekp_def_single}), and write $\tilde{\bf E}({\bf k}, p)$.
We can now solve (\ref{eq_phaeqnow_single}) for $\tilde h$, giving
\begin{eqnarray}
\label{f1_sol_laplace_single}
  &&
  (p + V)^{-1}\,\tilde h({\bf k}, {\bf p}, 0)
  \equiv
 \tilde h({\bf k}, {\bf p},p)
  \\[8pt] \hskip2.0cm
  && 
  =
 \frac{1}{p + i {\bf k}\cdot{\bf v}}\bigg[\tilde h({\bf k}, {\bf p}, 0)
  - 
  e \tilde{\bf E }({\bf k}, p) \cdot  \frac{\partial  f({\bf p})}{\partial {\bf p}}
 \bigg]
 \ .
 \nonumber
\end{eqnarray}
As with the point charge in (\ref{eq_E_point_ft}),  the self-consistent 
electric field $\tilde {\bf E}({\bf k},p)$ can be expressed in terms of a 
self-consistent potential $ \tilde \phi({\bf k}, p)$ defined by
\begin{eqnarray}
 \tilde {\bf E}({\bf k},p) = -i {\bf k} \, \tilde \phi({\bf k}, p)
 \ .
 \label{eq_tildE_tildephi_single}
\end{eqnarray}
In terms of this self-consistent potential, we have
\begin{eqnarray}
 \tilde h({\bf k}, {\bf p},p)
  &=&
 \frac{1}{p + i {\bf k}\cdot{\bf v}}\bigg[\tilde h({\bf k}, {\bf p}, 0)
  +
  e  \tilde\phi({\bf k}, p)
  (i {\bf k}) \cdot  \frac{\partial f({\bf p})}{\partial {\bf p}}
 \bigg]
\label{f1_sol_laplace_phi_single}
\\[11pt]
 \tilde\phi({\bf k},p)
  &=& 
   \frac{e}{k^2}\int\frac{d^\nu p^\prime}{(2\pi\hbar)^\nu} \
  \tilde h({\bf k}, {\bf p}^\prime, p)  
    \ ,
  \label{eq_Ekp_def_phi_single}
\end{eqnarray}
where we have changed integration variables from ${\bf p}_3$
to ${\bf p}^\prime$. 

Let us now substitute (\ref{f1_sol_laplace_phi_single}) for 
$\tilde h$ into (\ref{eq_Ekp_def_phi_single}) for the potential,  
\begin{eqnarray}
  \tilde \phi({\bf k}, p)
  &=& 
  \frac{e}{k^2}\int\frac{d^\nu p^\prime}{(2\pi\hbar)^\nu} \,
  \frac{1}{ p +  i {\bf k} \cdot {\bf v}^\prime }
  \left[
  \tilde h({\bf k}, {\bf p}^\prime, 0)
  +
 e \,\tilde \phi({\bf k}, p) \, i {\bf k}  \cdot  
 \frac{\partial  f({\bf p}^\prime)}{\partial {\bf p}^\prime}
  \right]
  \ ,
  \label{eq_E_p_new_a_single}
\end{eqnarray}
where ${\bf v}^\prime = {\bf p}^\prime/m$.
Note that $\tilde\phi({\bf k}, p)$ appears on both sides of this equation,
and upon isolating the $\tilde\phi({\bf k}, p)$ term, we find
\begin{eqnarray}
 \Bigg[
 1  -  
  \frac{e^2}{k^2} \int\frac{d^\nu p^\prime}{(2\pi\hbar)^\nu} \,
  \frac{1}{ p +  i {\bf k} \cdot {\bf v}^\prime } \, 
  i {\bf k}  \cdot  \frac{\partial  f({\bf p}^\prime)}{\partial {\bf p}^\prime}
  \Bigg]  \tilde \phi({\bf k}, p)
  &=& 
  \frac{e}{k^2} \int\frac{d^\nu p^\prime}{(2\pi\hbar)^\nu} \,
  \frac{\tilde h({\bf k}, {\bf p}^\prime, 0)}{ p +  i {\bf k} \cdot {\bf v}^\prime }
  \ .
  \label{eq_E_p_new_a_single}
\end{eqnarray}
Solving (\ref{eq_E_p_new_a_single}) for the self-consistent 
potential therefore gives
\begin{eqnarray}
  \tilde \phi({\bf k}, p)
  &=& 
  \frac{e}{\bar\epsilon({\bf k}, p) \,k^2} 
  \int\frac{d^\nu p^\prime}{(2\pi\hbar)^\nu} \,
  \frac{\tilde h({\bf k}, {\bf p}^\prime, 0)}{ p +  i {\bf k} \cdot {\bf v}^\prime } 
  \ ,
  \label{eq_E_p_new_c_single}
\end{eqnarray}
where the ``dielectric function'' in Laplace space is defined by 
\begin{eqnarray}
  \bar\epsilon({\bf k}, p)
  =
 1 - 
  \int\frac{d^\nu p^\prime}{(2\pi\hbar)^\nu} \,  \frac{e^2}{k^2} \,
  \frac{1}{ p +  i {\bf k} \cdot {\bf v}^\prime } \, 
  i {\bf k}  \cdot  \frac{\partial  f({\bf p}^\prime)}{\partial {\bf p}^\prime}
  \ ,
 \label{eq_E_p_new_b_single}
\end{eqnarray}
with $p$ lying on the contour $C$.
For future reference, we record the 
following identities:
\begin{eqnarray}
  \int \frac{d^\nu p^\prime}{(2\pi\hbar)^\nu} \,
  \frac{e^2}{k^2}\,
  \frac{ i {\bf  k} \cdot 
  \partial  f({\bf p}^\prime)/\partial {\bf p}^\prime}
  {p + i {\bf k} \cdot {\bf v}^\prime} \,
  &=&
  1 - \bar\epsilon({\bf k}, p)
\label{eq_formOne_single}
\\[8pt]
  \int \frac{d^\nu p^\prime}{(2\pi\hbar)^\nu} \,  
  \frac{e^2}{k^2}\,
  \frac{ i {\bf  k} \cdot 
  \partial  f({\bf p}^\prime)/\partial {\bf p}^\prime}
  {p - i {\bf k} \cdot {\bf v}^\prime} \,
  &=&
  \bar\epsilon(-{\bf k}, p) - 1
  \ .
\label{eq_formTwo_single}
\end{eqnarray}
We will use these expressions throughout. 
Now, upon substituting (\ref{eq_E_p_new_c_single}) back into 
(\ref{f1_sol_laplace_phi_single}), we find
\begin{eqnarray}
  && \hskip-1.5cm
  (p + V)^{-1} \, \tilde h({\bf k}, {\bf p}, 0)
  \equiv
  \tilde h({\bf k}, {\bf p}, p)
\nonumber \\[8pt] &&  \hskip1.0cm
  =
  \frac{1}{ p +  i {\bf k} \cdot {\bf v} }
  \left[
  \tilde h({\bf k}, {\bf p}, 0)
  +
  \frac{e^2}{\bar\epsilon({\bf k}, p)\, k^2}   \, \,
  i {\bf k} \cdot  \frac{\partial f({\bf p})}{\partial {\bf p}}
   \int\frac{d^\nu p^\prime}{(2\pi\hbar)^\nu} \,
  \frac{\tilde h({\bf k}, {\bf p}^\prime, 0)}{ p +  i {\bf k} \cdot {\bf v}^\prime }
  \right]
\label{eq_Vinverse_single}
  \ .
\end{eqnarray}
We will generalize this result to  the operators $V_1$ and $V_2$ 
shortly. To identify the quantity $\bar\epsilon({\bf k}, p)$ physically, 
we can analytically continue (\ref{eq_E_p_new_b_single}), allowing 
$p$ to lie anywhere in the complex plane.  When $p=-i\omega$, we 
note that 
\begin{eqnarray}
  \bar\epsilon({\bf  k}, -i \omega )
  &=&
  \epsilon({\bf k}, \omega)
\label{eq_epsilon_minus}
\end{eqnarray}
Thus, the analytically continued dielectric function in Laplace 
space is just the ordinary dielectric function in temporal Fourier
space. We also note that
\begin{eqnarray}
  \epsilon(-{\bf  k}, -\omega )
  &=&
  \epsilon^{\,*}({\bf k}, \omega)
  \ ,
\label{eq_epsilon_cc}
\end{eqnarray}
and therefore
\begin{eqnarray}
  \bar\epsilon(-{\bf  k}, i \omega )
  &=&
  \epsilon(-{\bf k}, -\omega)
  =
   \epsilon^{*}({\bf k}, \omega)
  \ .
\label{eq_epsilon_cc_minusk}
\end{eqnarray}
These complex conjugation properties will be useful in the
forthcoming calculation. 

\subsection{The Lenard-Balescu Equation}
\label{sec_LBE_alg}

We now return to the Lenard-Balescu formalism in $X_1$-$X_2$ 
space, and to the 2-point correlation $h(X_1, X_2, t)$. Bogoliubov's 
hypothesis 
means that the time scale of $h(X_1, X_2, t)$ is much shorter than the
time scale of $f(X, t)$, so we can replace $h$ by its $t \to \infty$ limit 
relative to $f$. Therefore,  we shall assume that $h(X_1, X_2, t)$ relaxes 
to its asymptotic value $h(X_1, X_2, \infty)$, and that the LBE kernel
is
\begin{eqnarray}
 L[h] 
 &\equiv&
    - \int dX_2 \,  e {\bf E}_1^{(2)} \cdot \frac{\partial 
    h(X_1, X_2, \infty)}{\partial {\bf p}_1} 
\label{eq_Lh_dotJ}
  =
  -\frac{\partial}{\partial{\bf p}_1} \cdot {\bf J}
  \\[8pt]
  {\bf J}(X_1)
  &\equiv&
  \int dX_2 \,\, e {\bf E}_1^{(2)} \, h(X_1, X_2,\infty)
\label{eq_J_hinf}
  \ .
\end{eqnarray}
In the single-particle distribution $f({\bf p},t)$, the time $t$ is
treated as a parameter. 
It is noteworthy that the approximation of replacing $h(t)$ by its
asymptotic value of $h(\infty)$ is where the non-reversibility in time 
enters the LBE kinetic equation. 
Because the distribution $f$ is uniform, the 2-point correlation 
function $h({\bf x}_1, {\bf x}_2,,{\bf p}_1, {\bf p}_2,\infty)$
reduces to a function of only a single spatial coordinate, 
$h({\bf x},{\bf p}_1, {\bf p}_2, \infty)$. The Fourier transform of this 
function is $\tilde h ({\bf k},{\bf p}_1, {\bf p}_2, \infty)$, and we
see that
\begin{eqnarray}
   h({\bf x}_1 - {\bf x}_2, {\bf p}_1, {\bf p}_2,\infty)
   =
    \int \frac{d^\nu k}{(2\pi)^\nu}\, 
    e^{i({\bf x}_1 - {\bf x}_2) \cdot {\bf k}} \,
   \tilde h({\bf k},{\bf p}_1, {\bf p}_2, \infty)
  \ .
  \label{eq_intp2_htilde}
\end{eqnarray}
In expression (\ref{eq_J_hinf}), let us write the inter-molecular as 
${\bf F}_1^{(2)} = e {\bf E}({\bf x}_1 - {\bf x}_2)$, where ${\bf E}({\bf x})$ 
is the electric field of a point charge at the origin. Recall that the 
Fourier transforms of ${\bf E}({\bf x})$ and the corresponding potential 
$\phi({\bf x})$ are
\begin{eqnarray}
  e \tilde {\bf E}({\bf k})
  &=&
  -i {\bf k} \tilde \phi({\bf k})
  \label{eq_etilde_phi}
  \\[5pt]
    \tilde\phi({\bf k}) 
  &=& 
  \frac{e^2}{ k^2}
  \label{eq_pot_energy_phi}
  \ ,
\end{eqnarray}
where I have temporarily changed conventions by including an extra 
factor of $e$ into the electric potential $\phi$. This is to avoid factors 
of $e \phi$, as the stray electric charge is cumbersome to track. For a 
multi-species  plasma, we would change $e^2$ in  (\ref{eq_pot_energy_phi})
to $e_a e_b$.  The Lenard-Balescu current (\ref{eq_J_hinf}) now becomes
\begin{eqnarray}
  {\bf J}
  &=&
  \int \frac{d^\nu p_2}{(2\pi\hbar)^\nu} \int d^\nu x_2 \,
  e{\bf E}({\bf x}_1 - {\bf x}_2) \, 
  h({\bf x}_1 - {\bf x}_2, {\bf p}_1, {\bf p}_2,\infty)
  \\[8pt]
  &=&
    \int \frac{d^\nu p_2}{(2\pi\hbar)^\nu}\, \int d^\nu x \,
  e{\bf E}({\bf x}) \,  h({\bf x}, {\bf p}_1, {\bf p}_2,\infty)
  \ ,
  \label{eq_J_nospace}
\end{eqnarray}
where we have made the change of variables ${\bf x} = {\bf x}_1 - {\bf x}_2$
in   (\ref{eq_J_nospace}), thereby illustrating that ${\bf J}$ is constant in 
space (independent of ${\bf x}_1$). We can express the $x$-integration in 
(\ref{eq_J_nospace}) as
\begin{eqnarray}
   \int d^\nu x \,
   e{\bf E}({\bf x}) \,  h({\bf x})
   &=&
    \int d^\nu x \, 
    \int \frac{d^\nu k_1}{(2\pi)^\nu}\, e^{i {\bf k}_1 \cdot {\bf x}} \,
    e\tilde {\bf E}({\bf k}_1) \,  
   \int \frac{d^\nu k_2}{(2\pi)^\nu}\, 
   e^{i {\bf k}_2 \cdot {\bf x}} \,\tilde h({\bf k}_2)
   \\[5pt]
   &=&
    \int \frac{d^\nu k_2}{(2\pi)^\nu}\, 
   \int \frac{d^\nu k_1}{(2\pi)^\nu}\, 
   (2\pi)^\nu\delta^{(\nu)}({\bf k}_2 + {\bf k}_1)\, 
   (- i {\bf k}_1) \tilde \phi({\bf k}_1) \, \tilde h({\bf k}_2)
   \\[5pt]
   &=&
    \int \frac{d^\nu k_2}{(2\pi)^\nu}\, 
    i {\bf k}_2\, \tilde \phi(-{\bf k}_2) \, \tilde h({\bf k}_2)
  \ ,
\end{eqnarray}
and since $\tilde\phi({\bf k})$ is even in ${\bf k}$, the current 
(\ref{eq_J_nospace}) becomes
\begin{eqnarray}
  {\bf J}({\bf p}_1)
  &=&
    \int \frac{d^\nu k}{(2\pi)^\nu}\,  \frac{d^\nu p_2}{(2\pi\hbar)^\nu}\, 
    i {\bf k} \tilde \phi({\bf k}) \, \tilde h({\bf k},{\bf p}_1, {\bf p}_2, \infty)
 \label{eq_J_alg}
 \\[8pt]
   &=&
    -\int \frac{d^\nu k}{(2\pi)^\nu}\,  
    {\bf k} \tilde \phi({\bf k}) \,
     {\rm Im}\,I({\bf k},{\bf p}_1, {\bf p}_2, \infty)
 \label{eq_J_alg_real}
  \ ,
\end{eqnarray}
where
\begin{eqnarray}
  I({\bf k}, {\bf p}_1)
  \equiv
    \int \frac{d^\nu p_2}{(2\pi\hbar)^\nu}\, 
    \tilde h({\bf k},{\bf p}_1, {\bf p}_2, \infty)
  \ .
\label{eq_htilde_int}
\end{eqnarray}
In (\ref{eq_J_alg_real}) we have used the fact that ${\bf J}$ must 
be real, although we will continue to employ the form  (\ref{eq_J_alg}) 
until the end of the calculation. Note that we only need to find a 
momentum integral of the correlation function, $I({\bf k}, {\bf p}_1)$,
and not the correlation function itself.  This is quite fortunate, since
the integral (\ref{eq_htilde_int}) turns out to simplify considerably
relative to the full perturbation $\tilde h({\bf k},{\bf p}_1, {\bf p}_2, \infty)$.
Using expression  (\ref{eq_h_inf}) relating the perturbation to the
source term,  the spatial Fourier transform of the correlation function
becomes
\begin{eqnarray}
  \tilde h({\bf k}, {\bf p}_1, {\bf p}_2,  \infty)
  &=&
  \int d^\nu x \, e^{-i {\bf k} \cdot {\bf x}} \, h({\bf x}, {\bf p}_1, {\bf p}_2, \infty)
  \\[5pt]
    &=&
  \lim_{p\to 0^+}
    \frac{1}{(2\pi i)^2}\int_{C_1} dp_1 \int_{C_2} dp_2 \,
    \frac{ 1}{(p - p_1 - p_2)(p_1 + V_1)
    (p_2 + V_2)}\,S({\bf k}, {\bf p}_1, {\bf p}_2)
  \ ,
  \nonumber\\
\end{eqnarray}
where the source term is
\begin{eqnarray}
  S({\bf k}, {\bf p}_1, {\bf p}_2)
  &=&
  \tilde\phi({\bf k}) \, i{\bf k} \cdot 
  \bigg[
  \underbrace{~
  \frac{\partial f}{\partial {\bf p}_1}\, f({\bf p}_2)
  ~}_{{\rm term B}}
  -
  \underbrace{~  
  \frac{\partial f}{\partial {\bf p}_2}\, f({\bf p}_1)
  ~}_{{\rm termA}}
  \bigg] 
  \ .
\label{eq_S_def_vOne}
\end{eqnarray}
For future reference, I have labeled the two terms of $S$ 
by the names termA and termB, and (\ref{eq_htilde_int})
becomes
\begin{eqnarray}
  &&
  I({\bf k}, {\bf p}_1) 
  \nonumber\\[5pt]
   && =
  \frac{1}{(2\pi i)^2}\lim_{p\to 0^+} \!
  \int \! \frac{d^\nu p_2}{(2\pi\hbar)^\nu} \! 
  \int_{C_1} \! dp_1 \! \int_{C_2} \! dp_2 \,
  \frac{ 1}{(p - p_1 - p_2)(p_1 + V_1)
  (p_2 + V_2)}\,S({\bf k}, {\bf p}_1, {\bf p}_2)
  \ .
  \nonumber\\[-3pt]
  \label{eq_intph_complete}
\end{eqnarray}

\begin{figure}[t!]
\includegraphics[scale=0.45]{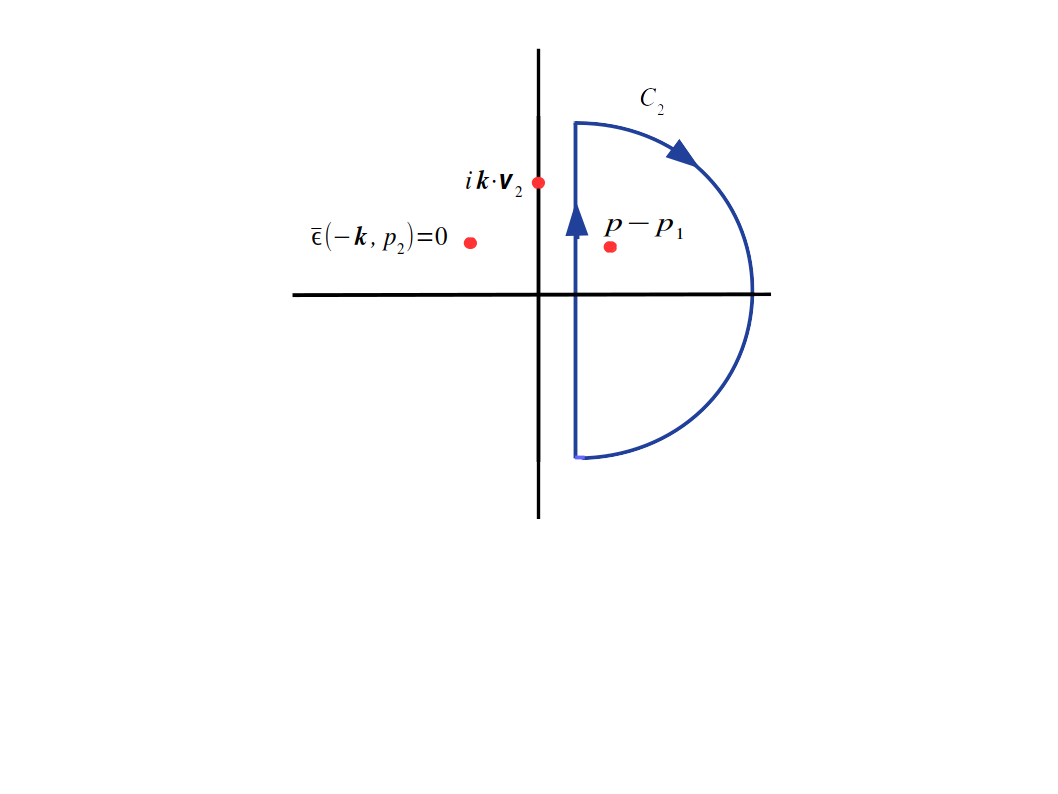} 
\vskip-3.0cm 
\caption{\footnoteskip  
  Contour $C_2$ in the $p_2$-plane. The contour can be closed
  in the right half-plane, oriented in the clockwise direction and 
  enclosing the simple pole $p_2 = p - p_1$. 
}
\label{fig_Cone}
\end{figure}

We must now calculate $(p_1 + V_1)^{-1}$  and $(p_2 + V_2)^{-1}$ 
on $S$. Expression (\ref{eq_Vinverse_single}) is the solution to 
the inverse problem in the simpler context of a single space-variable, 
and with this result in hand,  let us return to the full equation involving 
$V_1$ and $V_2$. Since ${\bf x}_1$ and ${\bf x}_2$ appear with 
opposite signs in  (\ref{eq_intp2_htilde}), the value of ${\bf k}$ in 
$V_2$ must be of the opposite sign as the corresponding value in 
$V_1$, and by referring back to (\ref{eq_Vinverse_single})
we can express
\begin{eqnarray}
  && \hskip-0.3cm
  (p_1 + V_1)^{-1} S({\bf p}_1, {\bf p}_2)
  \!=\!
  \frac{1}{p_1 + i {\bf k} \cdot {\bf v}_1} \!
  \left[
  S({\bf p}_1, {\bf p}_2) 
  +
  \frac{\tilde\phi({\bf k})}{\bar \epsilon({\bf k},p_1)}\,
  i{\bf k} \cdot \frac{\partial f({\bf p}_1)}{\partial {\bf p}_1} \!
  \int \! \frac{d^3p _1^\prime}{(2\pi\hbar)^\nu}  \,
   \frac{S({\bf p}_1^\prime,{\bf p}_2)}{p_1 + i {\bf k} \cdot {\bf v}_1^\prime}\,
  \right]
  \nonumber\\
  \label{p1Inverse_single}
\\[11pt]
  && \hskip-0.3cm
  (p_2 + V_2)^{-1} S({\bf p}_1, {\bf p}_2)
  \!=\!
  \frac{1}{p_2 - i {\bf k} \cdot {\bf v}_2} \!
  \left[
  S({\bf p}_1,{\bf p}_2) 
  -
  \frac{\tilde\phi({\bf k})}{\bar \epsilon(-{\bf k},p_2)}\,
  i{\bf k} \cdot \frac{\partial f({\bf p}_2)}{\partial {\bf p}_2} \!
  \int \! \frac{d^3p _2^\prime}{(2\pi\hbar)^\nu}  \,
   \frac{S({\bf p}_1,{\bf p}_2^\prime)}{p_2 - i {\bf k} \cdot {\bf v}_2^\prime}\,
  \right]
   \ .
  \nonumber\\
  \label{p2Inverse_single}
\end{eqnarray}
Returning to (\ref{eq_intph_complete}), we find
\begin{eqnarray}
 \int \frac{d^\nu p _2}{(2\pi\hbar)^\nu} \,
  (p_2 + V_2)^{-1} S({\bf p}_1, {\bf p}_2)
  &=&
  \int \frac{d^\nu p _2}{(2\pi\hbar)^\nu} 
  \frac{S({\bf p}_1, {\bf p}_2) }{p_2 - i {\bf k} \cdot {\bf v}_2}
  -
  \\[11pt] && \hskip-5.0cm
  \frac{1}{\bar \epsilon(-{\bf k},p_2)}\,  
 \underbrace{ \int \frac{d^\nu p _2}{(2\pi\hbar)^\nu} \,
 \frac{\tilde\phi({\bf k})}{p_2 - i {\bf k} \cdot {\bf v}_2} \,
  i{\bf k} \cdot \frac{\partial f({\bf p}_2)}{\partial {\bf p}_2}}_{
  \bar\epsilon(-{\bf k}, p_2)-1}  \,
  \int \frac{d^\nu p _2^\prime}{(2\pi\hbar)^\nu}
  \,
   \frac{S({\bf p}_1, {\bf p}_2^\prime)}{p_2 -  i {\bf k} \cdot {\bf v}_2^\prime}\,
  \nonumber
  \\[10pt]
  &=&
   \frac{1}{\bar \epsilon(-{\bf k},p_2)}\,
    \int \frac{d^\nu p _2^\prime}{(2\pi\hbar)^\nu} 
  \frac{ S({\bf p}_1, {\bf p}_2^\prime) }{p_2 - i {\bf k} \cdot {\bf v}_2^\prime}
  \ .
\end{eqnarray}
Thus, upon changing variables in the last integral from ${\bf p}_1^\prime$
to ${\bf p}_2$, we can write
\begin{eqnarray}
 \int \frac{d^\nu p _2}{(2\pi\hbar)^\nu} \,
  (p_2 + V_2)^{-1} S({\bf p}_1, {\bf p}_2)
  &=&
   \frac{1}{\bar \epsilon(-{\bf k},p_2)}\,
    \int \frac{d^\nu p _2}{(2\pi\hbar)^\nu} 
  \frac{ S({\bf p}_1, {\bf p}_2) }{p_2 - i {\bf k} \cdot {\bf v}_2}\,
  \ ,
\end{eqnarray}
and  (\ref{eq_intph_complete})  becomes
\begin{eqnarray}
  &&
    I({\bf k}, {\bf p}_1) 
\nonumber\\[5pt]
  && =
   -\frac{1}{(2\pi i)^2}\lim_{p \to 0^+}\int \! \frac{d^\nu p_2}{(2\pi\hbar)^\nu} \, 
   \int_{C_1} dp_1 \int_{C_2} dp_2 \,
    \frac{1}{p_2 + p_1 - p}\,
    %{p - p_1 - p_2}\, 
    \frac{ 1}{\bar\epsilon(-{\bf k}, p_2)} \,  \frac{1}{p_2 - i{\bf k}\cdot{\bf v}_2}\, \label{eq_inttwoh}
\nonumber \\[5pt]
   && \hskip7.1cm  \times \,
    (p_1 + V_1)^{-1} S({\bf k}, {\bf p}_1, {\bf p}_2)
    \ .
\end{eqnarray}
We now perform the integral over $p_2$. There 
are poles at $p_2 = i {\bf k} \cdot {\bf v}_2$ and $p_2 = p - p_1$, and the zeros 
of $\bar\epsilon({\bf k}, p_2)$. Recall that the contour $C_2$ runs parallel to 
the imaginary axis with all singularities of $(p_2 + V_2)^{-1}$ lying to the left 
of $C_2$. As illustrated in Fig.~\ref{fig_Cone}, we can complete the contour 
$C_2$ to include a large semicircle at infinity (as the integrand vanishes there).  
The contour now encloses the pole $p_2=p - p_1$ and is clock-wise oriented, 
and so the $p_2$ integral may be performed using the residue theorem,
\begin{eqnarray}
    I({\bf k}, {\bf p}_1) 
    =
    \frac{1}{2\pi i}\lim_{p \to 0^+}\int \frac{d^\nu p_2}{(2\pi\hbar)^\nu} \,  
    \int_{C_1} dp_1 \,
     \frac{ 1}{\bar\epsilon(-{\bf k}, p-p_1)} \, \frac{1}{p - p_1 - i{\bf k}\cdot{\bf v}_2}\, 
    (p_1 + V_1)^{-1} S({\bf k}, {\bf p}_1, {\bf p}_2)
    \ .
    \nonumber\\
\label{eq_intoneh}
\end{eqnarray}
Using  (\ref{p1Inverse_single}) to express $(p+V_1)^{-1}S$, we can now write
(\ref{eq_intoneh}) in the form
\begin{eqnarray}
  &&
  I({\bf k}, {\bf p}_1)
  \nonumber\\[5pt]
  &&=
  \frac{1}{2\pi i}\lim_{p \to 0^+}\int \frac{d^\nu p_2}{(2\pi\hbar)^\nu} \,  
  \int_{C_1} dp_1 \,
  \frac{ 1}{\bar\epsilon(-{\bf k}, p-p_1)} \, 
  \frac{1}{p - p_1 - i{\bf k}\cdot{\bf v}_2}\,   \frac{1}{p_1 + i {\bf k} \cdot {\bf v}_1}
\label{I_intCtwo}
  \\[8pt] && \hskip3.8cm
  \Bigg[
  \underbrace{~
  S({\bf p}_1, {\bf p}_2) 
    ~}_{{\rm term 1}}
  +
  \underbrace{~
  \frac{\tilde\phi({\bf k})}{\bar \epsilon({\bf k},p_1)}\,
  i{\bf k} \cdot \frac{\partial f({\bf p}_1)}{\partial {\bf p}_1}
  \int \frac{d^3p _1^\prime}{(2\pi\hbar)^\nu}  \,
  \frac{S({\bf p}_1^\prime, {\bf p}_2)}{p_1 + i {\bf k} \cdot {\bf v}_1^\prime}
      ~}_{{\rm term 2}}
  \Bigg]
  \ ,
\nonumber
\end{eqnarray}
where I have labeled the two terms of (\ref{I_intCtwo}) by term1 and term2.
The source $S$ given by (\ref{eq_S_def_vOne}) contains two terms, termA 
and termB, and we must therefore examine a total of four terms:
\begin{eqnarray}
  I({\bf k}, {\bf p}_1)
  &=&
  I_{\rm 1A} +  I_{\rm 1B} +  I_{\rm 2A} +  I_{\rm 2B}
  \ ,
\end{eqnarray}
where
\begin{eqnarray}
  I_{\rm 1A}
  &=&
  -
 \frac{1}{2\pi i} \lim_{p \to 0^+} \int_{C_1} dp_1 \int \frac{d^\nu p_2}{(2\pi\hbar)^\nu} \, 
  \frac{1}{\bar\epsilon(-{\bf k}, p-p_1)} \,   
  \frac{1}{p - p_1 - i {\bf k} \cdot {\bf v}_2} \, 
  \frac{1}{p_1 + i {\bf k} \cdot {\bf v}_1}
  \nonumber\\[5pt]
  && \hskip4.8cm
  \bigg[ \tilde\phi({\bf k}) i {\bf k} \cdot \partial f({\bf p}_2)/\partial {\bf p}_2 \, f({\bf p}_1)
  \bigg]
  \ ,
 \label{eq_term1A}
\end{eqnarray}
\begin{eqnarray}
  I_{\rm 1B}
  &=&
  \frac{1}{2\pi i} \lim_{p \to 0^+} \int_{C_1} dp_1 \int\frac{d^\nu p_2}{(2\pi\hbar)^\nu} \, 
  \frac{1}{\bar\epsilon(-{\bf k}, p-p_1)} \,   
  \frac{1}{p - p_1 - i {\bf k} \cdot {\bf v}_2} \, 
  \frac{1}{p_1 + i {\bf k} \cdot {\bf v}_1}
  \nonumber\\[5pt]
  && \hskip4.8cm
  \bigg[ \tilde\phi({\bf k}) i {\bf k} \cdot \partial f({\bf p}_1)/\partial {\bf p}_1 \, f({\bf p}_2)
  \bigg]
  \ ,
 \label{eq_term1B}
\end{eqnarray}
\begin{eqnarray}
  I_{\rm 2A}
  &=&
  -\frac{1}{2\pi i} \lim_{p \to 0^+} \int_{C_1} dp_1 \int \frac{d^\nu p_2}{(2\pi\hbar)^\nu} \, 
  \frac{1}{\bar\epsilon(-{\bf k}, p-p_1)} \,   
  \frac{1}{p - p_1 - i {\bf k} \cdot {\bf v}_2} \, 
  \frac{1}{p_1 + i {\bf k} \cdot {\bf v}_1}
  \nonumber\\[8pt]
  && \hskip2.00cm
  \bigg[ 
  \frac{\tilde\phi({\bf k}) (i {\bf k}) \cdot \partial f/
  \partial {\bf p}_1}{\bar\epsilon({\bf k}, p_1)} \,
  \int \frac{d^\nu p_1^\prime}{(2\pi\hbar)^\nu} \,
   \frac{\tilde\phi({\bf k}) (i {\bf k}) \cdot
  \partial f / \partial {\bf p}_2 \, f({\bf p}_1^\prime)}{p_1 + i {\bf k}
  \cdot {\bf v}_1^\prime}
  \bigg]
  \ ,
 \label{eq_term2A}
\end{eqnarray}
and
\begin{eqnarray}
  I_{\rm 2B}
  &=&
  \frac{1}{2\pi i} \lim_{p \to 0^+} \int_{C_1} dp_1 \int \frac{d^\nu p_2}{(2\pi\hbar)^\nu} \, 
  \frac{1}{\bar\epsilon(-{\bf k}, p-p_1)} \,   
  \frac{1}{p - p_1 - i {\bf k} \cdot {\bf v}_2} \, 
  \frac{1}{p_1 + i {\bf k} \cdot {\bf v}_1}
  \nonumber\\[8pt]
  && \hskip2.00cm
  \bigg[ 
  \frac{\tilde\phi({\bf k}) (i {\bf k}) \cdot \partial f/
  \partial {\bf p}_1}{\bar\epsilon({\bf k}, p_1)} \,
  \int \frac{d^\nu p_1^\prime}{(2\pi\hbar)^\nu} \, 
  \frac{\tilde\phi({\bf k}) (i {\bf k}) \cdot
  \partial f / \partial {\bf p}_1^\prime \, f({\bf p}_2)}{p_1 + i {\bf k}
  \cdot {\bf v}_1^\prime}
  \bigg]
  \ .
 \label{eq_term2B}
\end{eqnarray}
We combine the $A$-terms together and the $B$-terms together. 
Using (\ref{eq_formTwo_single}) in (\ref{eq_term1A}), we can
perform the ${\bf p}_2$-integral to give
\begin{eqnarray}
  I_{\rm 1A}
  &=&
  -\frac{1}{2\pi i} \lim_{p \to 0^+} \int_{C_1} dp_1
  \frac{ f({\bf p}_1)}{p_1 + i {\bf k} \cdot {\bf v}_1} \, 
  \frac{1}{\bar\epsilon(-{\bf k}, p-p_1)} \,
  {\scriptscriptstyle \times}
  \bigg[\bar\epsilon(-{\bf k}, p-p_1) - 1\bigg]
\nonumber\\[10pt]
  &=&
  \frac{1}{2\pi i} \lim_{p \to 0^+} \int_{C_1} dp_1
  \frac{1}{p_1 + i {\bf k} \cdot {\bf v}_1} \, 
  \bigg[\frac{1}{\bar\epsilon(-{\bf k}, p-p_1)} - 1 \bigg] f({\bf p}_1)
  \ ,
\label{eq_term1A_a}
\end{eqnarray}
Similarly, we use (\ref{eq_formTwo_single})  in (\ref{eq_term2A}) to
perform the ${\bf p}_2$-integral,  and after rearranging terms and 
changing the remaining integration variable from ${\bf p}_1^\prime$ 
to ${\bf p}_2$, we find
\begin{eqnarray}
  I_{\rm 2A}
  &=&
  \frac{1}{2\pi i} \lim_{p \to 0^+} \int_{C_1} dp_1\,
  \frac{1}{p_1 + i {\bf k} \cdot {\bf v}_1} \,
  \bigg[\frac{1}{\bar\epsilon(-{\bf k}, p-p_1)}-1 \bigg] \,
  \nonumber\\[5pt]
  && \hskip3.25cm
  \frac{1}{\bar\epsilon({\bf k}, p_1)}
  \bigg[
  \tilde\phi({\bf k})\, i {\bf k}\cdot \frac{\partial f}{\partial {\bf p}_1}
  \int \frac{d^\nu p_2}{(2\pi\hbar)^\nu} \, 
  \frac{ f({\bf p}_2)}{p_1 + i {\bf k}\cdot {\bf v}_2}
  \bigg]
  \ .
\label{eq_term2A_a}
\end{eqnarray}
Upon reorganizing the terms in (\ref{eq_term1B}) we can write 
\begin{eqnarray}
  I_{\rm 1B}
  &=&
  \frac{1}{2\pi i} \lim_{p \to 0^+} \int_{C_1} dp_1 \,
  \frac{1}{p_1 + i {\bf k} \cdot {\bf v}_1}
  \frac{1}{\bar\epsilon(-{\bf k}, p-p_1)} \,   
  \nonumber\\[5pt]
  && \hskip3.3cm
  \bigg[
  \tilde\phi({\bf k}) \, i {\bf k} \cdot \frac{\partial f}{\partial {\bf p}_1}
  \int \frac{d^\nu p_2}{(2\pi\hbar)^\nu} \, 
  \frac{f({\bf p}_2)}{p - p_1 - i {\bf k} \cdot {\bf v}_2} 
  \bigg]
  \ .
\label{eq_term1B_a}
\end{eqnarray}
Finally, upon using (\ref{eq_formOne_single}) to perform the
${\bf p}_1^\prime$-integral, expression (\ref{eq_term2B})
becomes
\begin{eqnarray}
  I_{\rm 2B}
  &=& 
  \frac{1}{2\pi i} \lim_{p \to 0^+} \int_{C_1} dp_1 \,
  \frac{1}{p_1 + i {\bf k} \cdot {\bf v}_1} \,
  \frac{1}{\bar\epsilon(-{\bf k}, p - p_1)} \,
  \bigg[\frac{1}{\bar\epsilon({\bf k}, p_1)} - 1\bigg] 
  \nonumber\\[11pt]
  && \hskip3.2cm
  \bigg[
  \tilde\phi({\bf k}) \,i {\bf k}\cdot \frac{\partial f}{\partial {\bf p}_1}
  \int \frac{d^\nu p_2}{(2\pi\hbar)^\nu} \, 
  \frac{f({\bf p}_2)}{p - p_1 - i {\bf k} \cdot {\bf v}_2} 
  \bigg]
  \ .
\label{eq_term2B_a}
\end{eqnarray}
Note that the  ${\bf k}\cdot {\bf v}_2$ term in the ${\bf p}_2$ integrals
of (\ref{eq_term2A_a}) and (\ref{eq_term2B_a}) have opposite 
signs, a fact that will be critical as the calculation proceeds.
Note that (\ref{eq_term1A_a}) and (\ref{eq_term2A_a}) give
\begin{eqnarray}
  I_{\rm 1A} + I_{\rm 2A}
  &=&
  \frac{1}{2\pi i} \lim_{p \to 0^+} \int_{C_1} dp_1
  \frac{1}{p_1 + i {\bf k} \cdot {\bf v}_1} \, 
  \bigg[\frac{1}{\bar\epsilon(-{\bf k}, p-p_1)} - 1 \bigg] \bigg[
  f({\bf p}_1) 
  +
  \nonumber \\[5pt] && \hskip2.5cm
  \tilde\phi({\bf k}) \, i {\bf k} \cdot \frac{\partial f}{\partial {\bf p}_1}\,
  \frac{1}{\bar\epsilon({\bf k}, p_1)} \,
  \int \frac{d^\nu p_2}{(2\pi\hbar)^\nu} \, 
  \frac{ f({\bf p}_2)}{p_1 + i {\bf k} \cdot {\bf v}_2}
  \bigg]
  \ ,
\label{eq_1Aplus2A}
\end{eqnarray}
while the final two terms become 
\begin{eqnarray}
  I_{\rm 1B} + I_{2B}
  &=&
  \frac{1}{2\pi i} \lim_{p \to 0^+} \int_{C_1} dp_1 \,
  \frac{1}{p_1 + i {\bf k} \cdot {\bf v}_1} \,
  \frac{1}{\bar\epsilon(-{\bf k}, p-p_1)\, \bar\epsilon({\bf k}, p_1)} 
\nonumber\\[8pt] && \hskip2.5cm
  \bigg[
  \tilde\phi({\bf k}) \, i {\bf k} \cdot \frac{\partial f}{\partial {\bf p}_1}
  \int\frac{d^\nu p_2}{(2\pi\hbar)^\nu} \, 
  \frac{f({\bf p}_2)}{p - p_1 - i {\bf k} \cdot {\bf v}_2} 
  \bigg]
  \ .
\label{eq_1Bplus2B}
\end{eqnarray}
Upon adding (\ref{eq_1Aplus2A}) and (\ref{eq_1Bplus2B}) we 
can write $I({\bf k}, {\bf p}_1)$ in the form
\begin{eqnarray}
  I
  &=& 
  \frac{1}{2\pi i} \lim_{p \to 0^+} \int_{C_1} dp_1
  \frac{1}{p_1 + i {\bf k} \cdot {\bf v}_1} \, \Bigg\{
  \underbrace{
  \frac{\tilde\phi({\bf k})\, i {\bf k} \cdot \partial f/\partial {\bf p}_1 }
  {\bar\epsilon(-{\bf k}, p-p_1)\, \bar\epsilon({\bf k}, p_1)} 
  \int \frac{d^\nu p_2}{(2\pi\hbar)^\nu} \, 
  \frac{f({\bf p}_2)}{p - p_1 - i {\bf k} \cdot {\bf v}_2}
  }_{{\rm (e)}} ~+
\nonumber\\[10pt] &&
  \bigg[\underbrace{
  \frac{1}{\bar\epsilon(-{\bf k}, p-p_1)}}_{{\rm (a)}} ~-~ \underbrace{~1~}_{{\rm (b)}} \bigg] %{\scriptstyle\times}
  \bigg[
  \underbrace{f({\bf p}_1)}_{{\rm (c)}}
  ~+~
  \underbrace{
  \frac{\tilde\phi({\bf k}) \, (i {\bf k}) \cdot \partial f/
  \partial {\bf p}_1}{\bar\epsilon({\bf k}, p_1)} \,
  \int \frac{d^\nu p_2}{(2\pi\hbar)^\nu} \,  \frac{ f({\bf p}_2)}{p_1 + i {\bf k} \cdot {\bf v}_2}
  }_{{\rm (d)}}
  \bigg]
  ~\Bigg\}
    \ ,
  \nonumber\\
\label{eq_I_final}
\end{eqnarray}
where I have labeled the terms as in Ref.~\cite{cd}. From (\ref{eq_J_alg_real}),
the Lenard-Balescu current can be expressed as
\begin{eqnarray}
  {\bf J}({\bf p}_1)
    &=&
     -\int \frac{d^\nu k}{(2\pi)^\nu}\,  {\bf k} \tilde \phi({\bf k}) 
     \bigg[
    {\rm Im}\,I_\text{[a+b]$\times$c} +   {\rm Im}\,I_{\rm b \times d} +    
    {\rm Im}\,I_\text{[a$\times$d]+e}
    \bigg]
  \ ,
 \label{eq_J_alg_real_Ib}
\end{eqnarray}
where $I_\text{[a+b]$\times$c}$ is the result of the (a)+(b) term times the 
(c) term in (\ref{eq_I_final}), $I_{\rm b \times d}$ is (b) times (d), and 
$I_\text{[a$\times$d]+e}$ is (a) times (d) plus term (e). 
\begin{figure}[t!]
\includegraphics[scale=0.50]{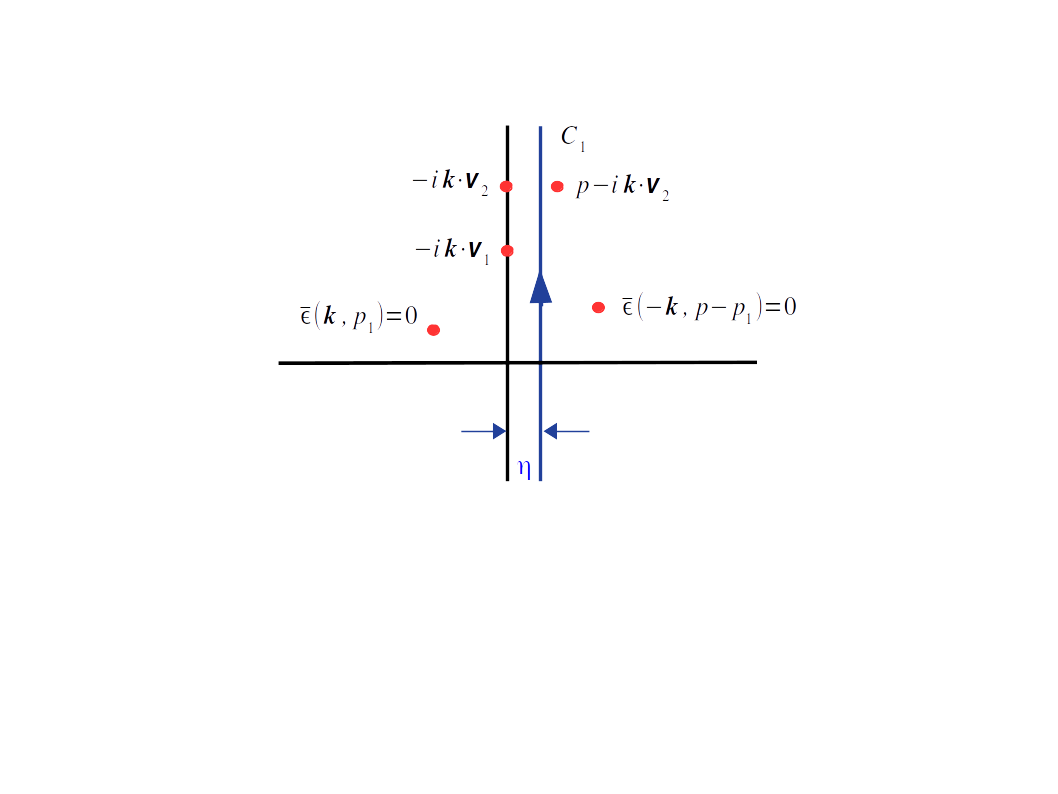} 
\vskip-4.0cm 
\caption{\footnoteskip  
  The Laplace contour $C_1$ and associated poles  
  $p_1=-i{\bf k} \cdot {\bf v}_1$, $p_1=-i{\bf k} \cdot {\bf v}_2$,
  and $p_1=p - i{\bf k} \cdot {\bf v}_2$, along with the
  singularities arising from the zeros of $\bar\epsilon({\bf k}, p_1)$
  and $\bar\epsilon(-{\bf k}, p - p_1)$. Not every pole or 
  singularity in the Figure is associated with every term
  in (\ref{eq_I_final}). 
  Since ${\rm Re}\,(p - p_1) > 0$, for real $p > 0$ we must 
  have $0 < \eta < p$.  
  This allows us take the limit $\eta \to 0^+$ before taking 
  $p \to 0^+$ (the order of limits cannot be reversed). 
}
\label{fig_C1_contour}
\end{figure}

\begin{figure}[t!]
\hskip-8.0cm
{
\begin{minipage}[c]{0.4\linewidth}
\includegraphics[scale=0.50]{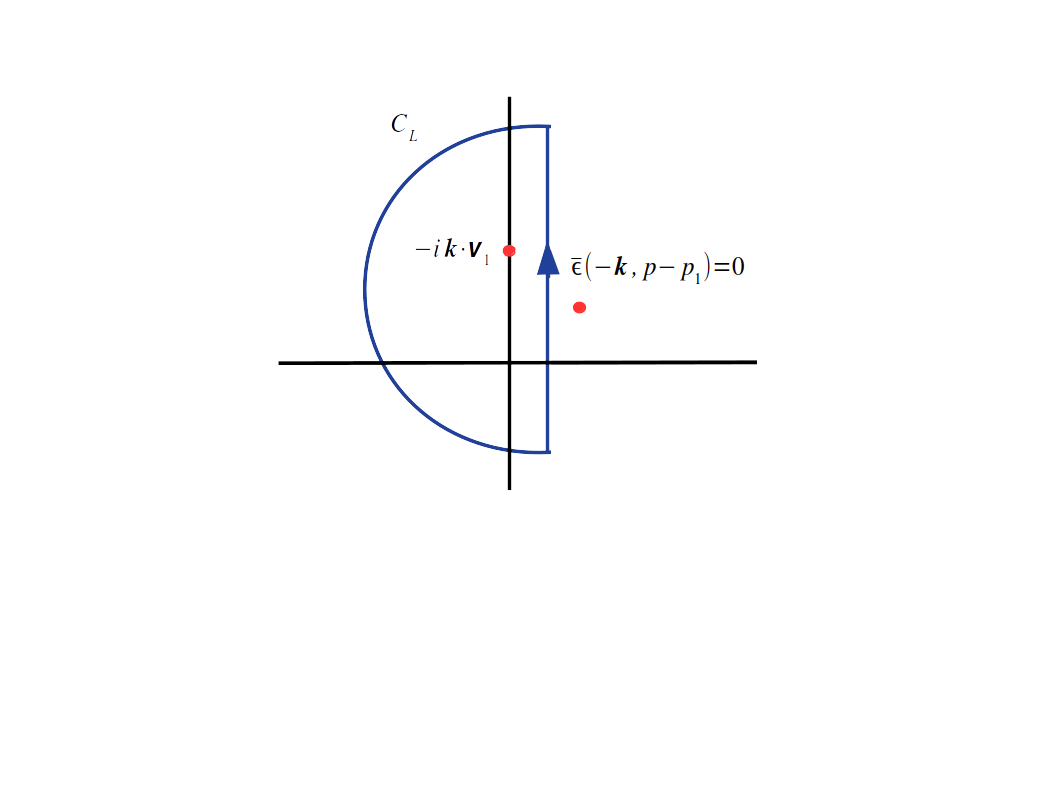} 
%\caption{Image A}
\end{minipage}
\hskip1.0cm
\begin{minipage}[c]{0.4\linewidth}
\includegraphics[scale=0.50]{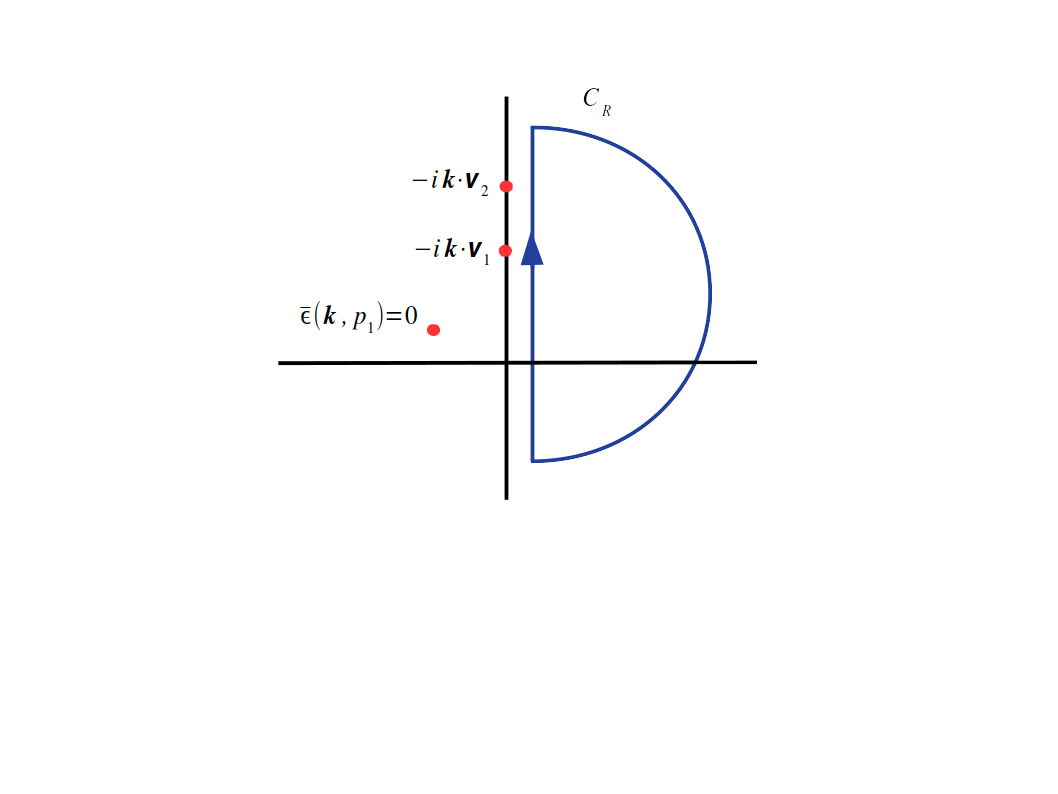} 
%\caption{Image B}
\end{minipage}
}
\vskip-3.5cm 
\caption{\footnoteskip  
  Closed contours $C_\smL$ and $C_\smR$ for the integrals
  $I_\text{[a+b]$\times$c}$ of Eq.~(\ref{eq_apbc}) and 
  $I_\text{b$\times$d}$ of Eq.~(\ref{eq_bd}), respectively. 
}
\label{fig_CLR}
\end{figure}

The contour $C_1$ lies parallel to the imaginary axis in complex $p_1$-plane 
such that the  singularities from $(p_1+V_1)^{-1}$ lie to the left of $C_1$. 
There are simple poles at $p_1 = -i {\bf k} \cdot{\bf v}_1$, $p_1 = -i {\bf k} 
\cdot {\bf v}_2$, $p_1 = p - i{\bf k} \cdot {\bf v}_2$. The latter pole, however,
is not associated with $(p + V_1)^{-1}$, as it is arose from the term
$1/(p - p_1 - p_2)$ in (\ref{eq_intph_complete}). There are also singularities 
arising from the zeros of $\bar\epsilon({\bf  k}, p_1)$  (the singularities and 
corresponding branch cut lie in the left half-plane for plasma stability) and 
the zeros of $\bar\epsilon(-{\bf  k}, p-p_1)$ (whose singularities and branch 
cut lie in the right half-plane). As shown in Fig.~\ref{fig_C1_contour}, we 
can offset $C_1$ by a small amount $\eta$ in the real direction, with 
$0 < \eta < p$. We must therefore take the $\eta \to 0^+$ limit before 
taking $p \to 0^+$.  Not every term in (\ref{eq_J_alg_real_Ib}) will involve
every singularity, so care must be taken when evaluating  (\ref{eq_I_final}). 
Reference~\cite{cd} emphasizes that the guiding principle in closing the 
contour $C_1$ is to avoid enclosing a singularity arising from the zeros 
of the dielectric function $\bar \epsilon$.  In this way, we avoid crossing
a branch-cut when closing the contour at infinity. The first term we
consider is
\begin{eqnarray}
  I_\text{[a+b]$\times$c}
  &=& 
  \frac{1}{2\pi i} \lim_{p \to 0^+} \int_{C_1} dp_1
  \frac{1}{p_1 + i {\bf k} \cdot {\bf v}_1} \, 
  \bigg[\frac{1}{\bar\epsilon(-{\bf k}, p-p_1)} - 1 \bigg]
  f({\bf p}_1) 
   \ .
\label{eq_apbc}
\end{eqnarray}
There is a simple pole at $p_1=-i{\bf k} \cdot {\bf v}_1$, so we 
complete the $C_1$ contour by a large semi-circle in the left-hand 
$p_1$-plane to form a closed contour $C_{\rm L}$, as illustrated
in the left panel of Fig.~\ref{fig_CLR}. 
This closed 
contour has a counter-clockwise orientation and encircles the 
pole $p_1 = -i {\bf k} \cdot {\bf v}_1$. The path $C_{\smL}$ avoids 
the singularity in the right half-plan arising from the zeros of the 
dielectric function $\bar\epsilon(-{\bf k}, p-p_1)$, and there are 
no such singularities in the left-half plane from $\bar\epsilon
({\bf k}, p_1)$, so upon applying the residue theorem we find
\begin{eqnarray}
  I_\text{[a+b]$\times$c}
    &=&
  \frac{1}{2\pi i} \int_{C_{\rm L}} dp_1
  \frac{1}{p_1 + i {\bf k} \cdot {\bf v}_1} \, 
  \bigg[\frac{1}{\bar\epsilon(-{\bf k}, -p_1)} - 1 \bigg] 
  f({\bf p}_1) 
  \\[5pt]
  &&
  \bigg[\frac{1}{\bar\epsilon(-{\bf k}, i {\bf k} \cdot {\bf v}_1)} - 1 \bigg] 
  f({\bf p}_1) 
  =
  \bigg[\frac{1}{\epsilon(-{\bf k}, -{\bf k} \cdot {\bf v}_1)} - 1 \bigg] f({\bf p}_1) 
  \ .
\end{eqnarray}
In the last equality we have used (\ref{eq_epsilon_minus}) to express the
result in terms of the ordinary dielectric function $\epsilon({\bf k}, \omega)$ 
in temporal Fourier space. We only need the imaginary component,
\begin{eqnarray}
  {\rm  Im}\,I_\text{[a+b]c}
    &=&
   \frac{{\rm Im }\,\epsilon({\bf k}, {\bf k} \cdot {\bf v}_1)}
   {\vert \epsilon({\bf k}, {\bf k} \cdot {\bf v}_1) \vert^2} \,
  f({\bf p}_1) 
  \ ,
\end{eqnarray}
where we have used the fact that $\epsilon(-{\bf k}, -\omega) = 
\epsilon^*({\bf k}, \omega)$. Recall that the sing-component
dielectric function takes the form
\begin{eqnarray}
  \epsilon({\bf k}, \omega)
  =
  1 + \int \frac{d^\nu p_2}{(2\pi\hbar)^\nu} \, \frac{\tilde\phi({\bf k})\, {\bf  k} \cdot 
  \partial f/\partial {\bf p}_2 }{\omega - {\bf k} \cdot {\bf v}_2 + i\eta} 
  \ .
\end{eqnarray}
\begin{eqnarray}
    I({\bf k}, {\bf p}_1) 
    =
    \frac{1}{2\pi i}\lim_{p \to 0^+}\int \frac{d^\nu p_2}{(2\pi\hbar)^\nu} \,  
    \int_{C_1} dp_1 \,
     \frac{ 1}{\bar\epsilon(-{\bf k}, p-p_1)} \, \frac{1}{p - p_1 - i{\bf k}\cdot{\bf v}_2}\, 
    (p_1 + V_1)^{-1} S({\bf k}, {\bf p}_1, {\bf p}_2)
    \ .
    \nonumber\\
%\label{eq_intoneh}
\end{eqnarray}
To find the imaginary part, we use the functional relation
\begin{eqnarray}
  \frac{1}{\omega - {\bf k} \cdot {\bf v}_2 + i \eta} 
  = 
  {\rm P}\, \frac{1}{\omega - {\bf k} \cdot {\bf v}_2}  
  - 
  i \pi \,\delta(\omega - {\bf k} \cdot {\bf v}_2)
  \ ,
\label{eq_omega_minus_omegaZero}
\end{eqnarray}
where the first term in (\ref{eq_omega_minus_omegaZero}) gives
the principle-part integral. We can therefore express
\begin{eqnarray}
  {\rm Im}\,\epsilon({\bf k}, \omega)
  =
  - \pi \int \frac{d^\nu p_2}{(2\pi\hbar)^\nu} \, 
  \delta(\omega - {\bf k} \cdot {\bf v}_2)\,
  \tilde\phi({\bf k})\, {\bf k} \cdot \frac{\partial f}{\partial {\bf p}_2}\,
  \ ,
\end{eqnarray}
and hence
\begin{eqnarray}
  {\rm  Im}\,I_\text{[a+b]$\times$c}
    &=&
    - \pi \int \frac{d^\nu p_2}{(2\pi\hbar)^\nu} \,  
    \frac{\tilde\phi({\bf k}) \,\delta({\bf k} \cdot {\bf v}_1 - {\bf k} \cdot {\bf v}_2)}
   {\vert \epsilon({\bf k}, {\bf k} \cdot {\bf v}_1) \vert^2} \,
   \, {\bf k} \cdot \frac{\partial f({\bf p}_2)}{\partial {\bf p}_2} \,   f({\bf p}_1) 
  \ .
\label{eq_ImIabc}
\end{eqnarray}
This gives the corresponding contribution to the Lenard-Balescu
current
\begin{eqnarray}
  {\bf J}_\text{[a+b]$\times$c}
  &=&
  -  \int \frac{d^\nu k}{(2\pi)^\nu} \, {\bf k} \, \tilde\phi({\bf k}) \,
  {\rm Im}\, I_\text{[a+b]c}({\bf k}, {\bf p}_1, {\bf p}_2)
  \\[5pt]
  &=&
   \pi \int \frac{d^\nu k}{(2\pi)^\nu}  \, \frac{d^\nu p_2}{(2\pi\hbar)^\nu} \,  
   \frac{[\tilde\phi({\bf k})]^2 {\bf k} \,\delta({\bf k} \cdot {\bf v}_1 - {\bf k} \cdot {\bf v}_2)}
   {\vert \epsilon({\bf k}, {\bf k} \cdot {\bf v}_1) \vert^2} \,
   \, {\bf k} \cdot \frac{\partial f({\bf p}_2)}{\partial {\bf p}_2} \,   f({\bf p}_1) 
   \ .
\label{eq_Jabc}
\end{eqnarray}

We now evaluate ${\rm b \times d}$ contribution. Interestingly, this 
term vanishes upon closing the contour $C_1$ to the right to form 
a closed contour $C_{\rm R}$ lying  in the right half-plane, 
\begin{eqnarray}
 %&&
% \int_{C_1} dp_1
%  \frac{1}{p_1 + i {\bf k} \cdot {\bf v}_1} \, 
%  \frac{ 1}{p_1 + i {\bf k} \cdot {\bf v}_2} \,
%  \frac{1}{\bar\epsilon({\bf k}, p_1)} \,
%  %\nonumber\\[8pt]
%  %&&=
%  =
I_\text{b$\times$d}
\propto
 \int_{C_\smR} dp_1
  \frac{1}{p_1 + i {\bf k} \cdot {\bf v}_1} \, 
  \frac{ 1}{p_1 + i {\bf k} \cdot {\bf v}_2} \,
  \frac{1}{\bar\epsilon({\bf k}, p_1)} \,
  = 0
  \ .
\label{eq_bd}
\end{eqnarray}
This is because the contour $C_\smR$ does not enclose the simple 
poles $p_1 = -i {\bf k} \cdot {\bf v}_1$ and \hbox{$p_1 = - i{\bf k} \cdot {\bf v}_2$}
on the imaginary axis, nor the zeros of $\bar\epsilon({\bf k}, p_1)$ 
in the left half-plane. The residue theorem therefore gives a vanishing 
integral. The final term involves $[{\rm a \times d}]+{\rm e}$. We can 
in fact take the limit $p \to 0^+$ inside $\bar\epsilon(-{\bf k}, p-p_1)$
to give $\bar\epsilon(-{\bf k}, -p_1)\, \bar\epsilon({\bf k}, p_1) 
= \vert \bar\epsilon({\bf k}, p_1)\vert^2$, and therefore
\begin{eqnarray}
   I_\text{[a$\times$d]+e}
  &=&
  \frac{1}{2\pi i} 
  \lim_{p \to 0^+}
  \int \frac{d^\nu p_2}{(2\pi\hbar)^\nu} \, f({\bf p}_2)\,
  \tilde\phi({\bf k}) \, i {\bf k} \cdot \frac{\partial f}{\partial {\bf p}_1}\,
  \int_{C_1} dp_1 \,
  \frac{1}{p_1 + i {\bf k} \cdot {\bf v}_1} \, 
  {\scriptstyle \times}
  \nonumber\\[5pt] && \hskip3.0cm
  \frac{1}{\vert\bar\epsilon({\bf k}, p_1) \vert^2} \,
  \bigg[ \frac{1}{p_1 + i {\bf k} \cdot {\bf v}_2} 
  +
  \frac{1}{p - p_1 - i {\bf k} \cdot {\bf v}_2 - p} \bigg]
\label{eq_Iade_def}
\\[11pt]
  &=&
  \lim_{p \to 0^+}
  \int \frac{d^\nu p_2}{(2\pi\hbar)^\nu} \,
  \tilde\phi({\bf k}) \, {\bf k} \cdot \frac{\partial f}{\partial {\bf p}_1} \,
  f({\bf p}_2)\,
  I_\smC({\bf k}, {\bf p}_1, {\bf p}_2)
  \ ,
\label{eq_Iade_Ic}
\end{eqnarray}
where we define the contour integral
\begin{eqnarray}
  I_{\rm\scriptscriptstyle C}
  &=&
  \frac{1}{2\pi}
  \int_{C_1} dp_1 \,
  \frac{1}{\vert \bar\epsilon({\bf k},p_1)\vert^2}\,
  \frac{1}{p_1 + i {\bf k}\cdot {\bf v}_1}\,
  \left[\frac{1}{p_1 + i {\bf k} \cdot {\bf v}_2} 
  -
  \frac{1}{p_1  + i {\bf k} \cdot {\bf v}_2 - p}
  \right]
 \  .
  \label{eq_I_C}
\end{eqnarray}
All factors of $i$ from and all signs in (\ref{eq_Iade_def}) have been 
placed in the contour integral $I_\smC$ of (\ref{eq_I_C}). We can 
parameterize points $p_1 \in C_1$ by $p_1 = -i \omega + \eta$ for 
arbitrary real $\omega$ and fixed real $\eta$, with $0 < \eta < p$.
We must therefore take the $\eta \to 0^+$ limit before the $p \to 0^+$ 
limit. The contour integral over $C_1$ in (\ref{eq_Iade_def}) can now be 
expressed an integral over real $\omega$, 
\begin{eqnarray}
  I_{\rm\scriptscriptstyle C}
  & =&
  -\frac{i}{2\pi}
  \int_{-\infty }^\infty  d\omega \,
  \frac{1}{\vert\epsilon({\bf k},\omega)\vert^2}\,
  \frac{1}{\omega - {\bf k}\cdot {\bf v}_1 + i \eta}\,
  \left[
  \frac{1}{\omega - {\bf k} \cdot {\bf v}_2 + i \eta} 
  -
  \frac{1}{\omega  - {\bf k} \cdot {\bf v}_2 - i p}
  \right]
   \ ,
   \nonumber\\[-5pt]
  \label{eq_I_C_omega}
\end{eqnarray}
where we taken the $\eta \to 0^+$ limit inside the dielectric function, 
and then used (\ref{eq_epsilon_minus}) to set $\bar\epsilon({\bf k}, 
-i\omega)=\epsilon({\bf k}, \omega)$. Furthermore, since the $p$-limit 
is taken at the end of the calculation, in performing the $\eta$-limit we 
have $p \ne 0$. Therefore, in the last term of (\ref{eq_I_C_omega}), we 
have taken $\eta \to 0^+$, leaving the small imaginary piece $-i p$ 
in the denominator. Since $\eta$ and $p$ are both infinitesimal,
we have
\begin{eqnarray}
  \frac{1}{\omega - {\bf k} \cdot {\bf v}_2 + i \eta} 
  -
  \frac{1}{\omega  - {\bf k} \cdot {\bf v}_2 - i p}
  = 
  -2\pi i \,\delta(\omega - {\bf k}\cdot {\bf v}_2)
  \ ,
\end{eqnarray}
where the principles parts cancel. Similarly, the second term in 
(\ref{eq_I_C_omega}) becomes
\begin{eqnarray}
  \frac{1}{\omega - {\bf k}\cdot {\bf v}_1 + i \eta}
  =
  {\rm P}\, \frac{1}{\omega - {\bf k}\cdot {\bf v}_1}
  -
  \pi i \,\delta( \omega - {\bf k}\cdot {\bf v}_1)
  \ .
\end{eqnarray}
The principle part is real and does not contribute to the imaginary piece 
of $I_\text{[a$\times$d]+e}$ (it also integrates to zero when performing 
the ${\bf k}$-integration), and therefore
\begin{eqnarray}
  I_{\rm\scriptscriptstyle C}
  &=&
  \pi i \int_{-\infty }^\infty  d\omega \,
  \frac{1}{\vert\epsilon({\bf k},\omega)\vert^2} \,
  \delta(\omega - {\bf k}\cdot {\bf v}_1)
  \delta(\omega - {\bf k}\cdot {\bf v}_2)
  =
  \pi i \,
  \frac{ \delta({\bf k}\cdot {\bf v}_1 - {\bf k}\cdot {\bf v}_2)}
  {\vert\epsilon({\bf k},{\bf k} \cdot {\bf v}_1)\vert^2}
  \ .
  \nonumber\\
  \label{eq_I_C_omega_final}
\end{eqnarray}
We therefore arrive at 
\begin{eqnarray}
  {\rm Im}\, I_\text{[a$\times$d]+e}
  &=&
  \pi \int \frac{d^\nu p_2}{(2\pi\hbar)^\nu} \, 
  \tilde\phi({\bf k}) \, {\bf k} \cdot  \frac{\partial f}{\partial {\bf p}_1}\,
  f({\bf p}_2)\,
  \frac{ \delta({\bf k}\cdot {\bf v}_1 - {\bf k}\cdot {\bf v}_2)}
  {\vert\epsilon({\bf k},{\bf k} \cdot {\bf v}_1)\vert^2}
  \ ,
\end{eqnarray}
which gives a contribution to the current
\begin{eqnarray}
  {\bf J}_\text{[a$\times$d]+e}
  &=&
  -  \int \frac{d^\nu k}{(2\pi)^\nu} \, {\bf k} \, \tilde\phi({\bf k}) \,
  {\rm Im}\, I_\text{[a$\times$d]+e}({\bf k}, {\bf p}_1, {\bf p}_2)
  \\[5pt]
  &=&
   -\pi \int \frac{d^\nu k}{(2\pi)^\nu}  \, \frac{d^\nu p_2}{(2\pi\hbar)^\nu} \,  
   \frac{[\tilde\phi({\bf k})]^2 {\bf k} \,\delta({\bf k} \cdot {\bf v}_1 - {\bf k} \cdot {\bf v}_2)}
   {\vert \epsilon({\bf k}, {\bf k} \cdot {\bf v}_1) \vert^2} \,
   \, {\bf k} \cdot \frac{\partial f({\bf p}_1)}{\partial {\bf p}_1} \,   f({\bf p}_2) 
   \ .
\label{eq_Jade}
\end{eqnarray}
Upon adding (\ref{eq_Jabc}) and (\ref{eq_Jade}) we find the
total Lenard-Balescu current,
\begin{eqnarray}
  {\bf J}
  &=&
  \! \pi\!
  \int \frac{d^\nu k}{(2\pi)^\nu}  \, \frac{d^\nu p_2}{(2\pi\hbar)^\nu} 
   \underbrace{~[\tilde\phi({\bf k})]^2 
   }_{(e^2/k^2)^2}\, {\bf k} \,
   \frac{\delta({\bf k} \cdot {\bf v}_2 - {\bf k} \cdot {\bf v}_1)}
   {\vert \epsilon({\bf k}, {\bf k} \cdot {\bf v}_1) \vert^2}
   \bigg[
    {\bf k} \cdot \frac{\partial }{\partial {\bf p}_2}
    -
    {\bf k} \cdot \frac{\partial}{\partial {\bf p}_1} 
    \bigg]  f({\bf p}_1) f({\bf p}_2)
    \ ,
    \nonumber\\[-5pt]
\end{eqnarray}
and the proof is complete for a single-component plasma.

%%
%\pagebreak
\subsection{Generalization to a Multi-species Plasma}

It is easy to generalize the previous result to a multi-species plasma.
The variables $X_i$ and the distribution functions must contain species 
indices, {\em e.g.} $f_1^{(a)}(X_a, t)$ and $  f_2^{(ab)}(X_a, X_b,t)$. The
latter gives the joint probability of finding species $a$ at $X_a$ and 
species $b$ at $X_b$.  The first order correction becomes
\begin{eqnarray}
  f_2^{(ab)}(X_a, X_b,t)
  =
  f_a(X_a, t) f_b(X_b, t) + h_{ab}(X_a, X_b, t)
  \ .
\end{eqnarray}
The Lenard-Balescu kernel  becomes
\begin{eqnarray}
  {\sum}_b\, L_{ab}[f]
  =
  -  {\sum}_b\,\frac{\partial}{\partial{\bf p}_a} \cdot {\bf J }_{ab}[f]({\bf p}_a)
  \ ,
\end{eqnarray}
where
\begin{eqnarray}
  {\bf J}_{ab}[f]({\bf p}_a)
  &=&
  \! \pi\!
  \int \frac{d^\nu k}{(2\pi)^\nu}  \, \frac{d^\nu p_b}{(2\pi\hbar)^\nu} 
   \left(\frac{e_a e_b}{k^2}\right)^2\, {\bf k} \,
   \frac{\delta[{\bf k} \cdot ({\bf v}_b -  {\bf v}_a)]}
   {\vert \epsilon({\bf k}, {\bf k} \cdot {\bf v}_a) \vert^2}
   \bigg[
    {\bf k} \cdot \frac{\partial }{\partial {\bf p}_b}
    -
    {\bf k} \cdot \frac{\partial}{\partial {\bf p}_a} 
    \bigg]  f_a({\bf p}_a) f_b({\bf p}_b)
    \ ,
    \nonumber\\[-3pt]
  \label{eq_Jrs_vone}
\end{eqnarray}
and $\epsilon({\bf k}, \omega)$ is the multi-component dielectric function.

%%%%%%%%%%%%%%%%%%%%%
\pagebreak
\section{Conclusions}
\label{sec_conclusions}

Calculating the rate of Coulomb energy exchange in a plasma is notoriously 
difficult, even for the case of a fully ionized weakly coupled plasma. Two
examples of experimental relevance are the charged particle stopping 
power and the temperature equilibration rate between electrons and 
ions in a non-equilibrium plasma. Naive calculations of these processes 
suffer from logarithmic divergences at both long- and short-distance 
scales, and we must therefore resort to more sophisticated methods 
of calculation. Corresponding to these divergences are two broad classes 
of kinetic equations, applicable in complementary regimes, represented 
by the Boltzmann equation (BE) and the Lenard-Balescu equation (LBE). 
The BE describes the short-distance effects of 2-body scattering, including
large angle scattering, while 
the LBE models 2-point long-distance correlations.  It is well known that 
the BE suffers a long-distance logarithmic divergence for Coulomb
scattering (in three spatial dimensions), confirming that it is indeed 
missing long-distance physics (correlations are being ignored). Conversely, 
the LBE suffers from a short-distance logarithmic divergence for Coulomb
interactions (in three dimensions), another indication that relevant physics 
is being overlooked (the short-distance scattering physics).  

The fact
that the BE and the LBE are relevant in complementary regimes allows 
us to capitalize on the lessons  physicists have learned from quantum 
field theory, a formalism developed by particle physicists for understanding
the fundamental interactions of nature. In quantum field theory, an array 
of divergences are encountered,  from logarithmic to quadratic and higher, 
and the so called  {\rm renormalization program} was developed to form 
meaningful and finite predictions from these divergent results. The first 
ingredient is to temporarily {\em regularize} the theory by rendering the 
integrals finite. At the end of the calculation, the regularization will be
removed, but in the interim, the finite expressions can be algebraically 
manipulated in a meaningful fashion. After regularization has been 
performed, one then  {\em renormalizes} the theory by reinterpreting 
physical properties like the electric charge and mass in such a way as to 
give finite predictions as the regularization scheme is removed. There 
are many regularization schemes in use, each with their own strengths 
and weaknesses. The simplest  one is to choose arbitrary 
large- and small-distance cutoffs in the integrals, such as the Debye 
wavelength and the classical distance of closest approach for the electrons 
or ions in the plasma. This is the regularization scheme first adopted by 
Landau and Spitzer, and it produces a scaling factor called the {\em Coulomb 
logarithm}, which is defined to be the natural 
logarithm of the ratio of the large- to small-distance scales. Much effort has 
been devoted to determining the precise value of the Coulomb logarithm.
However, such a crude regularization method is inherently uncertain in 
determining the exact value of the Coulomb logarithm, and this exercise 
is doomed to failure from the start. For example, one could just as correctly 
take twice the Debye length as the long distance cutoff, thereby leaving 
the constant inside the logarithm undetermined by this regularization 
method. It is interesting to note that if the divergence had been higher 
order rather than logarithmic, this crude cutoff method would not have 
been acceptable to plasma physicists. It is only because the divergence 
in Coulomb exchange processes is logarithmic that one  can get away 
with such a naive regularization scheme for so long. 

The most pertinent feature of relativistic quantum field theory is that 
it is a many-body theory. The non-relativistic limit of these theories
provides a rigorous treatment of plasma physics from which the 
framework of a non-relativistic many-body field theory\,\cite{lowell,by}.
One of the subtleties of the renormalization program is that the regularization 
scheme often breaks the symmetries of the system, thereby changing the
structure of the theory. For example, the cut-off method breaks Lorentz 
invariance, which is essential for electrodynamics. While the symmetries 
must return when the regularization is removed, the system becomes more 
complex and when its symmetries are broken, and the restoration of the
symmetries as the regularization is removed can often be nontrivial. Indeed,
some symmetries remain broken. For example, gauge symmetry and
particle number conservation cannot both be preserved in the standard 
model of particle physics. As gauge invariance is essential for defining 
the theory, it turns out that matter is not stable and decays by so-called 
nonperturbative {\em sphaleron} processes (albeit the proton is quite
long lived, with a decay rate  many times the age of the universe).
Since a plasma is a many-body 
system, it is not surprising that we encounter divergences similar to those 
in quantum field theory. Furthermore, since the renormalization program 
makes experimental predictions, we must take it seriously, and it is not
surprising that techniques developed in field theory are applicable to
plasma theory.  
Regularization methods are often chosen in such a way as to preserve 
as many symmetries as possible. The method of dimensional regularization 
is one such method that stands apart from most others in that it preserves 
the essential symmetries, such as Lorentz invariance and gauge invariance. 
The dimensional continuation formalism of Brown-Preston-Singleton (BPS)
relies on a technique adopted from quantum field theory called dimensional
regularization. Processes that are divergent in $\nu=3$ spatial dimensions
can often be regularized by looking at the system in a general number of
dimensions $\nu$. The three-dimensional divergences show up as simple
poles of the form $1/(\nu-3)$. Dimensional regularization soon led to the
insight that the physics of a system is critically dependent upon the 
dimension $\nu$ space. The general rule is that long distance fluctuations 
are greater in lower dimensions, while short distance physics is more 
important in higher dimensions. In fact, in $\nu=1$,  it has been shown 
that the quantum fluctuations are so large that spontaneous symmetry 
breaking cannot occur, even if it is permitted classically\,\cite{sydney}. 
Another interesting result is that in $\nu=1$ dimensions , the photon
acquires a mass via quantum loop corrections\,\cite{peskin}. Other 
phenomena are unique to $\nu=2$ dimensions, such as high
temperature superconductivity. BPS does not require the introduction 
of a Coulomb logarithm, as the regularization is performed by changing 
the dimension $\nu$ of space. The BPS method uses dimensional 
continuation to find the Coulomb energy exchange at the integer 
values $\nu=1,2,4, 5, \cdots$ (except for $\nu=3$). By applying 
Carlson's theorem\,\cite{Carlsonth}, we can define an analytically 
continued quantity 
for complex $\nu$, in a similar way that the factorial function on the 
positive integers can be analytically continued to the gamma function 
over the complex plane. For Coulomb energy exchange processes, 
the continuation to complex $\nu$ allows us to take the $\nu \to 3$
limit to obtain a {\em finite} result valid in $\nu=3$ dimensions. 
In this way, the BPS formalism regularizes the traditional
$\nu=3$ divergences, and allows us to define the theory in three 
dimensions. In this third 
installment of the BPS Explained lecture series, we have proven a 
pivotal result of process of dimensional continuation upon which the 
BPS formalism resides. Namely, that to leading order in the plasma 
coupling $g$, the BBGKY hierarchy of kinetic equations reduces to 
(i) the Boltzmann equation for spatial dimensions $\nu >3$, and (ii) 
the Lenard-Balescu equation for $\nu < 3$. The bulk of these notes 
were devoted to proving the latter. 

%%%%%%
%
\begin{acknowledgments}
  I would like to thank Jean-Etienne Sauvestre, Lowell Brown, and 
  Don Shirk for reading the manuscript and providing critical feed-back. 
  I would also like to thank Lowell Brown for a number of very useful 
  technical discussions. 
\end{acknowledgments}

%%%%%%%
\pagebreak
\appendix

\section{The Cross Section and Hyperspherical Coordinates}
\label{sec_cross}

To make these notes self contained, and to establish some notation, 
I shall give a quick review of the material required from Lectures I and II. 

\subsection{Hyperspherical Coordinates}
\label{sec:hypcoord} 

Kinematic quantities such as $\nu$-dimensional momentum or position 
vectors are elements of Euclidean space $\mathbb{R}^\nu$. We can decompose 
any element ${\bf x} \in \mathbb{R}^\nu$ in terms of a rectilinear orthonormal 
basis $\hat{\bm e}_\ell$, so that ${\bf x}=\sum_{\ell=1}^\nu x_\ell\,\hat{\bm e}_\ell$, 
or in component notation ${\bf x} = (x_1,\cdots,x_\nu)$. Each component 
is given by $x_\ell = \hat{\bm e}_\ell \cdot {\bf x}$, and a change $d{\bf x}$ 
in the vector ${\bf x}$ corresponds to a change $dx_\ell = \hat{\bm e}_\ell \cdot
d{\bf x}$ in the rectilinear coordinate $x_\ell$. Letting ${\bf x}$ vary successively 
along each independent direction $\hat{\bm e}_\ell$, we can trace out a small
$\nu$-dimensional hypercube with sides of length $dx_\ell$; therefore, the 
rectilinear volume element is given by the simple form
\begin{eqnarray}
  d^\nu x 
  = 
  \prod_{\ell=1}^\nu dx_\ell
  = 
  dx_1\, dx_2 \cdots dx_\nu \ .
\label{dnu:rec}
\end{eqnarray}
Similar considerations hold for momentum volume element $d^\nu p$. 
In performing integrals over the kinematic variables, however,
symmetry usually dictates the use of hyperspherical coordinates rather
than rectilinear coordinates. I will therefore review the
hyperspherical coordinate system in this section, deriving the
measure for a \hbox{$\nu$-dimensional} volume element $d^\nu x$ in
terms of hyperspherical coordinates.  It should be emphasized again
that this formalism also holds in momentum space for the momentum
volume element. For our purposes, the primary
utility of hyperspherical coordinates is that the volume element
$d^\nu x$ can be written as a product of certain conveniently chosen
dimensionless angles, which I will collectively refer to as
$d\Omega_{\nu-1}$, and an overall dimensionfull radial factor
$r^{\nu-1}\, dr$, so that
\begin{eqnarray}
  d^\nu x 
  =
  d\Omega_{\nu-1}\, r^{\nu-1}dr
  \ .
\label{Omegafactor}
\end{eqnarray}

\begin{figure}[t!]
\includegraphics[scale=0.45]{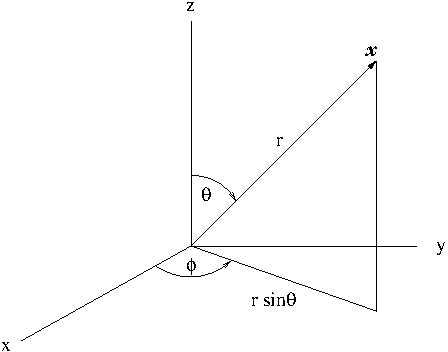}
%\vskip-0.1cm 
\caption{\captionskip
  Spherical coordinates $r,\theta,\phi$ of a point ${\bf x}$ in three
  dimensional space: radial distance $r$, polar angle $\theta$, and
  azimuthal angle $\phi$. The angles range over the values $0 \le
  \theta \le \pi$ and $0 \le \phi < 2\pi$.
}
\label{fig:coord1}
\end{figure}

\vskip0.2cm
\noindent
To prove this, let us recall why the three dimensional volume element 
takes the form $d^3 x = \sin\theta\, d\theta\, d\phi\, r^2 dr$
(with \hbox{$0 \le \theta \le \pi$},  \hbox{$0 \le \phi < 2\pi$}, 
and  $0 \le r < \infty$). As depicted in Fig.~\ref{fig:coord1},
the three dimensional vector ${\bf x}$ has length $r$, and subtends 
a polar angle $\theta$ relative to the \hbox{$z$-axis}, while its
projection onto the \hbox{$x$-$y$ plane} subtends an azimuthal angle
$\phi$ relative to the $x$-axis.  The two angles $\theta$ and $\phi$
specify completely the direction of the unit vector $\hat{\bf x}$,
while an additional coordinate $r$ determines the total vector 
${\bf x}= r \hat{\bf x}$. As we increase the polar angle $\theta$
by a small 
amount $d\theta$, the vector ${\bf x}$ sweeps out an arc of length 
$dR_1 = r d\theta$. Similarly, a change $d\phi$ in the azimuthal 
angle will cause ${\bf x}$ to sweep out an arc in the $x$-$y$ plane 
of length $dR_2 = r \sin\theta\,d\phi$, where the factor of $\sin\theta$ 
in $dR_2$ arises from the projection of ${\bf x}$ onto the $x$-$y$
plane. Moving along the radial direction gives the final independent
displacement $dR_3=dr$. For small displacements in $d\theta$, 
$d\phi$, and $dr$, the vector ${\bf x}$ sweeps out a small cubic 
volume element with sides of length $dR_1$, $dR_2$, and $dR_3$,  
and therefore  $d^3 x= d R_1\, d R_2\, dR_3 = r d\theta \cdot
r\sin\theta d\phi \cdot dr$.

\begin{figure}[t]
\includegraphics[scale=0.6]{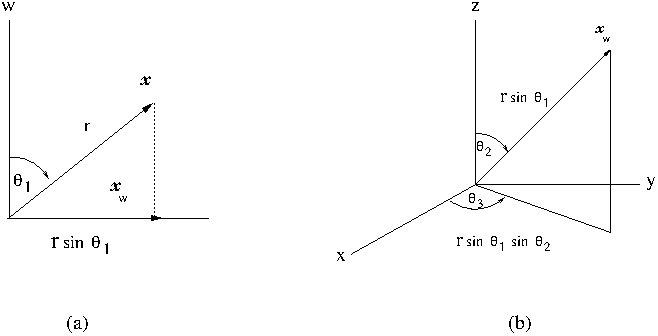}
\vskip-0.3cm 
\caption{\captionskip
  Hyperspherical coordinates $r,\theta_1,\theta_2,\theta_3$ of a point
  ${\bf x}$ in four dimensional space. As before, $r=\vert {\bf x}
  \vert$ is the radial distance. The angles are defined as follows.
  (a) First, let $\theta_1$ be the angle between ${\bf x}$ and the
  $w$-axis. Let us now project ${\bf x}$ onto the orthogonal three
  dimensional space, so that ${\bf x}=(w,x,y,z) \to {\bf x}_w =
  (0,x,y,z)$. The length of the projection ${\bf x}_w$ is $r_w = r \sin
  \theta_1$. (b) The vector ${\bf x}_w$ can be viewed as a three 
  dimensional vector ${\bf x}_w=(x,y,z)$, which then defines the 
  usual polar and azimuthal angles of Fig.~\ref{fig:coord1}, denoted 
  here by $\theta_2$ and $\theta_3$ respectively.
}
\label{fig:coord3}
\end{figure}

Let us now consider the volume element $d^4x$ in four dimensional
space, and denote the coordinate axes by $x, y, z, w$. Since we cannot 
visualize four dimensional space, let us examine this problem in two steps, 
each of which can be visualized in three dimensions.  As shown in 
Fig.~\hbox{\ref{fig:coord3}a}, let $\theta_1$ be the angle between the 
$w$-axis and the four dimensional vector ${\bf x}$. The $w$-axis and 
the vector ${\bf x}$ lie in a plane, and $\theta_1$ can therefore be visualized.
Let us now project ${\bf x}$ onto the $w=0$ hyperplane (a three dimensional 
slice of four-space), denoting the projected three-vector by ${\bf x}_w$. 
Since this vector lies in three-space, it too can be visualized. 
Since the three-plane $w=0$ lies perpendicular to each of the axes $x$, 
$y$, and $z$, the vector ${\bf x}_w$ lies in the three dimensional space 
shown in Fig.~\ref{fig:coord3}b, and its length is $ \vert {\bf x}_w \vert =
r\sin\theta_1$.  Let the angle $\theta_2$ be the polar angle between
the $z$-axis and the vector ${\bf x}_w$, while $\theta_3$ is the usual
azimuthal angle $\phi$. The last angle $\theta_3$ runs between $0$ 
and $2\pi$, while all previous angles run between $0$ and $\pi$. 
As we vary the three angles and the radial
coordinate, we sweep out a \hbox{four-dimensional} cube (or an
approximate cube) with sides of length $dR_1=r \,d\theta_1$, $dR_2=r
\sin\theta_1 d\theta_2$, $dR_3=r \sin\theta_1\sin\theta_2
\,d\theta_3$, and $dR_4=dr$. This gives a four dimensional volume
element
\begin{eqnarray}
  d^4x 
  \equiv 
  dR_1\,dR_2\,dR_3\,dR_4
  = 
  \sin^2\theta_1 d\theta_1~
  \sin\theta_2 d\theta_2 ~d\theta_3 ~
  r^3\,dr \ ,
\end{eqnarray}
where $0 \le \theta_\ell \le \pi$ for $\ell=1,2$ and $0 \le \theta_3 <2\pi$. 
It is amusing to calculate the four-volume of a four-dimensional ball of 
radius $r$ by integrating the volume element over the appropriate angles,
\begin{eqnarray}
  B_4
  =
  \int_0^\pi \!d\theta_1 \sin^2\theta_1\, 
  \int_0^\pi \! d\theta_2\sin\theta_2\, 
  \int_0^{2\pi} \!\! d\theta_3
  \,\int_0^r \! dr^\prime\,  r^{\prime\, 3}
  =
  \frac{1}{2}\, \pi^2 r^4 \ .
\end{eqnarray}
The derivative of $B_4$ with respect to $r$ gives the hypersurface
area of the enclosing \hbox{three-sphere},
\begin{eqnarray}
  S_3
  =
  \frac{dB_4}{dr}
  =
  2\pi^2 r^3 \ .
\end{eqnarray}
These are well known results, analogous to a three dimensional ball 
of radius $r$ and volume $B_3=4 \pi\, r^3/3$, which is of course bounded 
by the two-sphere of area $S_2=4\pi r^2$.

We can readily generalize this procedure to an arbitrary number of
dimensions. Consider a point ${\bf x} \in \mathbb{R}^\nu$ given by the
rectilinear coordinates ${\bf x}=(x_1, x_2, \cdots, x_\nu)$. Let
$\theta_1$ be the angle between the vector ${\bf x}$ and the
$x_1$-axis, in a manner similar to that of Figs.~\ref{fig:coord1}
and~\hbox{\ref{fig:coord3}a}. Note that $dR_1=r d\theta_1$ is the arc
length swept out by ${\bf x}$ as the angle $\theta_1$ is incremented
by $d\theta_1$.  Let us now project ${\bf x}$ onto the hyperplane
$x_1=0$, the $(\nu-1)$-plane normal to the $x_1$-axis and passing
through the origin. Denote this projection by ${\bf x}_1$, that
is to say, let $ {\bf x} \to {\bf x}_1=(0,x_2,\cdots,x_\nu)$, and  note
that the length of ${\bf x}_1$ is $r_1 = r \sin \theta_1$.  We proceed 
to the next step and define $\theta_2$ as the angle between 
the $x_2$-axis and the projection ${\bf x}_1$. Note that as  the angle 
$\theta_2$ is incremented by $d\theta_2$, the vector ${\bf x}_1$ 
sweeps out an arc of length $d R_2 = r_1 d\theta_2= r \sin\theta_1\, 
d\theta_2$. In a similar fashion, project ${\bf x}_1$ onto the 
\hbox{$x_2$-plane}, that is, the plane described by $x_1=0$ and 
$x_2=0$. This projection is given by ${\bf x} \to {\bf x}_2=(0,0,x_3,
\cdots,x_\nu)$, and the length of ${\bf x}_2$ is $r_2=r_1 \sin\theta_2
= r \sin\theta_1\sin\theta_2$.
For the general $\ell^{\rm th}$ iteration, let $\theta_\ell$ be the
angle between the $x_\ell$-axis and ${\bf x}_{\ell-1}$, so that $d
R_\ell = r_{\ell-1}\, d\theta_\ell = r \sin\theta_1\, \sin\theta_2\,
\cdots \sin\theta_{\ell-1}\, d\theta_\ell$, 
where we have used the fact that $r_{\ell-1} = r \sin\theta_1
\sin\theta_2 \cdots \sin\theta_{\ell-1}$. This gives
the \hbox{$\nu$-dimensional} volume element
\begin{eqnarray}
  d^\nu x 
  =
  \prod_{\ell=1}^\nu dR_\ell
  =
  \sin^{\nu-2}\theta_1 d\theta_1 \cdot 
  \sin^{\nu-3}\theta_2 d\theta_2 \,\cdots\,
  \sin\theta_{\nu-2} d\theta_{\nu-2} \cdot d\theta_{\nu-1}
  \cdot  r^{\nu-1} dr \ .
\label{dnux}
\end{eqnarray}
The angles $\theta_1, \cdots, \theta_{\nu-2}$ run from 0 to $\pi$, while 
$\theta_{\nu-1}$ runs from 0 to $2\pi$. For notational convenience, 
I will write the angular measure in (\ref{dnux}) as 
\begin{eqnarray}
 d\Omega_{\nu-1}
  =
  \sin^{\nu-2}\theta_1 d\theta_1 \, 
  \sin^{\nu-3}\theta_2 d\theta_2 \,\cdots\,
  \sin\theta_{\nu-2} d\theta_{\nu-2} \, d\theta_{\nu-1}
  \ , 
\label{dnuOmega}
\end{eqnarray}
so that $d^\nu x  = d\Omega_{\nu-1}\, r^{\nu-1} dr$, which establishes 
(\ref{Omegafactor}). Also note that
\begin{eqnarray}
 d\Omega_{\nu-1}
  =
  d\Omega_{\nu-2} \, \sin^{\nu-2}\theta_1 d\theta_1 
  \ .
\label{dnuOmegaMinus}
\end{eqnarray}
It is easy to show that the integration over all angles gives the total 
solid angle
\begin{eqnarray}
  \Omega_{\nu-1} \equiv \int \!d\Omega_{\nu-1} 
  =
  \frac{2\pi^{\nu/2}}{\Gamma(\nu/2)} \ .
\label{OmegaMinusOne}
\end{eqnarray}
To  prove this, first consider the one-dimensional Gaussian integral
\begin{eqnarray}
  \int_{-\infty}^\infty dx\, e^{-x^2} = \sqrt{\pi} \ .
\end{eqnarray}
If we multiply both sides together $\nu$ times, we find
\begin{eqnarray}
  (\sqrt{\pi}\,)^\nu 
  = 
  \int_{-\infty}^\infty \!\!dx_1\, e^{-x_1^2} 
  \int_{-\infty}^\infty \!\!dx_2\, e^{-x_2^2} 
  \, \cdots
  \int_{-\infty}^\infty \!\!dx_\nu\, e^{-x_\nu^2} 
  =
  \int \! d^\nu x\, e^{-{\bf x}^2} \ ,
\label{gaussnu}
\end{eqnarray}
where the vector ${\bf x}$ in the exponential of the last expression
is the $\nu$-dimensional vector ${\bf x}=(x_1, x_2, \cdots, x_\nu)$,
and ${\bf x}^2 = {\bf x} \cdot {\bf x} =\sum_{\ell=1}^\nu x_\ell^2$\,. 
As in (\ref{Omegafactor}), we can factor the angular
integrals out of the right-hand-side of (\ref{gaussnu}), and the
remaining one-dimensional integral can be converted to a Gamma
function with the change of variables $t=r^2$, 
\begin{eqnarray}
  \pi^{\nu/2}
  = \!\!
  \int \!\! d\Omega_{\nu-1} \cdot \!\!
  \int_0^\infty \!\!\! dr\, r^{\nu-1}\,e^{-r^2} 
  \! = \!\!
  \int \!\! d\Omega_{\nu-1} \cdot \frac{1}{2}\, \Gamma(\nu/2) 
  \ ,
\label{pinuomega}
\end{eqnarray}
and solving for $\int\!d\Omega_{\nu-1}$ gives (\ref{OmegaMinusOne}).  

A few general remarks on calculating physical quantities in the BPS 
program are in order. When we calculate the temperature equilibration 
rate between plasma species or the charged particle stopping power, 
we encounter integrals of the form
\begin{eqnarray}
  I_1(\nu) 
  &\equiv& 
  \int d^\nu x ~ F_1(r) ~\,
  = ~
  \Omega_{\nu-1} \int_0^\infty \! dr\, r^{\nu-1}\, F_1(r)
\label{fonedr}
\\[5pt]
  I_2(\nu) 
  &\equiv& 
  \int \! d^\nu x\, F_2(r,\theta) 
  = ~
  \Omega_{\nu-2} \int_0^\infty \! dr\, r^{\nu-1}
  \int_0^\pi \!d\theta\,\sin^{\nu-2}\theta\,F_2(r,\theta) 
  \ ,
\label{ftwodrdth}
\end{eqnarray}
respectively. 
The exact expressions for $F_1$ and $F_2$ are not important here, 
except that their angular dependence is determined by the following 
considerations. The integral (\ref{fonedr}) is spherically symmetric
because the energy exchange between plasma species is isotropic, 
while in integral (\ref{ftwodrdth}), the motion of the charged particle 
defines a preferred direction around which one must integrate, thereby
leaving a single angular dependence. The integrals $I_1(\nu)$ and 
$I_2(\nu)$ can be viewed as functions defined on the 
integers $\nu \in \mathbb{N}$, and as discussed at length in 
Lecture~I~\cite{bps1}, Carlson's Theorem\,\cite{Carlsonth} ensures 
that there are unique analytic continuations of $I_1(\nu)$ and $I_2(\nu)$ 
for $\nu \in \mathbb{C}$.  This is similar to extending 
the factorial function $n!$ on the integers to the gamma function 
$\Gamma(\nu)$ on the complex \hbox{$\nu$-plane}. Let us examine 
more closely how this analytic 
continuation to complex $\nu$ works in practice. First, the solid angles 
$\Omega_{\nu-1}$ and $\Omega_{\nu-2}$ are well defined for complex
arguments $\nu$, as they are composed of simple exponential factors
like $\pi^{\nu/2}$ and Gamma functions, whose analytic properties
are well known. As for the integrals, simply treat $\nu$ as an arbitrary 
integer dimension, and perform the integral for general $\nu$. The
integral will of course depend upon the value of $\nu$,  and once the
integral has been performed exactly (not approximately and not 
numerically), we are free to set the value of $\nu$ to a complex number
(presumably in a small neighborhood about $\nu=3$). This provides 
functions $I_1(\nu)$ and $I_2(\nu)$ with complex argument \hbox{$\nu
 \in \mathbb{C}$}.

\subsection{The Hypervolume of Spheres, Disks, and Cylinders}
\label{sec:areas} 

\begin{figure}[t]
\includegraphics[scale=0.35]{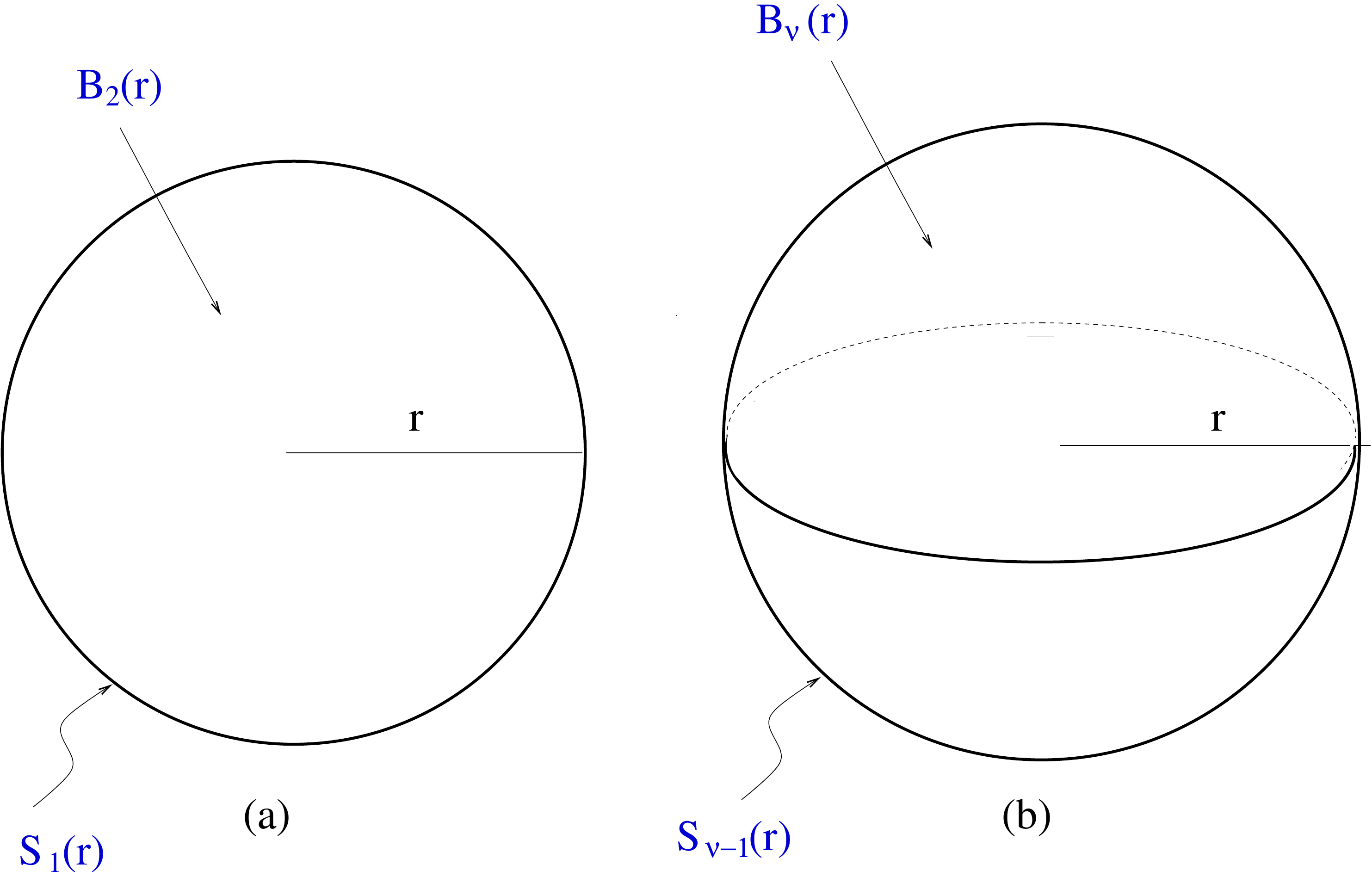}
\vskip-0.3cm 
\caption{\captionskip
  A $(\nu-1)$-dimensional sphere $S_{\nu-1}$ of radius $r$ bounds the
  $\nu$-dimensional ball $B_\nu(r)$ of radius $r$. By integrating over
  successive shells of area, we can find the volume by $B_\nu(r) =
  \int_0^r dr^\prime S_{\nu-1}(r^\prime)$; or conversely
  $S_\nu(r)=B_\nu^\prime(r)$.
}
\label{fig:sphere}
\end{figure}

We shall now calculate the hypervolume of several useful geometric
objects. Let us first consider a \hbox{$\nu$-dimensional} ball of
radius $r$, defined by the set of points \hbox{ ${\bf x} \in
\mathbb{R}^\nu$} for which $\vert {\bf x} \vert \le r$. We will denote
this object by $B_\nu(r)$, and in two and three dimensions this is a
disk and a ball, both volume centered at the origin.  We can find
the $\nu$-dimensional hypervolume of the ball $B_\nu(r)$ by 
simply integrating (\ref{dnux}) over all permissible values of the
coordinates. It should cause no confusion to denote the 
hypervolume of the region $B_\nu(r)$ by the same symbol, 
and using (\ref{OmegaMinusOne}), together with $x \Gamma(x) 
= \Gamma(x + 1)$, we find
\begin{eqnarray}
  B_\nu(r)
  &=&
  \int d\Omega_{\nu-1}
  \int_0^r dr^\prime \, r^{\prime\,\nu-1} 
  =
  \frac{\pi^{\nu/2}}{\Gamma(\nu/2+1)}\,r^\nu \ .
\label{Bnu}
\end{eqnarray}
The boundary of $B_\nu(r)$ is a $(\nu\!-\!1)$-dimensional sphere
$S_{\nu-1}(r)$ defined by $\vert {\bf x} \vert = r$, or
\hbox{$\sum_{\ell=1}^\nu x_\ell^2 = r^2$}.  By differentiating (\ref{Bnu})
with respect to the radius $r$, we can also find the hyperarea of a
\hbox{ $(\nu\!  -\!  1)$-dimensional} sphere $S_{\nu-1}(r)$ of radius
$r$ in ${\mathbb R}^\nu$,
\begin{eqnarray}
  S_{\nu-1}(r)   
  &=&
  \frac{d B_\nu(r)}{dr}
  =
  \frac{2\pi^{\nu/2}}{\Gamma(\nu/2)}\, r^{\nu-1} 
  =
  \Omega_{\nu-1}\,r^{\nu-1}\ .
\label{Snu}
\end{eqnarray}
For brevity, I have denoted the hyperarea by the same symbol
$S_{\nu-1}(r)$ as the sphere itself, which is simply the \hbox{
$(\nu\!-\!1)$-dimensional} boundary of the region $B_\nu(r)$. This is
illustrated in Fig.~\ref{fig:sphere}.  The distinction I am making
between ``hypervolume'' and ``hyperarea'' is somewhat arbitrary, since
these are both terms involving regions in a higher dimensional
space. When I wish to talk about a \hbox{ $\nu$-dimensional} subregion
of the hyperspace $\mathbb{R}^\nu$, such as $B_\nu(r)$, I will use the
term hypervolume. On the other hand, when I wish to emphasize a
boundary region of a hypervolume, such as $S_{\nu-1}(r)$, I will use
the term ``hyperarea.'' Regarding the usage of the term ``solid
angle,'' suppose we keep the radius $r$ fixed but vary the angles
$\theta_i$ over ranges $d\theta_i$. The region swept out by this
procedure lies on the $(\nu\!-\!1)$-dimensional sphere $S_{\nu-1}(r)$
with a hyperarea $dS_{\nu-1}= d\Omega_{\nu-1}\, r^{\nu-1}$. We are
therefore justified in calling $d\Omega_{\nu-1}$ the solid angle in
$\nu$ dimensions.

\begin{figure}[t]
\includegraphics[scale=0.3]{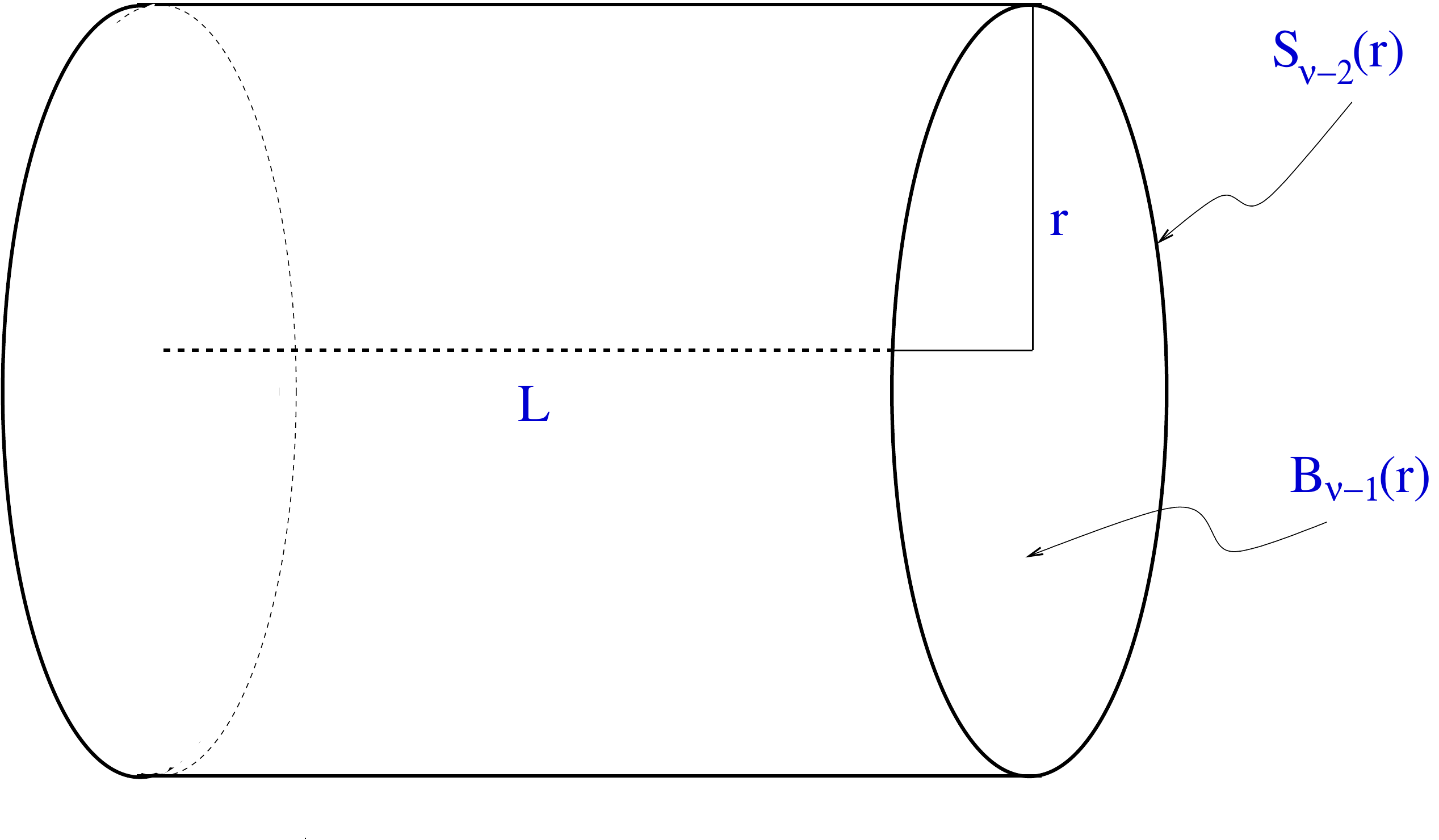}
\caption{\captionskip 
  The hyperarea of a hypercylinder $C_{\nu-1}(r,L)$ of length $L$ and
  radius $r$ is $C_{\nu-1}(r,L) = S_{\nu-2}(r)\cdot L$, and the hypervolume
  bounded by the cylinder is $V_\nu(r,L)= B_{\nu-1}(r)\cdot L$.
}
\label{fig:cylinder}
\end{figure}

Finally, let us discuss the $(\nu\!-\!1)$-dimensional cylindrical
$C_{\nu-1}(r,L)$ of radius $r$ and length $L$. Again, it is easiest to
argue from analogy in three dimensions.  To form a two-cylinder
$C_2(r,L)$ in $\mathbb{R}^3$, we let a two dimensional disk $B_2(r)$
sweep out a volume as it moves a distance $L$ in the orthogonal
direction, as illustrated in Fig.~\ref{fig:cylinder}. Similarly, a 
\hbox{ $(\nu\!-\!1)$-dimensional} cylinder can be formed by letting 
a \hbox{$(\nu\!-\!1)$-dimensional} ball $B_{\nu-1}(r)$ sweep out a 
distance $L$ along the orthogonal axis.  Therefore, the hyperarea 
of the \hbox{ $(\nu\!-\!1)$-dimensional} cylinder is
\begin{eqnarray}
  C_{\nu-1}(r,L) 
  &=&
  S_{\nu-2}(r) \cdot L  \ ,
\label{Cnu}
\end{eqnarray}
and the $\nu$-dimensional hypervolume enclosed by this cylinder is
\begin{eqnarray}
  V_\nu(r,L) 
  &=&
  B_{\nu-1}(r) \cdot L  \ .
\label{BCnu}
\end{eqnarray}
Note that this is the natural geometry of a scattering experiment to 
measure the cross section in $\nu$ dimensions, which leads us to the
next section. 

\pagebreak
\subsection{The Cross Section}
\label{sec_cross_section} 

\begin{figure}[h!]
\includegraphics[scale=0.45]{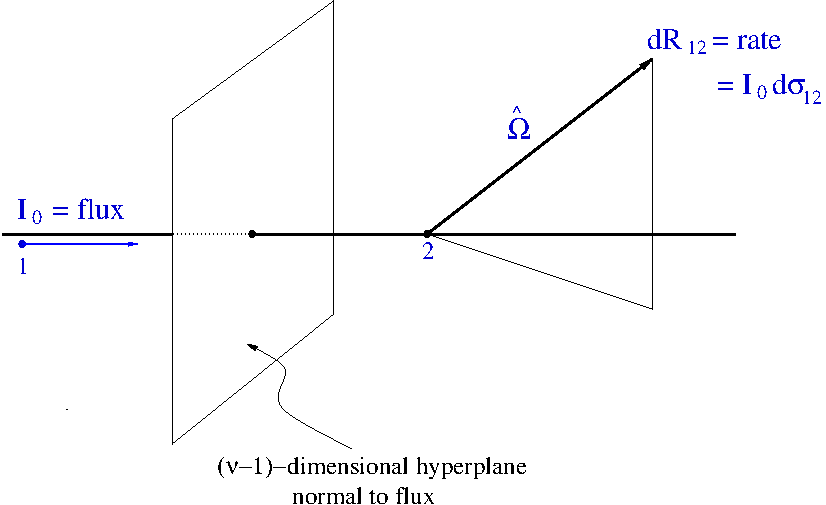}
\caption{\captionskip 
  Definition of the cross section in a general number of
  dimensions. The incident flux $I_0$ of species $1$ is the rate of
  particles per unit hyperarea normal to the beam. The units of $I_0$
  are $\text{L}^{1-\nu} \cdot \text{T}^{-1}$, where $\text{L}$ and
  $\text{T}$ denote the units of space and time. By definition, 
  the differential {\em cross section} $d\sigma_{12}$ is related 
  to the rate $d R_{12}$, each at angular position $\hat\Omega$, 
  by $d R_{12}(\hat\Omega)= I_0 \, d \sigma_{12}(\hat\Omega)$. 
  The cross section per unit solid angle about the direction 
  $\hat\Omega$ is denoted by $d\sigma_{12}/d\Omega$. The 
  engineering units of $d\sigma_{12}$ are $\text{L}^{\nu-1}$.
}
\label{fig:cross}
\end{figure}

Now that we have examined the Coulomb plasma in some
detail,  we should address two-body scattering and the cross section.
This is necessary formalism, since the Boltzmann equation contains 
the differential cross section for two-body scattering. For continuity, 
we review the notion of ``cross section''  in
$\nu$-dimensions.  As illustrated in  Fig.~\ref{fig:cross}, we consider 
a beam of projectiles $1$ with flux $I_0$ striking a fixed target $2$, 
although we can perform a similar analysis in the lab frame in which
the scattering centers are also moving. In $\nu$ dimensions, the 
spatial region normal 
to the beam axis is a $(\nu\!-\!1)$-dimensional hyperplane, and the
flux $I_0$ is the number of particles per second per unit hyperarea
passing through this plane. The engineering units of $I_0$ are therefore 
$L^{1-\nu} \cdot T^{-1}$.  In other words, the number of particles in 
a time interval $dt$ passing through a hyperarea $dA$ normal to the 
beam  is $dN = I_0 \cdot dA\cdot dt$; therefore, the differential rate 
through the normal area $dA$ is $dR = I_0 \cdot dA$. Let us now place 
a particle counter along position $\hat\Omega$ some distance away 
from the scattering center, and let us measure the rate 
$dR_{12}(\hat\Omega)$ at which the $1$-particles enter a given 
solid angle centered about direction $\hat{\Omega}$.  We can 
therefore define the differential cross section $d \sigma_{12}$ 
in the usual way,
\begin{equation}
  d \sigma_{12} \cdot I_0 = dR_{12} \ .
\end{equation}
Note that the engineering of $d \sigma_{12}$ are $L^{\nu-1}$. The
cross section is usually quite sensitive to the details of any given 
physical theory, thereby making  it  a good experimental probe. 
Indeed, in high energy physics, it is the primary diagnostic.

Suppose the scattering center arises from a central potential force, 
such as the \hbox{$\nu$-dimensional} Coulomb field. Then two-body 
particle motion is confined along a two-dimensional plane, and this 
holds true even in $\nu$ dimensions.  Let $b$ denote the impact 
parameter of the projectile relative to the scattering center.  As 
the particle traverses its plane of motion, its position
is uniquely characterized by a function $b=b(\theta)$, where $\theta$ 
is the angle between the beam direction and the projectile (with the 
scattering center defining the origin). The rate at which particles
pass through the hyperannulus of width $db$ and radius $b$ is 
$dR =  \Omega_{\nu-2}\, b^{\nu-2} \, db \cdot I_0$, 
and by particle number conservation, the same number
of scattered particles reaches the hyperannulus around $\hat\Omega$. 
This is the analog of $dR = 2\pi b \, db \cdot I_0$ in three dimensions.
It is actually better to consider a differential $d\Omega_{\nu-2}$
rather than the total angular extent $\Omega_{\nu-2}$. Again, this 
corresponds to $dR = d\theta \, b \,db \cdot I_0$ in three dimensions.
The cross
section in a $\nu$-dimensional central potential is therefore given by
\begin{equation}
  d \sigma_{12} =d\Omega_{\nu -2} \, b^{\nu -2} \,db 
  \ .
\label{eq_dsigma12_rel_b}
\end{equation}
This is the differential form of Eq.~(8.31) of Ref.~\cite{bps}. 
However, for include two-body quantum scattering effects, it is more 
convenient to replace the cross section $d\sigma_{12}$ by the quantum 
scattering amplitude  $T(1+2\to 1^\prime + 2^\prime) \equiv 
T_{1^\prime 2^\prime;\, 12}(W, q^2)$ by using the relation
\begin{eqnarray}
  \vert{\bf v}_1 - {\bf v}_2 \vert\, d\sigma_{12}
  &=&
 \int  
  \frac{d^\nu p_1}{(2\pi\hbar)^\nu}\,
  \frac{d^\nu p_2}{(2\pi\hbar)^\nu} \,
  \big\vert T_{1^\prime 2^\prime ; 1 2}(W, q^2) \big\vert^2
  (2\pi\hbar)^\nu\,\delta^\nu\!\Big( {\bf p}_1^\prime + 
  {\bf p}_2^\prime - {\bf p}_1 - {\bf p}_2 \Big) {\scriptstyle \times}
  \nonumber \\ && \hskip5.8cm
  (2\pi\hbar) \delta\Big(E_1^\prime + E_2^\prime - E_1 - E_2\Big) 
  \ .
\label{dSigmaa}
\end{eqnarray}
In the amplitude, $W$ is the total center-of-mass energy and $q^2$
is the square of the momentum exchange. This is just a rewriting
of the expression $I_0 \cdot d\sigma_{12} = dR_{12}$, since $\vert 
{\bf v}_1 - {\bf v}_2 \vert$ is proportional to the flux $I_0$, and the 
rate $dR_{12}$ is proportional to the square of the scattering amplitude 
$\vert T_{1^\prime 2^\prime;\, 12}\vert^2$. 
The integration is over all values of ${\bf p}_1$ and ${\bf p}_2$
consistent with energy and momentum conservation. In three
dimensions, for example, there are six momentum integrals
and four $\delta$-functions, leaving two differential degrees
of freedom, namely, the cross sectional area. This is why I
continue to use the differential cross section $d\sigma_{12}$ on
the left-hand-side of (\ref{dSigmaa}), to imply that some of the 
angular coordinates have not been integrated over. 

\pagebreak
\section{Center-of-Momentum Coordinates}
\label{sec_CM}

In calculating the convective terms in (\ref{ftwoequilib}), it is
useful to transform to center-of-momentum coordinates. We 
will generalize to two species for the scattering. We define the
 total and relative moment, and the center-of-mass and the 
 relative position by 
\begin{eqnarray}
  {\bf P} &=&m_1 {\bf v}_1 +  m_2 {\bf v}_2 =
  {\bf p}_1 + {\bf p}_2
\\[5pt]
  {\bf p} &=& m_{12}({\bf v}_1 - {\bf v}_2) 
  = \frac{m_2 {\bf p}_1 -  m_1\,{\bf p}_2}{M}
\\[5pt]
  {\bf R} &=& \frac{m_1 {\bf x}_1 + m_2 {\bf x}_2}{M}
\\[5pt]
  {\bf x} &=& {\bf x}_1 - {\bf x}_2
  \ ,
\end{eqnarray}
where
\begin{eqnarray}
 M &=& m_1 + m_2
\\[5pt]
  m_{12} &=& \frac{m_1 m_2}{m_1 + m_2}
  \ .
\end{eqnarray}
I am using the notation ${\bf x}$ for the relative position,
rather than usual notation ${\bf r}$, because in the text, 
the beam-axis in two-body scattering has been called $x$. 
The inverse transforms are
\begin{eqnarray}
  {\bf p}_2 &=& \frac{m_2}{M}\,{\bf P} -  {\bf p}
  \hskip1cm 
  {\bf p}_1 = \frac{m_1}{M}\,{\bf P} + {\bf p}
\\[5pt]
  {\bf x}_2 &=& {\bf R} - \frac{m_1}{M}\, {\bf x}
  \hskip1cm 
  {\bf x}_1 = {\bf R} +\frac{m_2}{M}\, {\bf x}
  \ ,
\end{eqnarray}
and it is easy to check that the gradients transform as,
\begin{eqnarray}
  \frac{\partial}{\partial {\bf p}_2} 
  &=& 
  \frac{\partial}{\partial {\bf P}} - \frac{m_1}{M}\,  \frac{\partial}{\partial {\bf p}} 
  \hskip1cm 
  \frac{\partial}{\partial {\bf p}_1} 
  =
  \frac{\partial}{\partial {\bf P}} + \frac{m_2}{M}\,  \frac{\partial}{\partial {\bf p}} 
\\[8pt]
 \frac{\partial}{\partial {\bf x}_2}  
 &=& 
 \frac{m_2}{M}\,\frac{\partial}{\partial {\bf R}}   - \frac{\partial}{\partial {\bf x}} 
  \hskip1cm
 \frac{\partial}{\partial {\bf x}_1}  
 =
 \frac{m_1}{M}\,\frac{\partial}{\partial {\bf R}}   + \frac{\partial}{\partial {\bf x}} 
 \ ,
\end{eqnarray}
and 
\begin{eqnarray}
  \frac{\partial}{\partial {\bf p}} 
  &=&
  \frac{\partial}{\partial {\bf p}_1} -  \frac{\partial}{\partial {\bf p}_2} 
  \hskip2.3cm
  \frac{\partial}{\partial {\bf P}} 
  =
  \frac{m_1}{M}\, \frac{\partial}{\partial {\bf p}_1} + 
  \frac{m_2}{M}\,  \frac{\partial}{\partial {\bf p}_2} 
  \hskip1cm 
\\[8pt]
 \frac{\partial}{\partial {\bf x}}  
 &=&
 \frac{m_2}{M}\,\frac{\partial}{\partial {\bf x}_1}   - 
 \frac{m_1}{M} \frac{\partial}{\partial {\bf x}_2}
 \hskip1.0cm
 \frac{\partial}{\partial {\bf R}}  
 =
 \frac{\partial}{\partial {\bf x}_1}   +  \frac{\partial}{\partial {\bf x}_2} 
 \ . 
\end{eqnarray}
I have recorded these formulae here for convenience.

\pagebreak
\section{The Multi-component Poisson-Vlasov Equation}

\label{sec_poisson_vlasov}

Let us restore the species index in this section, and consider a
collisionless plasma, 
\begin{eqnarray}
  &&
  \frac{\partial f_a}{\partial t} 
  +
  {\bf v} \cdot \frac{\partial f_a}{\partial {\bf x}}
  +
 e_a {\bf E} \cdot  \frac{\partial f_a}{\partial {\bf p}}
  = 0
  \label{eq_vlasov_f}
  \\[5pt]
  &&
  {\bf E}({\bf x}, t)
  = 
  {\sum}_b\int dX_b \, f_b(X_b, t) \, {\bf E}_{\bf x}^{(b)}
  \ ,
  \label{eq_vlasov_E}
\end{eqnarray}
where ${\bf E}({\bf x}, t)$ is the self-consistent electric field associated
with the $f_b$.  There is a charge $e_a$ at position ${\bf x}$,
and the quantity ${\bf E}_{\bf x}^{(b)}$ inside the integral is the static 
Coulomb field at ${\bf x}$ originating from a point-charge of type $b$ 
at position ${\bf x}_b$, so that
\begin{eqnarray}
  {\bf E}_{\bf x}^{(b)} 
  = 
  {\bf  E}_b({\bf x} - {\bf x}_b)
  =
  e_b \,\frac{\Gamma(\nu/2)}{2 \pi^{\nu/2}}\, \frac{{\bf x} - {\bf x}_b}
  {\vert {\bf x} - {\bf x}_b \vert^\nu}
  \ .
\label{E_xtwo_again}       
\end{eqnarray}
This means that the divergence of the self-consistent electric field is
\begin{eqnarray}
   {\bm\nabla} \cdot {\bf E}({\bf x}, t)
    =
   {\sum}_b \int \frac{d^\nu p}{(2\pi\hbar)^\nu}\,  
   e_b \, f_b({\bf x}, {\bf p}, t) 
   \ .
   \hskip-1.0cm && 
\label{E_first_plasma}       
\end{eqnarray}
To emphasize 
that ${\bf E}$ in (\ref{eq_vlasov_f}) is a functional of the distributions 
$f_b$, we shall often write ${\bf E}[f]$ in place of ${\bf E}({\bf x}, t)$. 
It is actually more convenient to write  (\ref{eq_vlasov_f}) and 
(\ref{E_first_plasma}) in terms of the electric potential $\phi$, 
where ${\bf E}= - \partial \phi/\partial {\bf x}$, so that the kinetic
equations become
\begin{eqnarray}
  &&
  \frac{\partial f_a}{\partial t} 
  +
  {\bf v} \cdot \frac{\partial f_a}{\partial {\bf x}}
  -
 e_a \, \frac{\partial \phi}{\partial {\bf x}} \cdot 
  \frac{\partial f_a}{\partial {\bf p}}
  = 0
  \label{eq_vlasov_f_phi}
  \\[5pt]
  &&
  \nabla^2 \phi({\bf x}, t)
  = 
  -{\sum}_b\int  \frac{d^\nu p}{(2\pi\hbar)^\nu}\,  
  e_b\, f_b(X, t)
  \ . 
  \label{eq_vlasov_E_phi}
\end{eqnarray}
This form of the kinetic equations is known as the Poisson-Vlasov equations, 
and we shall use it interchangeably with (\ref{eq_vlasov_f})--(\ref{E_first_plasma}).   
For visual clarity, I am using a mixed notation in which the Laplacian
is denoted by 
\begin{eqnarray}
 \frac{\partial}{\partial {\bf x}} \cdot  
 \frac{\partial \phi}{\partial {\bf x}} 
 =
  \nabla^2\phi 
 \ .
\end{eqnarray}

Let us now perform a perturbative analysis on the Vlasov 
equation (\ref{eq_vlasov_f}), or equivalently (\ref{eq_vlasov_f_phi}).
Rather than perturbing about an
equilibrium configuration, let us take the 0-th order starting
point as a solution to the Vlasov equation itself. In other words,
suppose $f_{a1}$ is a solution to 
\begin{eqnarray}
    \frac{\partial f_{a1}(X, t)}{\partial t} 
  +
  {\bf v} \cdot \frac{\partial f_{a1}(X,t)}{\partial {\bf x}} 
  +
    e_a \, {\bf E}^0
  \cdot \frac{\partial f_{a1}(X,t)}{\partial {\bf p}} 
  = 0
  \ ,
  \label{eq_self_fazero}
\end{eqnarray}
where the 0-th order self-consistence electric field at ${\bf x}$  is
\begin{eqnarray}
  {\bf E}^0({\bf x},t)
  =
  {\sum}_b
   \int dX_b \, f_{b1}({\bf x}_b ,{\bf p}_b, t) \, {\bf E}_b({\bf x} - {\bf x}_{b})
  \ .
  \label{eq_self_Eone}
\end{eqnarray}
Now suppose that the perturbation
\begin{eqnarray}
   f_a({\bf x}, {\bf p}, t) = f_{a 1}({\bf x}, {\bf p},t) + h_a({\bf x}, {\bf p}, t) 
  \label{f_h_pert}
\end{eqnarray}
satisfies the Vlasov equation (\ref{eq_vlasov_f}), and let us find the
corresponding equation satisfied by $h_a$. Upon substituting 
(\ref{f_h_pert}) for $h_a$ into the electric field (\ref{eq_vlasov_E}), 
the self-consistent electric field receives a 0-th order contribution 
from $f_{a1}$ and a 1-st order contribution from $h_a$, 
\begin{eqnarray}
  {\bf E}({\bf x},t)
  &=& 
  {\sum}_b
  \int dX_b \,
  %\frac{d^\nu p_b}{(2\pi\hbar)^\nu} \int d^\nu x_b \, 
  \Big[
  f_{b1}({\bf x}_b, {\bf p}_b,t) + h_b({\bf x}_b, {\bf p}_b, t)  
  \Big] 
  {\bf E}_b({\bf x} - {\bf x}_b)
  \nonumber\\[5pt]
  &=&
  {\bf E}^0({\bf x},t) + {\bf E}^1({\bf x},t)
  \ ,
\label{eq_E_first}
\end{eqnarray}
where the first-order self-consistent field is
\begin{eqnarray}
 {\bf E}^1({\bf x},t)
  &=& 
  {\sum}_b
  \int dX_b \,
  h_b({\bf x}_b, {\bf p}_b, t)  \,  {\bf E}_b({\bf x} - {\bf x}_b)
  \ .
  \label{eq_E_pert}
\end{eqnarray}
We also substitute the perturbation (\ref{f_h_pert}) back into the 
kinetic equation  (\ref{eq_vlasov_f}), using the electric field (\ref{eq_E_first}) 
and working only to first order in $h_a$. The 0-th order terms vanish 
because $f_{a1}$ is a solution to equation (\ref{eq_self_fazero}), 
and after some algebra we find that the perturbation satisfies
\begin{eqnarray}
  &&
  \frac{\partial h_a}{\partial t} 
  +
  {\bf v} \cdot \frac{\partial h_a}{\partial {\bf x}}
  +
  e_a {\bf E}^0[f_1] \cdot  \frac{\partial h_a}{\partial {\bf p}}
  +
 e_a {\bf E}^1[h] \cdot  \frac{\partial f_{a1}}{\partial {\bf p}}
  = 0
  \ .
  \label{eq_vlasov_h_first}
\end{eqnarray}
It should be reiterated that we have dropped the second-order term 
${\bf E}^1[h] \cdot (\partial h_a/\partial {\bf p}_a)$. We can also write 
(\ref{eq_vlasov_h_first}) in the form
\begin{eqnarray}
  &&
  \frac{\partial h_a}{\partial t} 
  +
  V h_a 
  = 0
  \ ,
  \label{eq_vlasov_h_first_V}
\end{eqnarray}
where the operator $V$ is defined by
\begin{eqnarray}
  V h_a
  &=&
  {\bf v} \cdot \frac{\partial h_a}{\partial {\bf x}}
  +
  e_a {\bf E}^0[f_1] \cdot  \frac{\partial h_a}{\partial {\bf p}}
  +
 e_a {\bf E}^1[h] \cdot  \frac{\partial f_{a1}}{\partial {\bf p}}
   \label{eq_vlasov_h_first_Vha_Eself}
 \\[8pt]
 &=&
   {\bf v} \cdot \frac{\partial h_a}{\partial {\bf x}}
  +
  e_a {\sum}_b \bigg[
   \int dX_b \, f_{b1}(X_b, t) \, {\bf E}_{\bf x}^{(b)} \cdot  
   \frac{\partial h_a}{\partial {\bf p}}
  +
  \int dX_b \, h_b(X_b, t) \, {\bf E}_{\bf x}^{(b)}\cdot  
  \frac{\partial f_{a1}}{\partial {\bf p}}
 \bigg]
 \ .
 \nonumber\\
  \label{eq_vlasov_h_first_Vha}
\end{eqnarray}
From Bogoliubov's hypothesis, the time dependence of $f_{a1}$ 
is much slower than that of $h_a$, and we can regard the operator 
$V$ as constant in time as far as its action on a perturbation $h_a$ 
is concerned. Note that equations  (\ref{eq_vlasov_h_first_Vha_Eself}) 
and (\ref{eq_vlasov_h_first_Vha}) serve a definition of the operator 
$V$ on {\em any} function $h_a(X,t)$, whether $h_a$ is the 
perturbation or not. 

As in Section~VI, to solve (\ref{eq_vlasov_h_first_V}) we take the 
temporal Laplace transform and the spatial Fourier transform,
\begin{eqnarray}
  (p + V) \tilde h({\bf k}, {\bf p}, p)
  =
  \tilde h({\bf k}, {\bf p}, 0)
  \ ,
\end{eqnarray}
giving the formal solution
\begin{eqnarray}
  \tilde h({\bf k}, {\bf p}, p)
  =
  (p + V)^{-1} \, \tilde h({\bf k}, {\bf p}, 0)
  \ .
\label{eq_lp_pV_inverse_formal}
\end{eqnarray}
We now preform the inversion (\ref{eq_lp_pV_inverse_formal}) for 
a specific example in which the unperturbed plasma is a function of 
momentum only, $f_{a1}=  f_{a}({\bf p})$. 
%Under this condition,
%note that the operator $V$ defined in (\ref{eq_vlasov_h_first_Vha_Eself}) 
%is  a constant in time without invoking Bogoliubov's hypothesis. Also 
%note that an arbitrary function of momentum $ f_a({\bf p})$ is 
%automatically a solution to (\ref{eq_self_fazero}), so that $f_{a1}$ 
%need not be a Maxwell-Boltzmann distribution. 
We then see that 
the 0-th order self-consistent field vanishes,
\begin{eqnarray}
  {\bf E}^0({\bf x},t)
  =
  {\sum}_b
   \int \frac{d^\nu p_b}{(2\pi\hbar)^\nu} \, 
     f_{b}({\bf p}_b) \, 
    \int d^\nu x_b \, {\bf E}_b({\bf x}_{a} - {\bf x}_{b})
  =
  0
  \ ,
  \label{eq_self_Eone_zero}
\end{eqnarray}
since the electric field
\begin{eqnarray}
  {\bf E}_b({\bf x}_{a} - {\bf x}_{b})
  =
  -\frac{\partial \phi({\bf x})}{\partial{\bf x}}
  \bigg\vert_{{\bf x}={\bf x}_a - {\bf x}_b} \!\!
  =
  ~\frac{\partial \phi({\bf x}_a - {\bf x}_b)}{\partial{\bf x}_b}
\end{eqnarray}
is a total derivative of the potential $\phi({\bf x})$. The operator $V$ 
now take the form 
\begin{eqnarray}
  V h_a
  &=&
  {\bf v} \cdot \frac{\partial h_a}{\partial {\bf x}}
  +
 e_a {\bf E} \cdot  \frac{\partial  f_{a}}{\partial {\bf p}}
 \label{Vha_rewrite}
 \\[5pt]
 &=&
   {\bf v} \cdot \frac{\partial h_a}{\partial {\bf x}}
  +
  e_a {\sum}_b 
  \int dX_b \, h_b({\bf x}_b, {\bf p}_b, t) \, {\bf E}_b({\bf x}-{\bf x}_b)\cdot 
   \frac{\partial  f_a({\bf p})}{\partial {\bf p}}
 \ .
  \label{eq_vlasov_h_first_Vha_constf}
\end{eqnarray}
Since ${\bf E}^0$ vanishes,  the first-order field ${\bf E}^1$ defined
in (\ref{eq_E_pert}) is the total self-consistent electric field, and I have 
therefore dropped the 1-superscript. 
Upon taking the temporal-Laplace 
transform and the spatial-Fourier transform of (\ref{eq_vlasov_h_first_V}) 
and (\ref{Vha_rewrite}), we find
\begin{eqnarray}
  p \tilde h_a + i {\bf k}\cdot{\bf v}_a \,\tilde h_a 
  +
  e_a \tilde{\bf E } \cdot  \frac{\partial  f_a}{\partial {\bf p}} 
  = 
  \tilde h_a( 0)
  \label{eq_phaeqnow}
  \ ,
\end{eqnarray}
where the Fourier transform $\tilde{\bf E}$ can be calculated
from (\ref{eq_vlasov_h_first_Vha_constf}) by the convolution theorem,
\begin{eqnarray}
 \tilde {\bf E}({\bf k},p)
  &=& 
 {\sum}_b \int\frac{d^\nu p_b}{(2\pi\hbar)^\nu} \
  \tilde h_b({\bf k}, {\bf p}_b, p)  \,  \tilde{\bf E}_b({\bf k})
  \ .
  \label{eq_Ekp_def}
\end{eqnarray}
By way of notation, the tilde over a function is used to denote
both Laplace and Fourier transforms, so care must be taken
when interpreting such terms. In other words, we take $\tilde h_a = 
\tilde h_a({\bf k}, {\bf p},  p)$, $\tilde {\bf E}=\tilde {\bf E}({\bf k}, p)$, 
and $\tilde h_a(0) = \tilde h_b({\bf k}, {\bf p}, t=0)$. That is to say, 
$\tilde h_a(0)$ is the spatial Fourier transform of $h_a({\bf x}, {\bf p}, t)$ 
evaluated at $t=0$, while $\tilde h_a$ and $\tilde {\bf E}$ are spatial 
Fourier transforms and temporal Laplace transforms. We can now 
solve (\ref{eq_phaeqnow}) for the perturbation, giving
\begin{eqnarray}
\label{f1_sol_laplace}
 \tilde h_a({\bf k}, {\bf p},p)
  &=&
  (p + V)^{-1}\,\tilde h_a({\bf k}, {\bf p}, 0)
  \\[8pt]
  &=&
 \frac{1}{p + i {\bf k}\cdot{\bf v}_a}\bigg[\tilde h_a({\bf k}, {\bf p}, 0)
  - 
  e_a \tilde{\bf E }({\bf k}, p) \cdot  \frac{\partial  f_a({\bf p})}{\partial {\bf p}}
 \bigg]
 \ .
\end{eqnarray}
Recall that $\tilde{\bf E}_b({\bf k})$ is the Fourier transform of 
the point Coulomb field ${\bf E}_b({\bf x})$, and can therefore 
be written in terms of the Fourier transform of the potential 
$\tilde\phi_b({\bf k})$ as
\begin{eqnarray}
  \tilde {\bf E}_b({\bf k}) = - i {\bf k}\,\tilde \phi_b({\bf k}) 
  =  - i {\bf k}\, \frac{e_b}{k^2}
  \ .
\end{eqnarray}
In like manner, the self-consistent electric field $\tilde {\bf E}({\bf k},p)$ 
can be expressed in terms of a self-consistent potential $ \tilde \phi({\bf k}, p)$ 
defined by
\begin{eqnarray}
 \tilde {\bf E}({\bf k},p) = -i {\bf k} \, \tilde \phi({\bf k}, p)
 \ .
 \label{eq_tildE_tildephi}
\end{eqnarray}
We can now write  (\ref{f1_sol_laplace}) and (\ref{eq_Ekp_def}) in
the Poisson-Vlasov form
\begin{eqnarray}
 \tilde h_a({\bf k}, {\bf p},p)
  &=&
 \frac{1}{p + i {\bf k}\cdot{\bf v}_a}\bigg[\tilde h_a({\bf k}, {\bf p}, 0)
  +
  e_a \tilde\phi({\bf k}, p)
  (i {\bf k}) \cdot  \frac{\partial  f_a({\bf p})}{\partial {\bf p}}
 \bigg]
\label{f1_sol_laplace_phi}
\\[8pt]
 \tilde\phi({\bf k},p)
  &=& 
   {\sum}_b \, \frac{e_b}{k^2}\int\frac{d^\nu p_b}{(2\pi\hbar)^\nu} \
  \tilde h_b({\bf k}, {\bf p}_b, p)  
    \ .
  \label{eq_Ekp_def_phi}
\end{eqnarray}
Let us substitute (\ref{f1_sol_laplace_phi}) for the perturbation
$\tilde h_a$ into (\ref{eq_Ekp_def_phi}) for the potential, thereby 
giving
\begin{eqnarray}
  \tilde \phi({\bf k}, p)
  &=& 
  {\sum}_b \,
  \frac{e_b}{k^2}\int\frac{d^\nu p_b}{(2\pi\hbar)^\nu} \,
  \frac{1}{ p +  i {\bf k} \cdot {\bf v}_b }
  \left[
  \tilde h_b({\bf k}, {\bf p}_b, 0)
  +
 e_b \,\tilde \phi({\bf k}, p) \, i {\bf k}  \cdot  \frac{\partial 
  f_b({\bf p}_b)}{\partial {\bf p}_b}
  \right]
  \ .
  \nonumber\\
  \label{eq_E_p_new_a}
\end{eqnarray}
Note that $\tilde\phi({\bf k}, p)$ appears on both sides of this equation,
and upon isolating the $\tilde\phi({\bf k}, p)$ term, we find
\begin{eqnarray}
 \Bigg[
 1  -  {\sum}_b \,
  \frac{e_b^2}{k^2} \int\frac{d^\nu p_b}{(2\pi\hbar)^\nu} \,
  \frac{1}{ p +  i {\bf k} \cdot {\bf v}_b }
  i {\bf k}  \cdot  \frac{\partial  f_b({\bf p}_b)}{\partial {\bf p}_b}
  \Bigg]  \tilde \phi({\bf k}, p)
  &=& 
  {\sum}_b \, \frac{e_b}{k^2} \int\frac{d^\nu p_b}{(2\pi\hbar)^\nu} \,
  \frac{\tilde h_b({\bf k}, {\bf p}_b, 0)}{ p +  i {\bf k} \cdot {\bf v}_b }
  \ .
  \nonumber\\
  \label{eq_E_p_new_a}
\end{eqnarray}
Solving for the self-consistent potential gives
\begin{eqnarray}
  \tilde \phi({\bf k}, p)
  &=& 
  {\sum}_b ~
  \frac{e_b}{\bar\epsilon({\bf k}, p) \,k^2} 
  \int\frac{d^\nu p_b}{(2\pi\hbar)^\nu} \,
  \frac{\tilde h_b({\bf k}, {\bf p}_b, 0)}{ p +  i {\bf k} \cdot {\bf v}_b } 
  \ ,
  \label{eq_E_p_new_c}
\end{eqnarray}
where the ``dielectric function'' in Laplace space is defined by 
\begin{eqnarray}
  \bar\epsilon({\bf k}, p)
  =
 1 - {\sum}_b \,
  \int\frac{d^\nu p_b}{(2\pi\hbar)^\nu} \,  \frac{e_b^2}{k^2} \,
  \frac{1}{ p +  i {\bf k} \cdot {\bf v}_b } \, 
  i {\bf k}  \cdot  \frac{\partial  f_b({\bf p}_b)}{\partial {\bf p}_b}
  \ ,
 \label{eq_E_p_new_b}
\end{eqnarray}
with $p$ lying on the contour $C$. In fact, we can analytically continue 
(\ref{eq_E_p_new_b}), and allow $p$ to lie anywhere in the complex plane
to the right of $C$.
%that is to say, ${\rm Re}\,p \ge p_0$, where $p_0$ is the real coordinate 
%along $C$ (recall that $C$ is parallel to the imaginary axis). 
For future reference, we record the following identities:
\begin{eqnarray}
  {\sum}_b\int \frac{d^\nu p_b}{(2\pi\hbar)^\nu} \,
  \frac{e_b^2}{k^2}\,
  \frac{ i {\bf  k} \cdot \partial  f_b/
  \partial {\bf p}_b}{p + i {\bf k} \cdot {\bf v}_b} \,
  &=&
  1 - \bar\epsilon({\bf k}, p)
\label{eq_formOne}
\\[11pt]
  {\sum}_b\int \frac{d^\nu p_b}{(2\pi\hbar)^\nu} \,  
  \frac{e_b^2}{k^2}\,
  \frac{ i {\bf  k} \cdot 
  \partial  f_b/\partial {\bf p}_b  }{p - i {\bf k} \cdot {\bf v}_b} \,
  &=&
  \bar\epsilon(-{\bf k}, p) - 1
  \ .
\label{eq_formTwo}
\end{eqnarray}

\vskip0.2cm
\noindent
Upon substituting (\ref{eq_E_p_new_c}) back into  (\ref{f1_sol_laplace_phi})
we find the inverse
\begin{eqnarray}
  \label{eq_Vinverse}
  && \hskip-1.0cm
  (p + V)^{-1} \, \tilde h_a({\bf k}, {\bf p}, 0)
  \equiv
  \tilde h_a({\bf k}, {\bf p}, p)
  \\[8pt]
  &=&
  \frac{1}{ p +  i {\bf k} \cdot {\bf v} }
  \left[
  \tilde h({\bf k}, {\bf p}, 0)
  +
  {\sum}_b \,
  \frac{e_a e_b}{\bar\epsilon({\bf k}, p)\, k^2}   \, \,
  i {\bf k} \cdot  \frac{\partial  f_a({\bf p})}{\partial {\bf p}}
   \int\frac{d^\nu p^\prime}{(2\pi\hbar)^\nu} \,
  \frac{\tilde h_b({\bf k}, {\bf p}^\prime, 0)}{ p +  i {\bf k} \cdot {\bf v}^\prime }
  \right]
  \nonumber
  \ .
\end{eqnarray}
We have use expressions (\ref{eq_formOne}),  (\ref{eq_formTwo}),
and  (\ref{eq_Vinverse}) in Section~\ref{sec_LBE_alg} in their
single-component forms. This analysis shows that the results
in Section~\ref{sec_LBE_alg} also hold for a multi-component 
plasma.

Let us pause now to understand these results physically. We see 
from (\ref{epsilonmulti})  that the dielectric function 
$\epsilon({\bf k},\omega)$ can be analytically continued to complex 
values of $\omega$, thereby taking the form

\begin{equation}
  \epsilon({\bf k},\omega) 
  = 
  1 + {\sum}_b \, \int \! \frac{d^\nu p}{(2\pi \hbar)^\nu}  \,
  \frac{e_b^2}{k^2} \,
  \frac{1}{\omega - {\bf k}\!\cdot\!{\bf v}_b}\, 
  {\bf k} \cdot  \frac{\partial  f_b}{\partial {\bf p} } 
  \ ,
\label{f1_sol_e}
\end{equation}

\vskip0.2cm
\noindent
where ${\rm Re}\,\omega > 0$.  The quantity $\bar\epsilon({\bf k}, p)$
in (\ref{eq_E_p_new_b}) is just the analytically continued dielectric 
function $\epsilon({\bf k},\omega)$ to a complex frequency $\omega 
= i p$ (for real $p$),\footnote{
  \footnoteskip
  We should actually set $\omega = i p + \eta$ with $\eta > 0$, so that
  ${\rm Re}\, \omega > 0$. We can then take the limit $\eta \to 0^+$,
  which gives~(\ref{eq_epsilon_cont}). 
} % footnote
and we see that
\begin{equation}
 \epsilon({\bf k}, \omega= ip) = \bar\epsilon({\bf k}, p)
  \ ,
\label{eq_epsilon_cont}
\end{equation}
or equivalently, $\bar\epsilon({\bf k}, p=-i\omega)=\epsilon({\bf k}, \omega)$. 
We can also analytically continue
the self-consistent potential (\ref{eq_E_p_new_c}) to Fourier space by 
setting $p=-i\omega + \eta$, where $\omega$ is real and $\eta > 0$.
Upon taking the limit $\eta \to 0^+$, we find

\begin{eqnarray}
  \tilde\phi({\bf k}, \omega)
 = 
   {\sum}_b \, \frac{e_b}{\epsilon({\bf k}, \omega)\, k^2}
   \int \frac{d^\nu p_b}{(2\pi\hbar)^\nu}\,
   \frac{i \,\tilde h_b({\bf k}, {\bf p}_b, 0)}{\omega - {\bf k}\cdot{\bf v}_b + i \eta}
  \ .
\label{f1_sol_h}
\end{eqnarray}

\vskip0.2cm
\noindent
Note that the $k^2$ term in the denominator of the self-consistent potential 
(\ref{f1_sol_h}) is accompanied by a factor of $\epsilon({\bf k}, \omega)$
relative to  the static Coulomb potential of a point charge, $\tilde \phi_a({\bf k})
=e_a/k^2$. This means that the self-consistent field is accompanied by Landau 
screening,  and in fact we could write the equations using only the 
{\em screened potential}, 

\begin{eqnarray}
  \tilde\phi^{\,{\rm landau}}_a({\bf k}, \omega)
  =
  \frac{e_a}{\epsilon({\bf k}, \omega)\, k^2}
  \ .
\end{eqnarray}
We choose, however, to keep the factors of $\epsilon({\bf k},\omega)$ 
explicit. The Fourier components of the electric field are determined 
by $\tilde{\bf E}({\bf k},\omega) = -i {\bf k} \, \tilde \phi({\bf k},\omega)$, 
so that

\begin{eqnarray}
  \tilde{\bf E}({\bf k}, \omega)
 = 
  {\sum}_b \, 
   \frac{{\bf k}}{\epsilon({\bf k}, \omega)} \, \frac{e_b}{k^2}
   \int \frac{d^\nu p_b}{(2\pi\hbar)^\nu}\,
   \frac{\tilde h_b({\bf k}, {\bf p}_b, 0)}{\omega - {\bf k}\cdot{\bf v}_b + i \eta}
  \ ,
\label{eq_Ek_omega}
\end{eqnarray}
where ${\bf v}_b = {\bf p}_b/m_b$. If we wanted to find the electric
field ${\bf E}({\bf x}, t)$ as a function of space and time, it is more 
convenient to revert back to Laplace space by setting $p = -i \omega$
in (\ref{eq_Ek_omega}), 

\begin{eqnarray}
  \tilde{\bf E}({\bf k}, p)
 = 
  {\sum}_b \, 
   \frac{-i{\bf k}}{\bar\epsilon({\bf k}, p)} \, \frac{e_b}{k^2}
   \int \frac{d^\nu p_b}{(2\pi\hbar)^\nu}\,
   \frac{\tilde h_b({\bf k}, {\bf p}_b, 0)}{p + i {\bf k}\cdot{\bf v}_b}
  \ ,
%\label{f1_sol_h}
\end{eqnarray}
and then taking the inverse Laplace transform. We will, however, 
not work through this algebra, and remain content to have found
the electric field and the perturbation in Laplace and Fourier space. 
Chapter~24 of Ref.~\cite{pl_phys} does a good job of finding 
${\bf E}({\bf x}, t)$ in various physical cases of interest.

\pagebreak

\end{document}